\documentclass[11pt]{article}
\pdfoutput=1

\usepackage{amssymb}
\usepackage{amsmath}
\usepackage{bbold}
\usepackage{color}
\usepackage{dsfont}
\usepackage{dsfont}
\usepackage{colordvi}
\usepackage{fancybox}
\usepackage[footnotesize]{caption2}
\usepackage{graphicx}
\usepackage[center,footnotesize,hang]{subfigure}
\usepackage{bbm}
\usepackage{url}
\usepackage{multirow}
\usepackage{array}
\usepackage{arydshln}
\usepackage{pifont}
\usepackage{mathrsfs}
\usepackage{mathtools}
\usepackage{fancybox}
\usepackage[colorlinks]{hyperref}
\usepackage{cite}
\usepackage{enumitem}
\usepackage{float}
\newcommand{\PreserveBackslash}[1]{\let\temp=\\#1\let\\=\temp}
\newcolumntype{C}[1]{>{\PreserveBackslash\centering}p{#1}}
\newcolumntype{R}[1]{>{\PreserveBackslash\raggedleft}p{#1}}
\newcolumntype{L}[1]{>{\PreserveBackslash\raggedright}p{#1}}
\newcommand{\xmark}{\ding{55}}

\allowdisplaybreaks \allowdisplaybreaks[2]

\makeatletter
\@addtoreset{equation}{section}
\makeatother

\begin{document}

\title{
\begin{flushright}
\ \hfill\mbox{\small USTC-ICTS-17-01}\\[5mm]
\begin{minipage}{0.2\linewidth}
\normalsize
\end{minipage}
\end{flushright}
%{\Large \bf Implications of residual CP symmetry for leptogenesis in two %right-handed neutrino model \\[2mm]}}
{\Large \bf Implications of residual CP symmetry for leptogenesis in a model with two right-handed neutrinos \\[2mm]}}

\date{}

\author{
Cai-Chang Li~\footnote{E-mail: {\tt
lcc0915@mail.ustc.edu.cn}},  \
Gui-Jun Ding~\footnote{E-mail: {\tt
dinggj@ustc.edu.cn}}  \
\\*[20pt]
\centerline{
\begin{minipage}{\linewidth}
\begin{center}
\it{\small Interdisciplinary Center for Theoretical Study and  Department of Modern Physics, }\\
\it{\small University of Science and
    Technology of China, Hefei, Anhui 230026, China}\\[4mm]
\end{center}
\end{minipage}}
\\[10mm]}
\maketitle
\thispagestyle{empty}

\begin{abstract}
\noindent

We analyze the interplay between leptogenesis and residual symmetry in the framework of two right-handed neutrino model. Working in the flavor basis, we show that all the leptogenesis CP asymmetries are vanishing for the case of two residual CP transformations or a cyclic residual flavor symmetry in the neutrino sector. If a single remnant CP transformation is preserved in the neutrino sector, the lepton mixing matrix is determined up to a real orthogonal matrix multiplied from the right side. The $R$-matrix is found to depend on only one real parameter, it can take three viable forms, and each entry is either real or purely imaginary. The baryon asymmetry is generated entirely by the CP violating phases in the mixing matrix in this scenario. We perform a comprehensive study for the $\Delta(6n^2)$ flavor group and CP symmetry which are broken to a single remnant CP transformation in the neutrino sector and an abelian subgroup in the charged lepton sector. The results for lepton flavor mixing and leptogenesis are presented.

\end{abstract}
\newpage

\section{\label{sec:Int}Introduction}

A large amount of experiments with solar, atmospheric, reactor and accelerator neutrinos have provided compelling evidence for oscillations of neutrinos caused by nonzero neutrino masses and neutrino mixing~\cite{Kajita:2016cak,McDonald:2016ixn,nobel_NPB}. Both three flavor neutrino and antineutrino oscillations can be described by three lepton mixing angles $\theta_{12}$, $\theta_{13}$ and $\theta_{23}$, one leptonic Dirac CP violating phase $\delta$, and two independent mass-squared splittings $\delta m^2\equiv m^2_2-m^2_1>0$ and $\Delta m^2\equiv m^2_3-(m^2_1+m^2_2)/2$, where $m_{1,2,3}$ are the three neutrino masses, $\Delta m^2>0$ and $\Delta m^2<0$ correspond to normal ordering (NO) and inverted ordering (IO) mass spectrum respectively. All these mixing parameters except $\delta$ have been measured with good accuracy~\cite{Capozzi:2013csa,Forero:2014bxa,Gonzalez-Garcia:2014bfa,Capozzi:2016rtj,Esteban:2016qun}, the experimentally allowed regions at $3\sigma$ confidence level (taken from Ref.~\cite{Capozzi:2013csa}) are:
\begin{eqnarray}
\nonumber 0.259\leq&\sin^2\theta_{12}&\leq0.359,\\
\nonumber 1.76(1.78)\times 10^{-2}\leq&\sin^3\theta_{13}&\leq2.95(2.98)\times 10^{-2},\\
\nonumber0.374(0.380)\leq&\sin^2\theta_{23}&\leq0.626(0.641),\\
\nonumber 6.99\times 10^{-5} \text{eV}^2\leq&\delta m^2&\leq8.18\times 10^{-5} \text{eV}^2,\\
2.23(-2.56)\times 10^{-3} \text{eV}^2\leq&\Delta m^2&\leq2.61(-2.19)\times 10^{-3} \text{eV}^2
\end{eqnarray}
for NO (IO) neutrino mass spectrum. At present, both T2K~\cite{Abe:2015awa,T2K_delta_CP,Abe:2017uxa} and NO$\nu$A~\cite{NovA_delta_CP,Adamson:2017gxd} report a weak evidence for a nearly maximal CP violating phase $\delta\sim-\pi/2$, and hits of $\delta\sim-\pi/2$ also show up in the global fit of neutrino oscillation data~\cite{Capozzi:2013csa,Forero:2014bxa,Gonzalez-Garcia:2014bfa,Capozzi:2016rtj,Esteban:2016qun}. Moreover, several experiments are being planned to look for CP violation in neutrino oscillation, including long-baseline facilities, superbeams, and neutrino factories. The above structure of lepton mixing, so different from the the small mixing in the quark sector, provides a great theoretical challenge. The idea of flavor symmetry has been extensively exploited to provide a realistic description of the lepton masses and mixing angles. The finite discrete non-abelian flavor symmetries have been found to be particularly interesting as they can naturally lead to certain mixing patterns~\cite{Lam:2007qc}, please see Refs.~\cite{Altarelli:2010gt,Ishimori:2010au,King:2013eh} for review.

Although the available data are not yet able to determine the individual neutrino mass $m_i$, the neutrino masses are known to be of order eV from tritium endpoint, neutrinoless double beta decay and cosmological data. The smallness of neutrino masses can be well explained within the see-saw mechanism~\cite{Minkowski:1977sc}, in which the Standard Model (SM) is extended by adding new heavy states. The light
neutrino masses are generically suppressed by the large masses of the new states. In type I seesaw model~\cite{Minkowski:1977sc} the extra states are right-handed (RH) neutrinos which have Majorana masses much larger than the electroweak scale, unlike the standard model fermions which acquire mass proportional to electroweak symmetry breaking. Apart from elegantly explaining the tiny neutrino masses, the seesaw mechanism provides a simple and attractive explanation for the observed baryon asymmetry of the Universe, one of the most longstanding cosmological puzzles. The CP violating decays of heavy RH neutrinos can produce a lepton asymmetry in the early universe, which is then converted into a baryon asymmetry through $B+L$ violating anomalous sphaleron processes at the electroweak scale. This is the so-called leptogenesis mechanism~\cite{Fukugita:1986hr}.

It is well-known that in the paradigm of the unflavored thermal leptogenesis the CP phases in the neutrino Yukawa couplings in general are not related to the the low energy leptonic CP violating parameters (i.e. Dirac and Majorana phases) in the mixing matrix. However, the low energy CP phases could play a crucial role in the flavored thermal leptogenesis~\cite{flavored_leptogenesis} in which the flavors of the charged leptons produced in the heavy RH neutrino decays are relevant. In models with flavor symmetry, the total number of free parameters is greatly reduced, therefore the observed baryon asymmetry could possibly be related to other observable quantities~\cite{Jenkins:2008rb_leptogenesis}. In general, the leptogenesis CP asymmetries would vanish if a Klein subgroup of the flavor symmetry group is preserved in the neutrino sector~\cite{Chen:2016ptr}.

Recent studies show that the extension of discrete flavor symmetry to include CP symmetry is a very predictive framework~\cite{Feruglio:2012cw,Ding:2013bpa,Li:2015jxa,Branco:2015hea,Ding:2013nsa,King:2014rwa,Ding:2014ssa,Hagedorn:2014wha,Ding:2014ora,Ding:2015rwa,Yao:2016zev,Lu:2016jit,Chen:2014wxa}. If the given flavor and CP symmetries are broken to an abelian subgroup and $Z_2\times CP$ in the charged lepton and neutrino sectors respectively, the resulting lepton mixing matrix would be determined in terms of a free parameter $\theta$ whose value can be fixed by the reactor angle $\theta_{13}$. Hence all the lepton mixing angles, Dirac CP violating phase and Majorana CP phases can be predicted~\cite{Chen:2014wxa}. Moreover, other phenomena involving CP phases such as neutrinoless double beta decay and leptogenesis are also strongly constrained in this approach~\cite{Chen:2016ptr,Yao:2016zev,Hagedorn:2016lva}. In fact, we find that the leptogenesis CP asymmetries are exclusively due to the Dirac and Majorana CP phases in the lepton mixing matrix, and the $R$-matrix depends on only a single real parameter in this scenario~\cite{Chen:2016ptr}.

In this paper we shall extend upon the work of~\cite{Chen:2016ptr} in which the SM is extended to introduce three RH neutrinos. Here we shall study the interplay between residual symmetry and leptogenesis in seesaw model with two RH neutrinos. We find that all the leptogenesis CP asymmetries would be exactly vanishing if two residual CP transformations or a cyclic residual flavor symmetry are preserved by the seesaw Lagrangian. On the other hand, if only one remnant CP transformation is preserved in the neutrino sector,
all mixing angles and CP phases are then fixed in terms of three real parameters $\theta_{1,2,3}$ which can take values between 0 and $\pi$, and the $R$-matrix would be constrained to depend on only one free parameter. The total CP asymmetry $\epsilon_1\equiv\epsilon_{e}+\epsilon_{\mu}+\epsilon_{\tau}$ in leptogenesis is predicted to be zero. Hence our discussion will be entirely devoted to the flavored thermal leptogenesis scenario in which the lightest RH neutrino mass is typically in the interval of $10^9$ GeV $\leq M_{1}\leq10^{12}$ GeV. Our approach is quite general and it is independent of the explicit form of the residual symmetries and how the vacuum alignment achieving the residual symmetries is dynamically realized. In order to show concrete examples, we apply this general formalism to the flavor group $\Delta(6n^2)$ combined with CP symmetry which is broken down to an abelian subgroup in the charged lepton sector and a remnant CP transformation in the neutrino sector. The expressions for lepton mixing matrix as well as mixing parameters in each possible cases are presented. We find that for small values of the flavor group index $n$, the experimental data on lepton mixing angles can be accommodated for certain values of the parameters $\theta_{1,2,3}$. The corresponding predictions for the cosmological matter-antimatter asymmetry are discussed.

The rest of the paper is organized as follows. In section~\ref{sec:framework} we briefly review some generic aspects of
leptogenesis in two RH model and present some analytic approximations
which will be used later. In section~\ref{sec:LepG_one_CP} we study the scenario that one residual CP transformation is preserved in the neutrino sector. The lepton mixing matrix is determined up to an arbitrary real orthogonal matrix multiplied from the right hand side. The $R$-matrix contains only one free parameter, and each element is either real or purely imaginary. The total CP asymmetry $\epsilon_1$ is vanishing, consequently the unflavored leptogenesis is not feasible unless subleading corrections are taken into account. The scenario of two remnant CP transformations or a cyclic residual flavor symmetry is discussed in section~\ref{sec:Lep_CP_flavor}. All leptogenesis CP asymmetries $\epsilon_{e, \mu, \tau}$ are found to vanish in both cases. Leptogenesis could become potentially viable only when higher order contributions lift the postulated residual symmetry. In section~\ref{sec:example} we apply our general formalism to the case that the single residual CP transformation of the neutrino sector arises from the breaking of the most general CP symmetry compatible with $\Delta(6n^2)$ flavor group which is broken down to an abelian subgroup in the charged lepton sector. The predictions for lepton flavor mixing and baryon asymmetry are studied analytically and numerically. Finally, in section~\ref{sec:Conclusions} we summarize our main results and draw the conclusions.

\section{\label{sec:framework}General set-up of leptogenesis in two right-handed neutrino model}

The seesaw mechanism is a popular extension of the Standard Model (SM) to explain the smallness of neutrino masses. In the famous type I seesaw mechanism~\cite{Minkowski:1977sc}, one generally introduces additional three right-handed neutrinos which are singlets under the SM gauge group. Although the seesaw mechanism describes qualitatively well the observations in neutrino oscillation experiments, it is quite difficult to make quantitative predictions for neutrino mass and mixing without further hypothesis for underlying dynamics. The reason is that the seesaw mechanism involves a large number of undetermined parameters at high energies whereas much less parameters could be measured experimentally.

A intriguing way out of this problem is to simply reduce the
number of right-handed neutrinos from three to two~\cite{King:1999mb,Frampton:2002qc,Raidal:2002xf}. The two right-handed neutrino (2RHN) model can be regarded as a limiting case of three right-handed neutrinos where one of the RH neutrinos decouples from the seesaw mechanism either because it is very heavy or because its Yukawa couplings are very weak. Since the number of free parameters is greatly reduced, the 2RHN model is more predictive than the standard scenario involving three RH neutrinos. Namely, the lightest left-handed neutrino mass automatically vanishes, while the masses of the other two neutrinos are fixed by $\delta m^2$ and $\Delta m^2$. Hence only two possible mass spectrums can be obtained
\begin{eqnarray}
\nonumber&&\text{NO}:~m_1=0, \quad m_2=\sqrt{\delta m^2},\quad m_3=\sqrt{\Delta m^2+\delta m^2/2}\,,\\
\label{eq:mass_relation}&&\text{IO}:~m_1=\sqrt{-\delta m^2/2-\Delta m^2},\quad m_2=\sqrt{\delta m^2/2-\Delta m^2},\quad  m_3=0\,.
\end{eqnarray}
Moreover there is only one Majorana CP violating phase corresponding to the phase difference between these two nonzero mass eigenvalues. The Lagrangian responsible for lepton masses in the 2RHN model takes the following form
\begin{equation}\label{eq:lagragian}
\mathcal{L}= -y_{\alpha}\bar{L}_\alpha H l_{\alpha R} -\lambda_{i\alpha}\bar{N}_{iR}\widetilde{H}^\dag L_\alpha-\frac{1}{2}M_i\bar{N}_{iR}N_{iR}^c+h.c.~\,,
\end{equation}
where $L_\alpha\equiv(\nu_{\alpha L}, l_{\alpha L})^{T}$ and $l_{\alpha R}$ indicate the lepton doublet and singlet fields with flavor $\alpha=e,\mu,\tau$ respectively, $N_{iR}$ is the RH neutrino with mass $M_{i}$ ($i=1,2$), and $H\equiv(H^+,H^0)^{T}$ is the Higgs doublet with $\widetilde{H}\equiv i\sigma_{2}H^{\ast}$. The Yukawa couplings $\lambda_{i\alpha}$ form an arbitrary complex $2\times 3$ matrix, here we have worked in the basis in which both the Yukawa couplings for the charged leptons and the Majorana mass matrix for the RH neutrinos are diagonal and real. After electroweak symmetry breaking, the light neutrino mass matrix is given by the famous
seesaw formula
\begin{equation}\label{eq:mnu}
m_{\nu}=-v^2\lambda^{T}M^{-1}\lambda=U^{\ast}mU^{\dagger}\,,
\end{equation}
where $v=175$ GeV refers to the vacuum expectation value of the Higgs field $H^{0}$, $M\equiv\text{diag}(M_1, M_2)$ and $m\equiv\text{diag}(m_1, m_2, m_3)$ with $m_1=0$ for NO and $m_3=0$ for IO, and $U$ is the lepton mixing matrix. It is convenient to express the Yukawa coupling $\lambda$ in terms of the neutrino mass eigenvalues, mixing angles and CP violation phases as\footnote{For other parameterizations of the neutrino Yukawa coupling, see Ref.~\cite{Pascoli:2003uh}.}
\begin{equation}\label{eq:CI_para}
\lambda=iM^{1/2} R m^{1/2} U^{\dagger }/v\,,
\end{equation}
where $R$ is a $2\times 3$ complex orthogonal matrix having the following structure~\cite{Casas:2001sr,Ibarra:2003xp}
\begin{subequations}
\begin{eqnarray}
\label{eq:R_NO} \text{NO}:~ R&=&\left(
\begin{array}{ccc}
0 ~&~ \cos\hat{\theta} ~&~ \xi \sin\hat{\theta} \\
0 ~&~ -\sin\hat{\theta} ~&~ \xi \cos\hat{\theta}
\end{array} \right)\,, \\
\label{eq:R_IO}\text{IO}:~\, R&=&\left(
\begin{array}{ccc}
\cos\hat{\theta} ~&~ \xi \sin\hat{\theta} ~&~ 0\\
-\sin\hat{\theta} ~&~ \xi \cos\hat{\theta}  ~&~ 0
\end{array} \right)\,,
 \end{eqnarray}
\end{subequations}
where $\hat{\theta}$ is an arbitrary complex number and $\xi=\pm 1$. From Eqs.~(\ref{eq:R_NO}, \ref{eq:R_IO}) we can check that the $R$-matrix satisfies
\begin{equation}
\label{eq:orthogonal_R}
\begin{aligned}
RR^T&=\text{diag}(1,1),~~\text{for}~~\text{NO~and~IO}\,,\\
R^TR&=\text{diag}(0,1,1),~~\text{for}~~\text{NO}\,,\\
R^TR&=\text{diag}(1,1,0),~~\text{for}~~\text{IO}\,.
\end{aligned}
\end{equation}
Leptogenesis is a natural consequence of the seesaw mechanism, and it provides an elegant explanation for the baryon asymmetry of the Universe~\cite{Fukugita:1986hr}. For illustration, we shall work in the typical $N_1$-dominated scenario, and we assume that right-handed neutrinos are hierarchical $M_{2}\gg M_1$ such that the asymmetry is dominantly produced from the decays of the lightest RH neutrino $N_1$. The approach of this paper can also be applied to discuss the resonant leptogenesis~\cite{Flanz:1996fb_resonant_LepG}. The naturalness of the electroweak scale restricts the heavy RH neutrino mass to be $M_1\le10^{7}$ GeV~\cite{Vissani:1997ys_bound}. This bound arises from the naturalness requirement that the RH neutrino loops don't lead to unnaturally large radiative corrections to the Higgs mass. However, the unknown dynamics of quantum gravity at the Planck scale $M_{\mathrm{P}}$ would always introduce an unavoidable naturalness problem. In addition, the theoretical criterium of naturalness requires the presence of new physics at the TeV scale. But no any signal of new physics has been observed at the LHC or elsewhere. The argument for naturalness has failed so far as a guiding principle, and Nature does not too much care about our notion of naturalness. Therefore we don't require that the Vissani bound $M_1\le10^{7}$ GeV must be fulfilled in this paper. Actually we shall work in the two flavored leptogenesis regime, that is at $10^9$ GeV $\leq M_{1}\leq10^{12}$ GeV.

The phenomenology of leptogenesis in 2RHN model has been comprehensively studied~\cite{Frampton:2002qc,Raidal:2002xf,Ibarra:2003xp,Endoh:2002wm}. The flavored CP asymmetries in the decays of $N_1$ into leptons of different flavors are of the form~\cite{Covi:1996wh,Buchmuller:2004nz,Buchmuller:2005eh,Davidson:2008bu}
\begin{eqnarray}
\nonumber \epsilon_{\alpha}&\equiv&\frac{\Gamma(N_1\rightarrow l_{\alpha} H)-\Gamma(N_1\rightarrow \bar{l}_{\alpha}\bar{H})}{\sum_{\alpha}\Gamma(N_1\rightarrow l_{\alpha} H)+\Gamma(N_1\rightarrow \bar{l}_{\alpha}\bar{H})}\\
 \label{eq:epsilon_alpha_gen} &=&\frac{1}{8\pi(\lambda\lambda^\dag)_{11}}\sum_{j\neq 1}\bigg\{\mathrm{Im}\big[(\lambda\lambda^\dag)_{1j}\lambda_{1\alpha}\lambda^*_{j\alpha}\big]g(x_j)+\mathrm{Im}\big[(\lambda\lambda^\dag)_{j1}\lambda_{1\alpha}\lambda^*_{j\alpha}\big]\frac{1}{1-x_j}\bigg\} \,,
\end{eqnarray}
where $\Gamma(N_1\rightarrow l_{\alpha} H)$ and $\Gamma(N_1\rightarrow \bar{l}_{\alpha}\bar{H})$ with $\alpha=e, \mu, \tau $
denote the flavored decay rates of $N_1$ into lepton $l_{\alpha}$ and anti-lepton $\bar{l}_{\alpha}$ respectively, the parameter $x_{j}$ is defined as $x_{j}\equiv M^2_{j}/M^2_{1}$ and the loop function $g(x)$ is
\begin{equation}
g(x)=\sqrt{x}\Big[\frac{1}{1-x}+1-(1+x)\ln\big(\frac{1+x}{x}\big)\Big]\simeq -\frac{3}{2\sqrt{x}}+\mathcal{O}(x^{-3/2})\quad \text{for} \quad x\gg1\,.
\end{equation}
In the hierarchical limit $M_{2}\gg M_1$, i.e., $x_{2}\gg1$, the CP asymmetries can be written as\footnote{The flavored CP asymmetry $\epsilon_{\alpha}$ contains two terms: the lepton-number-violating (LNV) piece $\epsilon_\alpha^{\rm LNV} \propto\Im\left[\big(\lambda\lambda^\dagger\big)_{1j}\,\lambda_{1\alpha}^{\vphantom{*}}\lambda_{j\alpha}^*\right]$ and the lepton-flavor-violating (LFV) piece $\epsilon_\alpha^{\rm LFV} \propto\Im\left[\big(\lambda\lambda^\dagger\big)_{j1}\,\lambda_{1\alpha}^{\vphantom{*}}
\lambda_{j\alpha}^*\right]$. Since $\epsilon_\alpha^{\rm LNV}\sim\mathcal{O}(x^{-1/2}_{j})$ and $\epsilon_\alpha^{\rm LFV}\sim\mathcal{O}(x^{-1}_{j})$ in the limit $x_j\gg1$, LFV term is suppressed with respect to the LNV one, hence we shall neglect the LFV
contribution in this work.}
\begin{eqnarray}
\nonumber \epsilon_{\alpha}&\simeq&-\frac{3}{16\pi}\sum^{2}_{j
=1}\frac{M_1}{M_j}\frac{\Im\left[(\lambda\lambda^{\dagger})_{1j}\lambda_{1\alpha}\lambda^*_{j\alpha}\right]}{(\lambda\lambda^{\dagger})_{11}}\\
\label{eq:epsilon_alpha}&=&-\frac{3M_1}{16\pi v^2}\frac{\Im\big(\sum_{ij} \sqrt{m_im_j}m_jR_{1i}R_{1j}U^*_{\alpha i}U_{\alpha j}\big)}{\sum_j m_j|R_{1j}|^2}\,.
\end{eqnarray}
Actually, only the $j=2$ term is relevant in the first line of Eq.~\eqref{eq:epsilon_alpha}, since $\Im\left[(\lambda\lambda^{\dagger})_{1j}\lambda_{1\alpha}\lambda^*_{j\alpha}\right]=0$ for the case of $j=1$. Here the summation over $j$ allows us to straightforwardly derive the compact expression of Eq.~\eqref{eq:epsilon_alpha}. We notice that $\epsilon_{\alpha}$ is invariant under the transformation $\xi\rightarrow-\xi$ and $\hat{\theta}\rightarrow -\hat{\theta}$. Consequently we shall choose $\xi=1$ as an illustration in the following numerical analysis. Inserting the expression for the Yukawa coupling in Eqs.~(\ref{eq:R_NO}, \ref{eq:R_IO}) into Eq.~\eqref{eq:epsilon_alpha}, we obtain the CP asymmetry
\begin{eqnarray}
\nonumber&&\hskip-0.4in \epsilon_{\alpha}\simeq-\frac{3}{16\pi v^2}\frac{M_1}{{m_2|\cos\hat{\theta}|^2+m_3|\sin\hat{\theta}|^2}}\bigg\{
(m^2_3|U_{\alpha3}|^2-m^2_2|U_{\alpha2}|^2)\;\Im \sin^2\hat{\theta}\\
\label{CP-flavour-2RHN-conv_NO}&&\hskip-0.1in+\xi\sqrt{m_{2}m_{3}}\;\Big[(m_2+m_3)\Re(U^{*}_{\alpha2}U_{\alpha3})\Im(\sin\hat{\theta}\cos\hat{\theta})+
(m_3-m_2)\Im(U^{*}_{\alpha2}U_{\alpha3})\Re(\sin\hat{\theta}\cos\hat{\theta})
\Big]\bigg\}\,,
\end{eqnarray}
for NO and
\begin{eqnarray}
\nonumber &&\hskip-0.4in \epsilon_{\alpha}\simeq-\frac{3}{16\pi v^2}\frac{M_1}{m_1|\cos\hat{\theta}|^2+m_2|\sin\hat{\theta}|^2}\bigg\{(m^2_2|U_{\alpha2}|^2-m^2_1|U_{\alpha1}|^2)\;\Im\sin^2\hat{\theta}\\
\label{CP-flavour-2RHN-conv_IO}&&\hskip-0.1in+\xi\sqrt{m_{1}m_{2}}\;\Big[
(m_1+m_2)\Re(U^{*}_{\alpha1}U_{\alpha2})\Im(\sin\hat{\theta}\cos\hat{\theta})+(m_2-m_1)\Im(U^{*}_{\alpha1}U_{\alpha2})\Re(\sin\hat{\theta}\cos\hat{\theta})\Big]
\bigg\}\,,
\end{eqnarray}
for IO neutrino mass spectrum. If the RH neutrino mass $M_1$ is large enough (e.g. $M_1>10^{12}$ GeV), the interactions mediated by all the three charged lepton Yukawa couplings are out of equilibrium. As a result, the one flavor approximation rigorously holds, and the total CP asymmetry is
\begin{equation}
\label{eq:epsilon1}\epsilon_{1}\equiv\sum_{\alpha}\epsilon_{\alpha}=-\frac{3M_1}{16\pi v^2}\frac{\Im\left(\sum_{i} m^2_iR^2_{1i}\right)}{\sum_j m_j|R_{1j}|^2}\,,
\end{equation}
which is completely independent of the lepton mixing matrix $U$. For the parametrization of the $R$-matrix in Eqs.~(\ref{eq:R_NO}, \ref{eq:R_IO}), we have
\begin{subequations}
\begin{eqnarray}
\label{CP-flavour-2RHN-conv_NO} ~~~~ \text{NO}:~~\epsilon_{1}&=& -\frac{3M_1}{16\pi v^2}
\frac{(m^2_3-m^2_2)\;\Im\sin^2\hat{\theta}}
{m_2|\cos\hat{\theta}|^2+m_3|\sin\hat{\theta}|^2}\,, \\
\label{CP-flavour-2RHN-conv_IO} \text{IO}:~~\epsilon_{1}&=& -\frac{3M_1}{16\pi v^2}
\frac{(m^2_2-m^2_1)\; \Im\sin^2\hat{\theta}}
{m_1|\cos\hat{\theta}|^2+m_2|\sin\hat{\theta}|^2}\,.
\end{eqnarray}
\end{subequations}
We see that the total CP  asymmetry $\epsilon_{1}$ would vanish when the parameter $\hat{\theta}$ is real or purely imaginary up to $\pi/2$. The total baryon asymmetry is the sum of each individual lepton asymmetry. In the present paper we will be concerned with temperature window ($10^9\leq T\sim M_1\leq10^{12})$ GeV. In this range only the $\tau$ charged lepton Yukawa interaction is in equilibrium, the $e$ and $\mu$ flavors are indistinguishable, and the final baryon asymmetry is well approximated by~\cite{Abada:2006ea,Abada:2006fw,Pascoli:2006ci,Nardi:2006fx}
\begin{equation}
\label{eq:Yb}Y_{B}\simeq -\frac{12}{37\,g^*}\left[\epsilon_{2}\eta\left(\frac{417}{589}{\widetilde{m}_2}\right)\,+\,\epsilon_{\tau}\eta\left(\frac{390}{589}{\widetilde{m}_{\tau}}\right)\right]\,,
\end{equation}
where $\epsilon_2\equiv\epsilon_{e}+\epsilon_{\mu}$, $\widetilde{m}_2\equiv\widetilde{m}_{e}+\widetilde{m}_{\mu}$, $g_{*}$ is the number of relativistic degrees of freedom, and $\eta$ is the efficiency factor which depends on the initial abundance of $N_{1}$. The washout mass $\widetilde{m}_{\alpha}$ parametrizes the decay rate of $N_1$ into the leptons of flavor $\alpha$ with
\begin{equation}\label{eq:mtilde_alpha}
\widetilde{m}_\alpha\equiv\frac{|\lambda_{1\alpha}|^2v^2}{M_1}=\Big|\sum_i m_i^{1/2} R_{1i} U_{\alpha i}^*\Big|^2\,, ~\quad \alpha=e,\mu,\tau.
\end{equation}
Plugging Eqs.~\eqref{eq:R_NO} and \eqref{eq:R_IO} into above equation we find the explicit expression of the washout mass is
\begin{equation}
\widetilde{m}_{\alpha}=
\left\{\begin{array}{ll}
\left|\sqrt{m_2}\,U^*_{\alpha 2}\cos\hat{\theta}+\xi\sqrt{m_3}\,U^*_{\alpha 3}\sin\hat{\theta}\right|^2\,, ~~&~\text{for}~~ \text{NO}\,,  \\
\left|\sqrt{m_1}\,U^*_{\alpha 1}\cos\hat{\theta}+\xi\sqrt{m_2}\,U^*_{\alpha 2}\sin\hat{\theta}\right|^2\,, ~~&~\text{for}~~ \text{IO}\,.
\end{array}\right.
\end{equation}
Then the washout parameter $K$ defined as $K=\sum_{\alpha}\widetilde{m}_\alpha/\widetilde{m}^*$ with $\widetilde{m}^*\sim 10^{-3}$ eV takes the form
\begin{equation}
K=
\left\{\begin{array}{ll}\frac{m_{2}|\cos\hat{\theta}|^2+m_3|\sin\hat{\theta}|^2}{\widetilde{m}^*}\geq\frac{m_2}{\widetilde{m}^*}\simeq 8.683\,, ~~&~\text{for}~~ \text{NO}\,,  \\
\frac{m_{1}|\cos\hat{\theta}|^2+m_2|\sin\hat{\theta}|^2}{\widetilde{m}^*}\geq\frac{m_1}{\widetilde{m}^*}\simeq48.397\,, ~~&~\text{for}~~ \text{IO}\,.
\end{array}
\right.
\end{equation}
Therefore the two right-handed models are always in the strong washout regime. As a consequence, the initial $N_1$ abundance is almost irrelevant~\cite{Giudice:2003jh}, and the right-handed neutrinos are brought to thermal equilibrium by inverse decays and by $\Delta L=1$ scatterings. To a good accuracy, the efficiency factor $\eta(\widetilde{m}_{\alpha})$ is approximately given by~\cite{Abada:2006ea}
\begin{equation}
\label{eq:efficency_factor}\eta(\widetilde{m}_{\alpha})\simeq\left[\left(\frac{\widetilde{m}_{\alpha}}{8.25\times 10^{-3}\,{\rm eV}}\right)^{-1}+\left(\frac{0.2\times 10^{-3}\,{\rm eV}}{\widetilde{m}_{\alpha}}\right)^{-1.16}\ \right]^{-1}\,.
\end{equation}

\section{\label{sec:LepG_one_CP}Leptogenesis with one residual CP transformation}

In a series of papers~\cite{Feruglio:2012cw,Ding:2013bpa,Li:2015jxa,Branco:2015hea,Ding:2013nsa,King:2014rwa,Ding:2014ssa,Hagedorn:2014wha,Ding:2014ora,Ding:2015rwa,Yao:2016zev,Lu:2016jit,Chen:2014wxa}, it has been shown that the residual CP symmetry of the light neutrino mass matrix can quite efficiently predict the lepton mixing angles as well as CP violation phases. If the residual CP symmetry is preserved by the seesaw Lagrangian, leptogenesis would be also strongly constrained~\cite{Chen:2016ptr,Hagedorn:2016lva,Chen:2016ica}. We assume that the flavor and CP symmetries are broken at a scale above the leptogenesis scale. As a consequence, leptogenesis occurs in the standard framework of the SM plus two heavy RH neutrinos without involving any additional state in its dynamics. Otherwise if the flavor and CP symmetries are broken close or below the leptogenesis scale, the additional interactions and new particles related to flavor and CP symmetries should be considered~\cite{AristizabalSierra:2007ur}, and the resulting scenarios would be quite different from the standard one. In this section, we shall study the implications of residual CP for leptogenesis in 2RHN model, and we assume that both the neutrino Yukawa coupling and the Majorana mass term in Eq.~\eqref{eq:lagragian} are invariant under one generic residual CP transformation defined as
\begin{equation}\label{eq:res_CP_v2}
\nu_{L}\stackrel{\text{CP}}{\longmapsto} iX_{\nu}\gamma_0C\bar{\nu}^{T}_{L}\,,~\quad N_{R}\stackrel{\text{CP}}{\longmapsto} i\widehat{X}_{N}\gamma_{0}C\bar{N}^{T}_{R}\,, \\
\end{equation}
where $\nu_{L}\equiv (\nu_{eL}, \nu_{\mu L}, \nu_{\tau L})^{T}$, $N_{R}\equiv(N_{1R}, N_{2R})^{T}$, $C$ denotes the charge-conjugation matrix, $X_{\nu}$ is a $3\times3$ symmetric unitary matrix to avoid degenerate neutrino masses and $\widehat{X}_{N}$ is a $2\times2$ symmetric unitary matrix. For the symmetry to hold, $\lambda$ and $M$ have to fulfill
\begin{equation}\label{eq:cons_rcp_v2}
\widehat{X}^{\dagger}_{N}\lambda X_{\nu}=\lambda^{\ast},~~\quad \widehat{X}^{\dagger}_{N}M\widehat{X}^{\ast}_{N}=M^{\ast}\,.
\end{equation}
As we work in the basis in which the RH neutrino mass matrix $M$ is real and diagonal, the residual CP transformation $\widehat{X}_{R}$ should be diagonal with elements equal to $\pm1$, i.e.,
\begin{equation}\label{eq:XR12_v2}
\widehat{X}_{N}=\text{diag}(\pm1,\pm1)\,,
\end{equation}
Notice that the conclusion would not be changed even if $M$ is non-diagonal in a concrete flavor symmetry model~\cite{Chen:2016ptr}, and the reason is explained in Appendix~\ref{sec:Basis_independence}. Thus we can find that the light neutrino mass matrix $m_{\nu}$ given by the seesaw formula satisfies
\begin{equation}\label{eq:constraint_mnu}
X^{T}_{\nu}m_{\nu}X_{\nu}=m^{\ast}_{\nu}\,,
\end{equation}
which means (as expected) $m_{\nu}$ is invariant under the residual CP transformation $X_{\nu}$. The light neutrino mass matrix can be diagonalized by a unitary transformation $U_{\nu}$ with $m_{\nu}=U^{\ast}_{\nu}\text{diag}(m_1, m_2, m_3)U^{\dagger}_{\nu}$.  Then from Eq.~\eqref{eq:constraint_mnu} we can obtain
\begin{eqnarray}
\left(U^{\dagger}_{\nu}X_{\nu}U^{\ast}_{\nu}\right)^{T}\text{diag}(m_1, m_2, m_3)\left(U^{\dagger}_{\nu}X_{\nu}U^{\ast}_{\nu}\right)=\text{diag}(m_1, m_2, m_3)\,.
\end{eqnarray}
Note $m_1=0$ for NO and $m_3=0$ for IO in the 2RHN model. Hence $U_{\nu}$ is subject to the following constraint from the residual CP transformation $X_{\nu}$,
\begin{equation}\label{eq:constraint_resodual_CP_v2}
U^{\dagger}_{\nu}X_{\nu}U^{\ast}_{\nu}=\widehat{X}_{\nu}\,,
\end{equation}
with
\begin{equation}
\begin{array}{ll}
\widehat{X}_{\nu}=\text{diag}\left(e^{i\alpha}, \pm1, \pm1\right) ~ &~\text{for}~~\text{NO}\,, \\
\widehat{X}_{\nu}=\text{diag}\left(\pm1, \pm1,e^{i\alpha}\right) ~ &~\text{for}~~ \text{IO}\,, \\
\end{array}
\end{equation}
where $\alpha$ is a real parameter in the interval between 0 and $2\pi$.
Then it is easy to check that $X_{\nu}$ is a symmetric and unitary matrix for both NO and IO cases. Moreover, with the definition of $R$-matrix in Eq.~\eqref{eq:CI_para}, we can derive that the postulated residual symmetry leads to the following constraint on $R$ as
\begin{equation}\label{eq:constr_R}
\widehat{X}_{N}R^{\ast}\widehat{X}^{-1}_{\nu}=R\,,
\end{equation}
Obvioulsy $-\widehat{X}_{N}$ and $-\widehat{X}_{\nu}$ give rise to the same constraint on $R$ as $\widehat{X}_{N}$ and $\widehat{X}_{\nu}$, therefore it is sufficient to only consider the cases of $\widehat{X}_{N}=\text{diag}(1,\pm1)$,  $\widehat{X}_{\nu}=\text{diag}(e^{i\alpha},\pm1,\pm1)$ for NO and $\widehat{X}_{\nu}=\text{diag}(\pm1,\pm1,e^{i\alpha})$ for IO. The explicit forms of the $R$-matrix for all possible values of $\widehat{X}_{N}$ and $\widehat{X}_{\nu}$ are collected in table~\ref{tab:orth_R_v2}. We see that there are three admissible forms of the $R$-matrix summarized as follows
\begin{equation}\label{eq:R-matrix_three_cases}
\begin{array}{ll}
\text{R-1st:}~~& \left\{\begin{array}{l}
R=\begin{pmatrix}
0  ~&~ \cos\vartheta  ~&~ \xi\sin\vartheta \\
0  ~&~ -\sin\vartheta ~&~ \xi\cos\vartheta
\end{pmatrix}~~\text{for}~~\text{NO},\\
R=\begin{pmatrix}
 \cos\vartheta  ~&~ \xi\sin\vartheta ~&~ 0\\
 -\sin\vartheta ~&~ \xi\cos\vartheta ~&~ 0
\end{pmatrix}~~~\text{for}~~\text{IO}\,,
\end{array}\right. \\[8mm]
\text{R-2nd:}~~& \left\{\begin{array}{l}
R=\pm\begin{pmatrix}
0  ~&~ \cosh\vartheta  ~&~ i\xi\sinh\vartheta \\
0  ~&~ -i\sinh\vartheta ~&~ \xi\cosh\vartheta
\end{pmatrix}~~\text{for}~~\text{NO},\\
R=\pm\begin{pmatrix}
 \cosh\vartheta  ~&~ i\xi\sinh\vartheta ~&~ 0 \\
 -i\sinh\vartheta ~&~ \xi\cosh\vartheta ~&~ 0
\end{pmatrix}~~~\text{for}~~\text{IO},
\end{array}\right.\\[8mm]
\text{R-3rd:}~~&
\left\{\begin{array}{l}
R=\pm\begin{pmatrix}
0  ~&~ i\sinh\vartheta  ~&~  -\xi\cosh\vartheta  \\
0  ~&~ \cosh\vartheta   ~&~ i\xi \sinh\vartheta
\end{pmatrix}~~\text{for}~~\text{NO},\\
R=\pm\begin{pmatrix}
 i\sinh\vartheta  ~&~  -\xi\cosh\vartheta ~&~ 0  \\
 \cosh\vartheta   ~&~ i\xi \sinh\vartheta ~&~ 0
\end{pmatrix}~~\text{for}~~\text{IO}\,.
\end{array}\right.
\end{array}
\end{equation}
We would like to point out that the $R$-matrix is constrained to depend on a single real parameter $\vartheta$ in this setup.
\begin{table}[t!]
\begin{center}
\begin{tabular}{|c|c|c|c|c|c|} \hline\hline
  &   &&  \\ [-0.16in]
 $\widehat{X}_{N} $ & $ \widehat{X}_{\nu} $ & $R$ (NO) & $R$ (IO)\\ \hline
    &   &&    \\ [-0.16in]
 $\text{diag}(1,1)$ & $\mathcal{D}(1,1)$   & $\begin{pmatrix}
0  ~&~ \cos\vartheta  ~&~ \xi\sin\vartheta \\
0  ~&~ -\sin\vartheta ~&~ \xi\cos\vartheta
\end{pmatrix}$ & $\begin{pmatrix}
 \cos\vartheta  ~&~ \xi\sin\vartheta ~&~ 0\\
 -\sin\vartheta ~&~ \xi\cos\vartheta ~&~ 0
\end{pmatrix}$ \\
  &  &&    \\ [-0.17in]\hline
    &   &&    \\ [-0.16in]
 $\text{diag}(1,1)$ & $\mathcal{D}(1,-1)$ & \xmark & \xmark\\ \hline
 $\text{diag}(1,1)$ & $\mathcal{D}(-1,1)$ & \xmark & \xmark  \\ \hline
 $\text{diag}(1,1)$ & $\mathcal{D}(-1,-1)$ & \xmark & \xmark  \\  \hline
 $\text{diag}(1,-1)$   & $\mathcal{D}(1,1)$   & \xmark & \xmark \\ \hline
   &   &&    \\ [-0.16in]
 $\text{diag}(1,-1)$   & $\mathcal{D}(1,-1)$ & $\pm\begin{pmatrix}
0  ~&~ \cosh\vartheta  ~&~ i\xi\sinh\vartheta \\
0  ~&~ -i\sinh\vartheta ~&~ \xi\cosh\vartheta
\end{pmatrix}$  & $\pm\begin{pmatrix}
 \cosh\vartheta  ~&~ i\xi\sinh\vartheta ~&~ 0 \\
 -i\sinh\vartheta ~&~ \xi\cosh\vartheta ~&~ 0
\end{pmatrix}$  \\
  &  &&    \\ [-0.17in]\hline
    &   &&    \\ [-0.16in]
 $\text{diag}(1, -1)$   & $\mathcal{D}(-1,1)$ & $\pm\begin{pmatrix}
0  ~&~ i\sinh\vartheta  ~&~  -\xi\cosh\vartheta  \\
0  ~&~ \cosh\vartheta   ~&~ i\xi \sinh\vartheta
\end{pmatrix}$
& $\pm\begin{pmatrix}
 i\sinh\vartheta  ~&~  -\xi\cosh\vartheta ~&~ 0  \\
 \cosh\vartheta   ~&~ i\xi \sinh\vartheta ~&~ 0
\end{pmatrix}$  \\
   &  &&    \\ [-0.17in]\hline
 $\text{diag}(1, -1)$   & $\mathcal{D}(-1,-1)$ & \xmark & \xmark  \\  \hline \hline
\end{tabular}
\end{center}
\renewcommand{\arraystretch}{1.0}
\caption{\label{tab:orth_R_v2}The explicit form of $R$-matrix for all possible independent values of $\widehat{X}_{N}$ and $\widehat{X}_{\nu}$, where $\vartheta$ is a real free parameter. The symbol ``\xmark'' denotes that the solution for $R-$matrix does not exist since it has to fulfill the equality of Eq.~\eqref{eq:orthogonal_R}. The notation $\mathcal{D}(x,y)$ with $x,y=\pm1$ refers to $\text{diag}(e^{i\alpha},x,y)$ and $\text{diag}(x,y,e^{i\alpha})$ for NO and IO respectively.}
\end{table}
Moreover, from Eq.~\eqref{eq:epsilon1} we can see that the total lepton asymmetry $\epsilon_1$ is vanishing, i.e.
\begin{equation}\label{eq:total_CP}
\epsilon_1=\epsilon_{e}+\epsilon_{\mu}+\epsilon_{\tau}=0\,.
\end{equation}
As a result, the net baryon asymmetry can not be generated in the one flavor approximation which is realized when the mass of the lightest right-handed neutrino $M_1$ is larger than about $10^{12}$ GeV, unless the residual CP symmetry is further broken by subleading order corrections. This result is quite general, it is independent of the explicit form of the residual CP transformation and how the residual symmetry is dynamically realized.

Next we proceed to determine the lepton mixing matrix from the postulated remnant CP transformation. Since $X_\nu$ must be a symmetric unitary matrix to avoid degenerate neutrino masses, by performing the Takagi factorization $X_{\nu}$ can be written as~\cite{Feruglio:2012cw,Chen:2014wxa}
\begin{equation}
\label{eq:Takagi_fact}X_{\nu}=\Sigma_{\nu}\Sigma^{T}_{\nu}\,,
\end{equation}
where $\Sigma_{\nu}$  is a unitary matrix and it can be expressed in terms of the eigenvalues and eigenvectors of $X_{\nu}$~\cite{Chen:2014wxa}. Thus the constraint on the neutrino diagonalization matrix $U_{\nu}$ in Eq.~\eqref{eq:constraint_resodual_CP_v2} can be simplified into
\begin{equation}
\label{eq:constraint_single_RCP}\Sigma^{T}_{\nu}U^{*}_{\nu}\widehat{X}^{-\frac{1}{2}}_{\nu}=\Sigma^{\dagger}_{\nu}U_{\nu}\widehat{X}^{\frac{1}{2}}_{\nu}\,.
\end{equation}
The matrices on the two sides of this equation are unitary and
complex conjugates of each other. Therefore the combination $\Sigma^{\dagger}_{\nu}U_{\nu}\widehat{X}^{\frac{1}{2}}_{\nu}$ is a generic real orthogonal matrix, and consequently the unitary transformation $U_{\nu}$ takes the form~\cite{Chen:2014wxa,Chen:2015siy,Chen:2016ica}
\begin{equation}\label{eq:Unu_one_CP}
U_{\nu}=\Sigma_{\nu}O_{3\times3}\widehat{X}^{-\frac{1}{2}}_{\nu}\,,
\end{equation}
where $O_{3\times3}$ is a three dimensional real orthogonal matrix, and it can be generally parameterized as
\begin{equation}\label{eq:Orthogonal_matrix}
O_{3\times3}(\theta_{1},\theta_{2},\theta_{3})=\begin{pmatrix}
1 &~ 0 ~& 0 \\
0 &~ \cos\theta_1   ~&   \sin\theta_1 \\
0 &~ -\sin\theta_1  ~&   \cos\theta_1
\end{pmatrix}
\begin{pmatrix}
\cos\theta_2   &~   0    ~&    \sin\theta_2 \\
0   &~   1   ~&   0 \\
-\sin\theta_2   &~   0   ~&    \cos\theta_2
\end{pmatrix}
\begin{pmatrix}
\cos\theta_3     &~    \sin\theta_3    ~&    0 \\
-\sin\theta_3    &~    \cos\theta_3    ~& 0   \\
0    &~    0     ~&    1
\end{pmatrix}\,,
\end{equation}
where $\theta_{i}$ ($i=1,2,3$) are real free parameters in the range of $[0, \pi)$. In our working basis (usually called leptogenesis basis) where the charged lepton mass matrix is diagonal, lepton flavor mixing completely arises from the neutrino sector, and therefore the lepton mixing matrix $U$ coincides with $U_{\nu}$. Hence we conclude that the mixing matrix and all mixing angles and CP phases would depend on three free continuous parameters $\theta_{1,2,3}$ if only one residual CP transformation is preserved in the neutrino sector. In order to facilitate the discussion of leptogenesis, we separate out the CP parity matrices $\widehat{X}_{N}$ and $\widehat{X}_{\nu}$ and define the following three parameters
\begin{equation}
U^\prime\equiv U\widehat{X}^{\frac{1}{2}}_{\nu}, ~\quad R^\prime\equiv \widehat{X}^{-\frac{1}{2}}_{N}R\widehat{X}^{\frac{1}{2}}_{\nu},
~\quad K_{i}\equiv(\widehat{X}_{N})_{11}(\widehat{X}^{-1}_{\nu})_{ii}\,,~~ i=1,2,3\,.
\end{equation}
We see that $R'$ is real and the parameter $K_i$ is equal to $+1$, $-1$ or $\pm e^{-i\alpha}$. As a consequence, the flavored CP asymmetry $\epsilon_{\alpha}$ can be expressed as
\begin{equation}\label{eq:epsilon_alpha_p}
\epsilon_{\alpha}
=-\frac{3M_1}{16\pi v^2}\frac{\Im\big(\sum_{ij} \sqrt{m_im_j}m_jR^\prime_{1i}R^\prime_{1j}U^{\prime*}_{\alpha i}U^\prime_{\alpha j}K_j\big)}{\sum_j m_jR^{\prime2}_{1j}} \,,
\end{equation}
and the washout mass $\tilde{m}_{\alpha}$ is given by
\begin{equation}\label{eq:mtilde_p}
\widetilde{m}_{\alpha}=\left|\sum_{i}\sqrt{m_i}\,R^\prime_{1i}U^\prime_{\alpha i}\right|^2\,.
\end{equation}
Taking into account that the lightest neutrino is massless in 2RHN model, we find $\epsilon_{\alpha}$ and $\tilde{m}_{\alpha}$ can be written into a rather simple form
\begin{subequations}
\begin{eqnarray}
\text{NO}:\, && \epsilon_{\alpha}=-\frac{3M_1}{16\pi v^2}
W_{\text{NO}}I_{\text{NO}}^{\alpha},
\quad \tilde{m}_{\alpha}=\left|\sqrt{m_3}R^\prime_{13}U^\prime_{\alpha 3}+\sqrt{m_2}R^\prime_{12}U^\prime_{\alpha 2}\right|^2,   \\
\text{IO}:\, && \epsilon_{\alpha}=-\frac{3M_1}{16\pi v^2}
W_{\text{IO}}I_{\text{IO}}^{\alpha},
\quad \tilde{m}_{\alpha}=\left|\sqrt{m_2}R^\prime_{12}U^\prime_{\alpha 2}+\sqrt{m_1}R^\prime_{11}U^\prime_{\alpha 1}\right|^2,
\end{eqnarray}
\end{subequations}
with
\begin{eqnarray}
\nonumber && W_{\text{NO}}=\frac{\sqrt{m_2m_3}R^{\prime}_{12}R^{\prime}_{13}(m_3K_3-m_2K_2)}{m_2R^{\prime2}_{12}+m_3R^{\prime2}_{13}}, \qquad I^{\alpha}_{\text{NO}}=\Im(U^\prime_{\alpha 3}U^{\prime*}_{\alpha 2})\,, \\
\label{eq:W_I_alpha}&& W_{\text{IO}}=\frac{\sqrt{m_1m_2}R^{\prime}_{11}R^{\prime}_{12}(m_2K_2-m_1K_1)}{m_1R^{\prime2}_{11}+m_2R^{\prime2}_{12}}, \qquad
I^{\alpha}_{\text{IO}}=\Im(U^\prime_{\alpha 2}U^{\prime*}_{\alpha 1})\,.
\end{eqnarray}
The explicit expressions of $W_{\text{NO}}$ and $W_{\text{IO}}$ for the three viable forms of the $R$-matrix are shown in table~\ref{tab:R_W_para}.
Notice that $W_{\text{NO}, \text{IO}}$ are fixed by the light neutrino masses $m_{2,3}$ and $\vartheta$ which parametrizes the $R$-matrix, and the bilinear invariants $I^{\alpha}_{\text{NO}, \text{IO}}$ depend on the low energy CP phases contained in the mixing matrix $U$. As a result, if the signal of CP violation was observed in future neutrino oscillation experiments or neutrinoless double beta ($0\nu\beta\beta$) decay experiments, a nonzero baryon asymmetry is expected to be generated through leptogenesis in this framework. In the following, we shall perform a general analysis of leptogenesis in the 2RHN model with a generic residual CP transformation, and the lepton mixing matrix can be parameterized as~\cite{Olive:2016xmw}
\begin{equation}\label{eq:PMNS_parameterized}
U=\left(\begin{array}{ccc}
c_{12}c_{13}  &   s_{12}c_{13}   &   s_{13}e^{-i\delta}  \\
-s_{12}c_{23}-c_{12}s_{13}s_{23}e^{i\delta}   &  c_{12}c_{23}-s_{12}s_{13}s_{23}e^{i\delta}  &  c_{13}s_{23}  \\
s_{12}s_{23}-c_{12}s_{13}c_{23}e^{i\delta}   & -c_{12}s_{23}-s_{12}s_{13}c_{23}e^{i\delta}  &  c_{13}c_{23}
\end{array}\right)\text{diag}(1,e^{i\frac{\phi}{2}},1)\,,
\end{equation}
where $c_{ij}\equiv \cos\theta_{ij}$, $s_{ij}\equiv \sin\theta_{ij}$,
$\delta$ and $\phi$ and the Dirac type and Majorana type CP violating phases respectively. Note that there is only one Majorana CP phase $\phi$ in the presence of one massless light neutrino.

\begin{table}[t!]
\begin{center}
\begin{tabular}{|c|c|c|c|c|}\hline\hline
   &   &    & &     \\ [-0.16in]
& \texttt{Mass ordering} & $K_i$ &$(R^{\prime}_{11}, R^{\prime}_{12}, R^{\prime}_{13})$   &  $W_{\text{NO}}$ ($W_{\text{IO}}$) \\\hline
   &   &    & &     \\ [-0.16in]
\multirow{2}{*}[-5pt]{R-1st} & NO &  $K_2=K_3=1$&$(0,\cos\vartheta,\xi\sin\vartheta)$
& \begin{large}$\frac{\xi\sqrt{m_2m_3}(m_3-m_2)\sin2\vartheta}{2(m_2\cos^2\vartheta+m_3\sin^2\vartheta)}$\end{large} \\[0.08in]\cline{2-5}

   &   &    & &    \\ [-0.16in]

& IO &   $K_1=K_2=1$   &  $(\cos\vartheta,\xi\sin\vartheta,0)$
&  \begin{large}$\frac{\xi\sqrt{m_1m_2}(m_2-m_1)\sin2\vartheta}{2(m_1\cos^2\vartheta+m_2\sin^2\vartheta)}$\end{large} \\[0.08in]\hline

 &   &    & &     \\ [-0.16in]
\multirow{2}{*}[-5pt]{R-2nd} & NO &  $K_2=-K_3=1$   &    $\pm(0,\cosh\vartheta,-\xi\sinh\vartheta)$
& \begin{large}$\frac{\xi\sqrt{m_2m_3}(m_2+m_3)\sinh2\vartheta}{2(m_2\cosh^2\vartheta+m_3\sinh^2\vartheta)}$\end{large} \\[0.08in]\cline{2-5}

   &   &    & &    \\ [-0.16in]

& IO &   $K_1=-K_2=1$   &  $\pm(\cosh\vartheta,-\xi\sinh\vartheta,0)$
&  \begin{large}$\frac{\xi\sqrt{m_1m_2}(m_1+m_2)\sinh2\vartheta}{2(m_1\cosh^2\vartheta+m_2\sinh^2\vartheta)}$\end{large} \\[0.08in]\hline

&   &    & &     \\ [-0.16in]
\multirow{2}{*}[-5pt]{R-3rd} & NO &  $-K_2=K_3=1$   &    $\pm(0,-\sinh\vartheta,-\xi\cosh\vartheta)$
& \begin{large}$\frac{\xi\sqrt{m_2m_3}(m_2+m_3)\sinh2\vartheta}{2(m_2\sinh^2\vartheta+m_3\cosh^2\vartheta)}$\end{large} \\[0.08in]\cline{2-5}

   &   &    & &    \\ [-0.16in]

& IO &   $-K_1=K_2=1$   &  $\pm(-\sinh\vartheta,-\xi\cosh\vartheta,0)$
&  \begin{large}$\frac{\xi\sqrt{m_1m_2}(m_1+m_2)\sinh2\vartheta}{2(m_1\sinh^2\vartheta+m_2\cosh^2\vartheta)}$\end{large} \\[0.08in]\hline\hline

\end{tabular}
\end{center}
\renewcommand{\arraystretch}{1.0}
\caption{\label{tab:R_W_para}The parametrization of the first row of $R^{\prime}$ and the corresponding expressions of $W_{\text{NO}}$ and $W_{\text{IO}}$ for the three viable forms of the $R-$matrix.}
\end{table}

\begin{figure}[t!]
\centering
\begin{tabular}{c}
\includegraphics[width=1\linewidth]{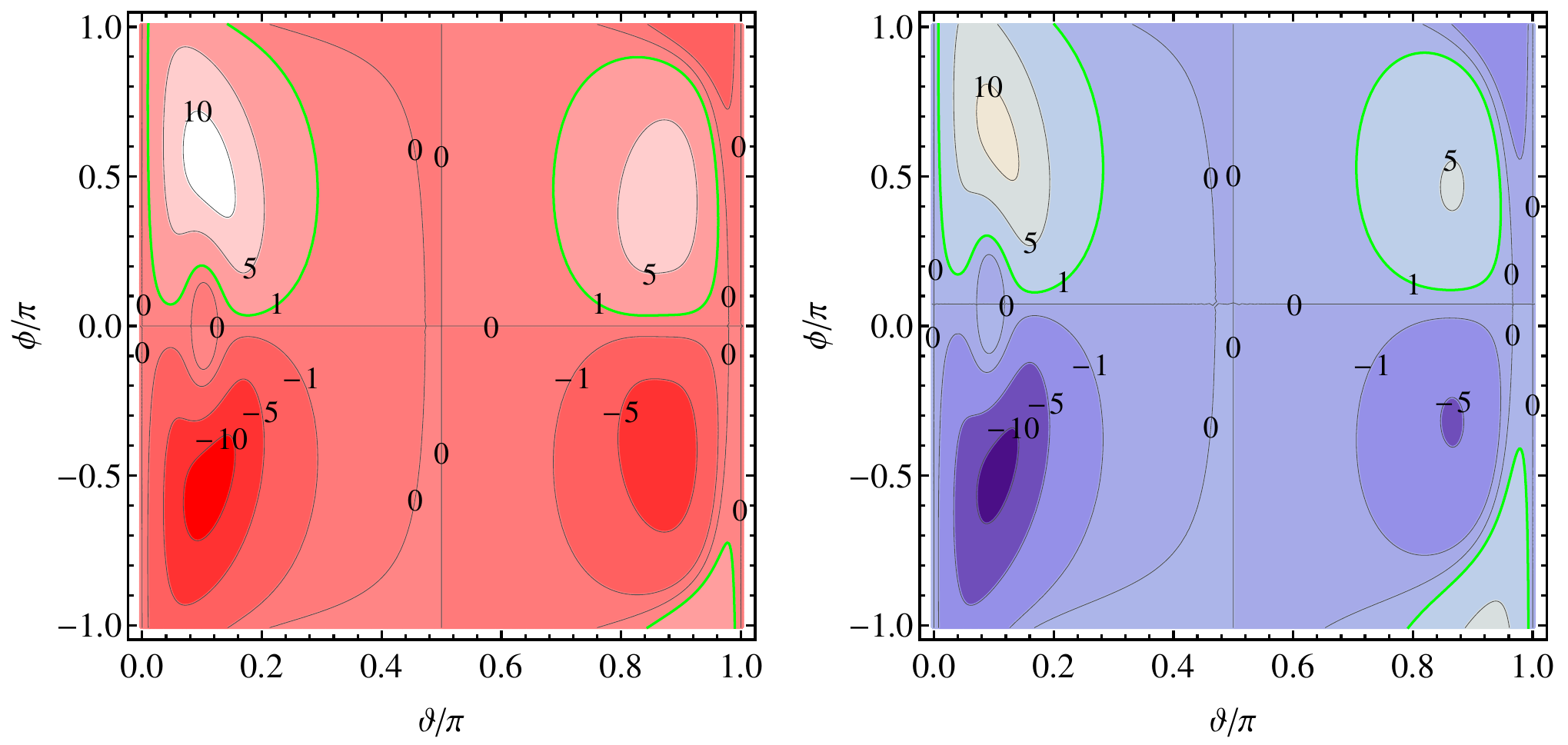}\\
\includegraphics[width=1\linewidth]{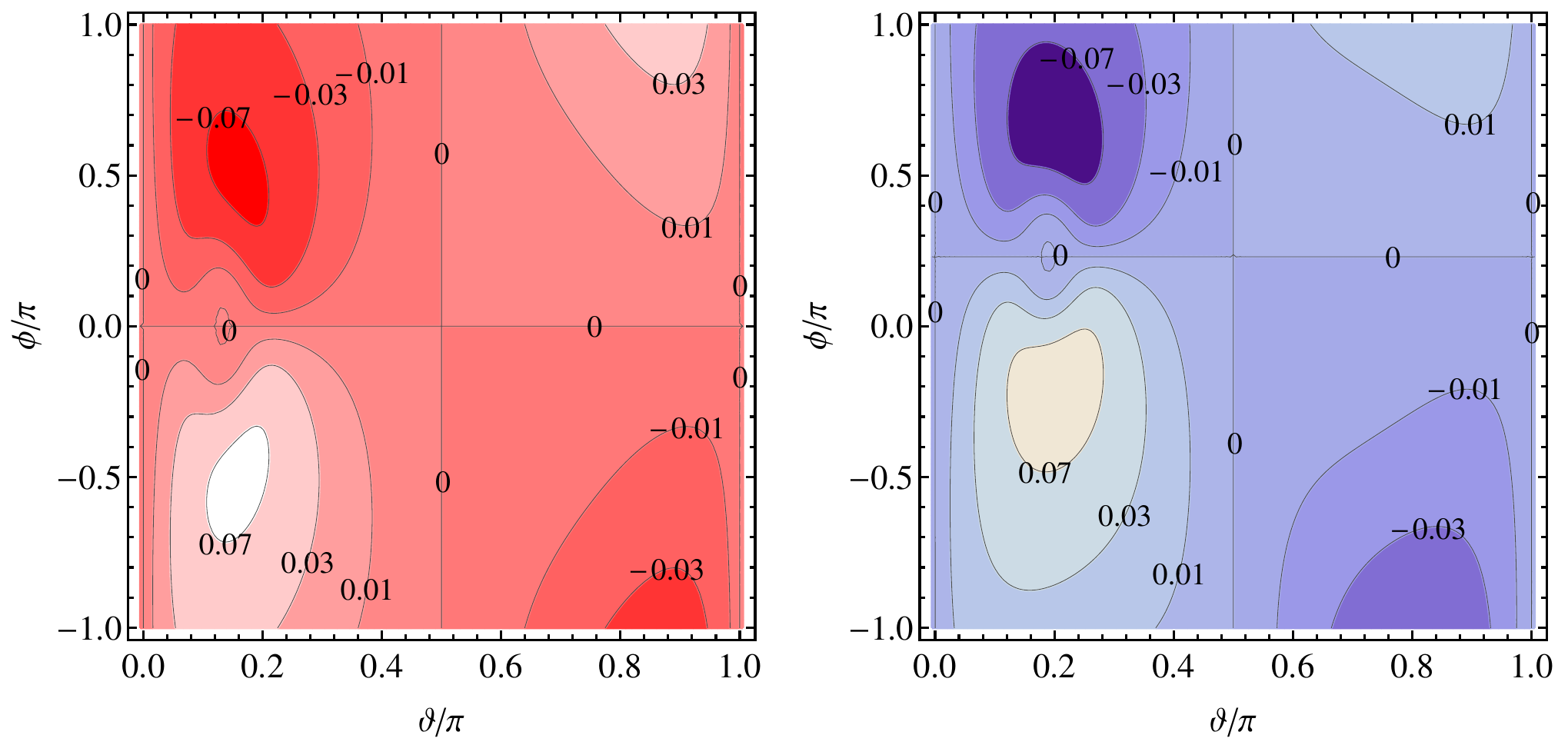}
\end{tabular}
\caption{\label{fig:YB_C11} The contour plots of $Y_{B}/Y^{obs}_B$ in the $\vartheta-\phi$ plane for the case of R-1st. Here we choose $M_1=5\times10^{11}\,\mathrm{GeV}$, so that only the tau Yukawa couplings are in equilibrium. The first row and the second row are for NO and IO spectrums respectively, and the Dirac CP phase $\delta$ is taken to be $0$ on the left panels and $-\pi/2$ on the right panels. The neutrino oscillation parameters $\theta_{12}$, $\theta_{13}$, $\theta_{23}$, $\delta m^2$ and $\Delta m^2$ are fixed at their best fit values~\cite{Capozzi:2013csa}. The thick green curve represents the experimentally observed values of the baryon asymmetry $Y_B^{obs}=8.66\times10^{-11}$~\cite{Ade:2015xua}.}
\end{figure}

\begin{figure}[t!]
\centering
\begin{tabular}{c}
\includegraphics[width=1\linewidth]{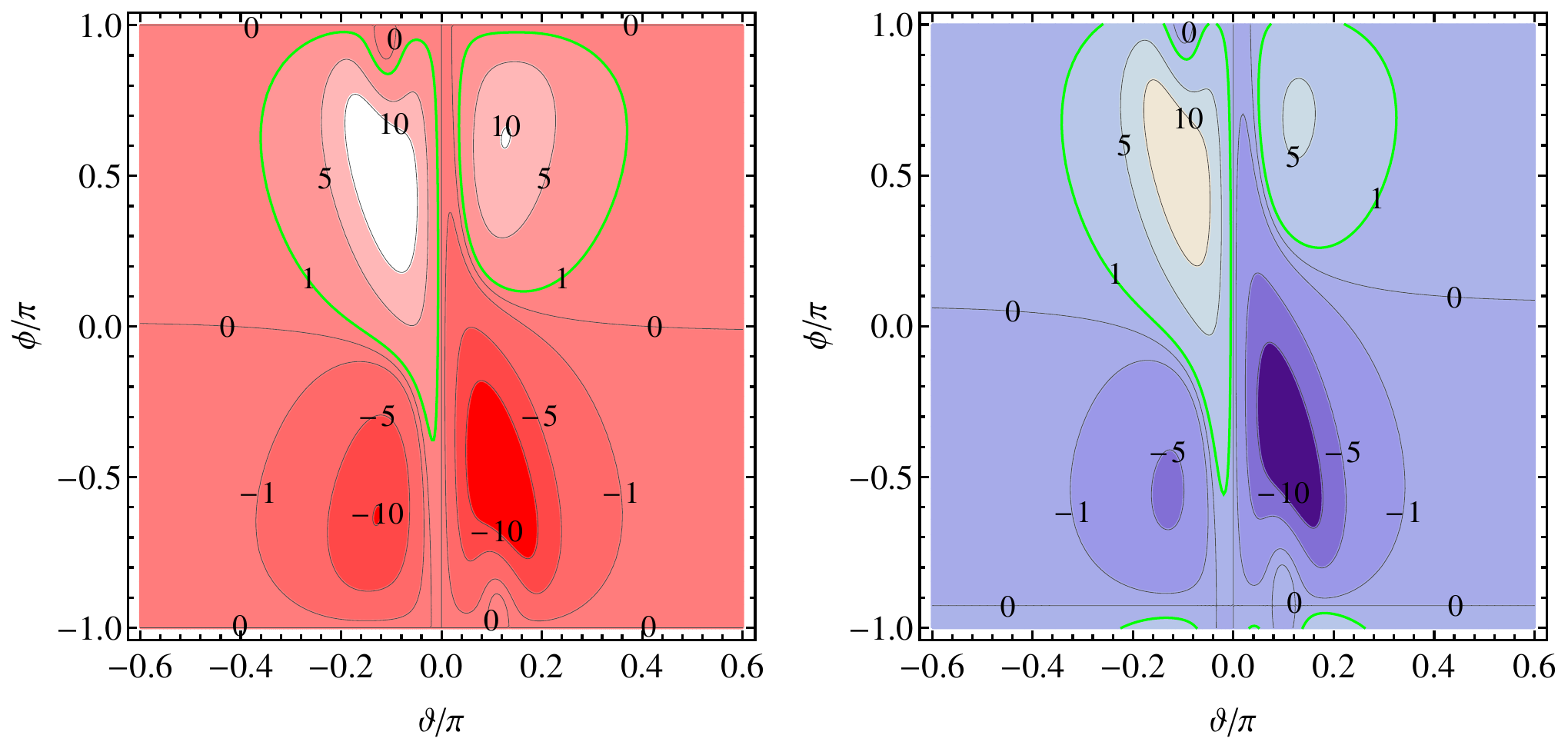}\\
\includegraphics[width=1\linewidth]{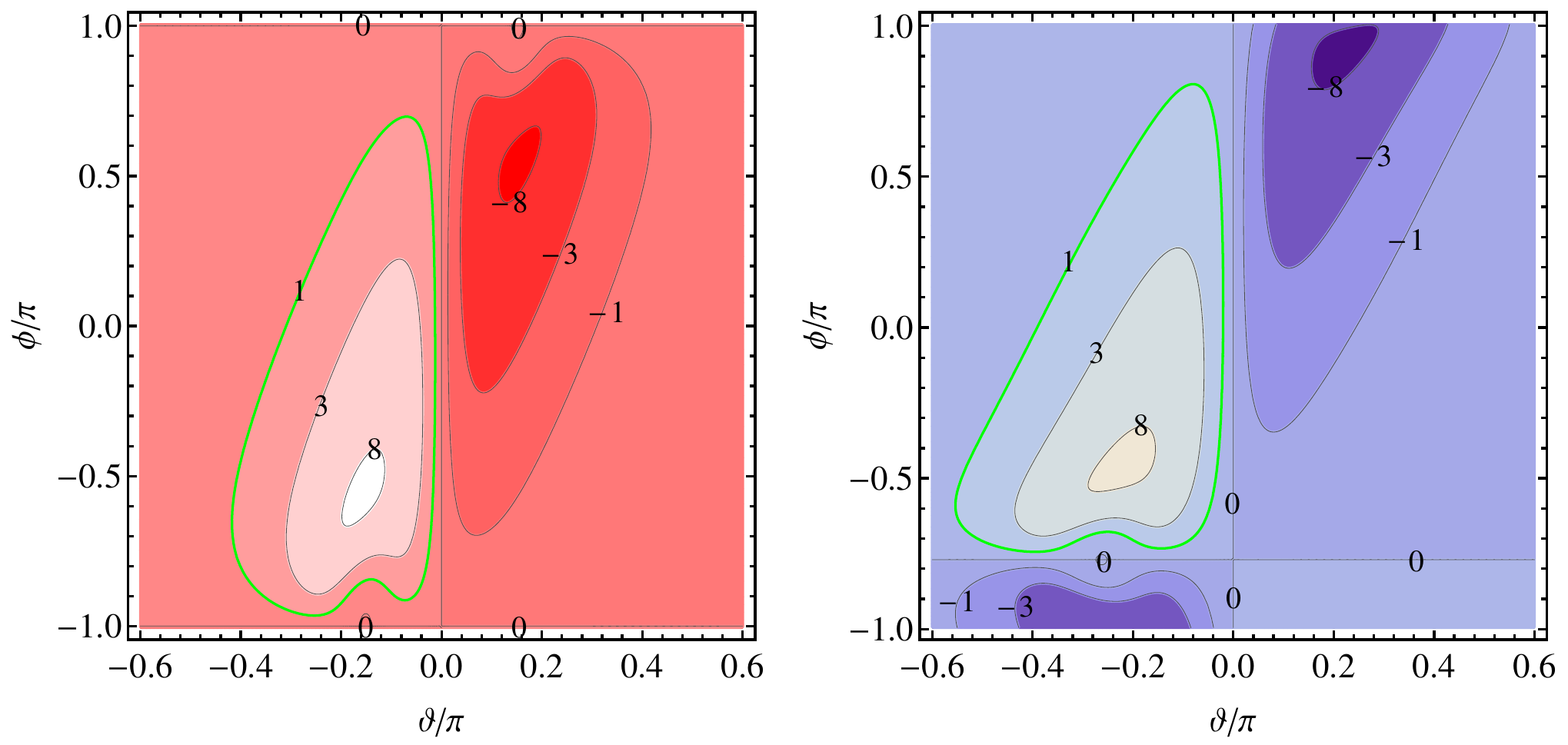}
\end{tabular}
\caption{\label{fig:YB_C22} The contour plots of $Y_{B}/Y^{obs}_B$ in the $\vartheta-\phi$ plane for the case of R-2nd. Here we choose $M_1=5\times10^{11}\,\mathrm{GeV}$, so that only the tau Yukawa couplings are in equilibrium. The first row and the second row are for NO and IO spectrums respectively, and the Dirac CP phase $\delta$ is taken to be $0$ on the left panels and $-\pi/2$ on the right panels. The neutrino oscillation parameters $\theta_{12}$, $\theta_{13}$, $\theta_{23}$, $\delta m^2$ and $\Delta m^2$ are fixed at their best fit values~\cite{Capozzi:2013csa}. The thick green curve represents the experimentally observed values of the baryon asymmetry $Y_B^{obs}=8.66\times10^{-11}$~\cite{Ade:2015xua}.}
\end{figure}

\begin{figure}[t!]
\centering
\begin{tabular}{c}
\includegraphics[width=1\linewidth]{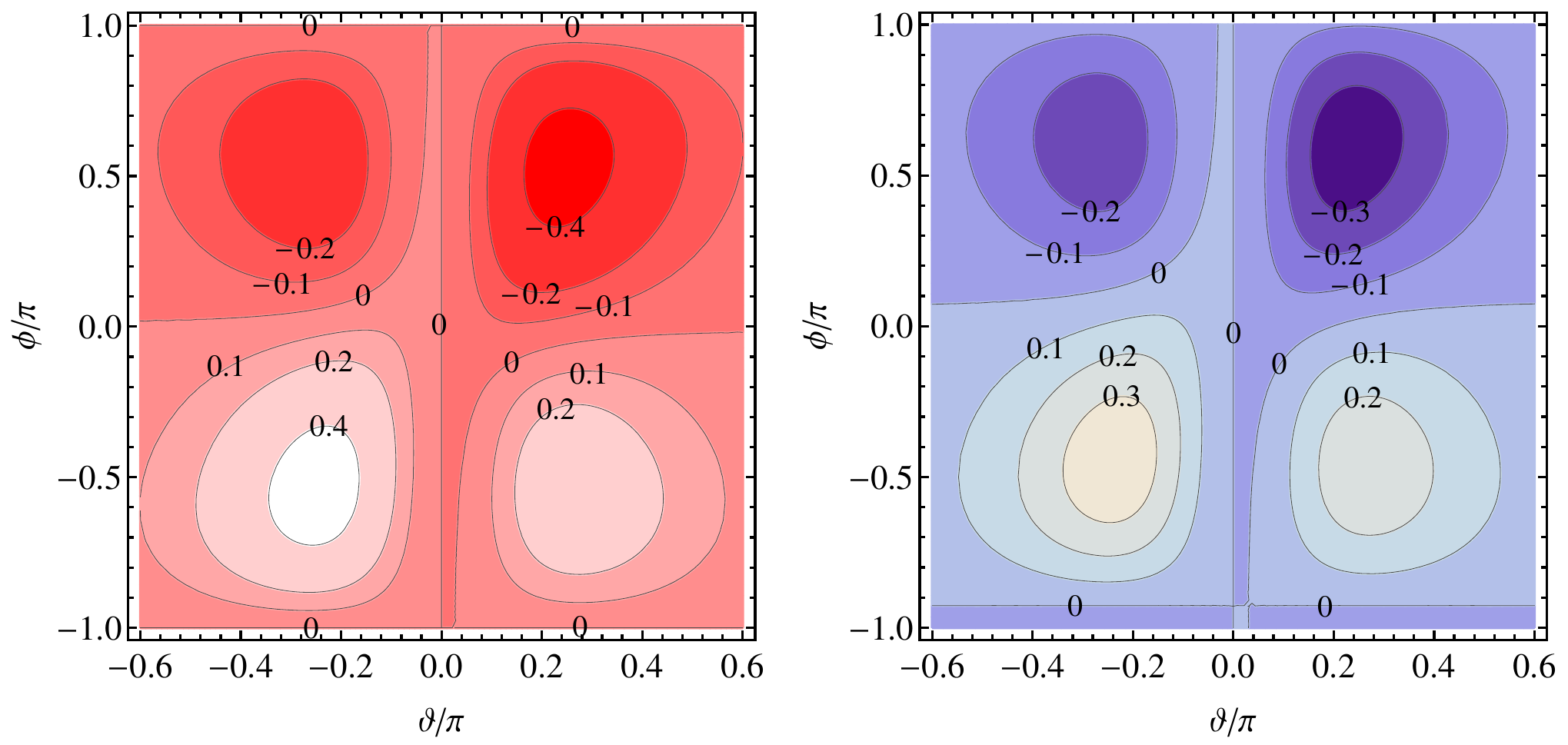} \\
\includegraphics[width=1\linewidth]{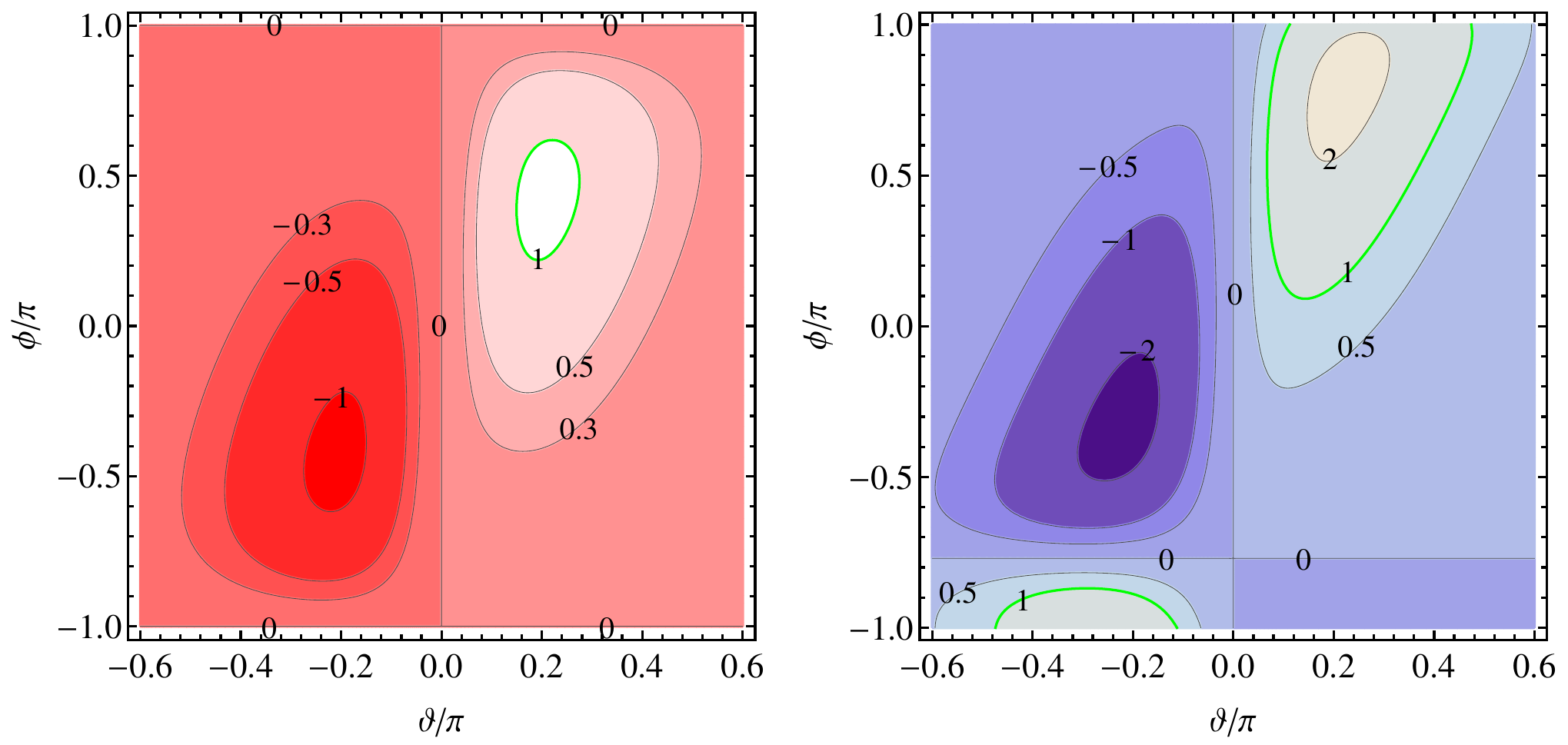}
\end{tabular}
\caption{\label{fig:YB_C23} The contour plots of $Y_{B}/Y^{obs}_B$ in the $\vartheta-\phi$ plane for the case of R-3rd. Here we choose $M_1=5\times10^{11}\,\mathrm{GeV}$, so that only the tau Yukawa couplings are in equilibrium. The first row and the second row are for NO and IO spectrums respectively, and the Dirac CP phase $\delta$ is taken to be $0$ on the left panels and $-\pi/2$ on the right panels. The neutrino oscillation parameters $\theta_{12}$, $\theta_{13}$, $\theta_{23}$, $\delta m^2$ and $\Delta m^2$ are fixed at their best fit values~\cite{Capozzi:2013csa}. The thick green curve represents the experimentally observed values of the baryon asymmetry $Y_B^{obs}=8.66\times10^{-11}$~\cite{Ade:2015xua}.}
\end{figure}

Now we discuss the predictions for matter/antimatter asymmetry for each admissible $R$-matrix. The explicit expressions of the $CP$ asymmetry parameter $\epsilon_{\alpha}$ and the washout mass $\widetilde{m}_{\alpha}$ are given in Appendix~\ref{sec:con_one_CP}. The contour regions of $Y_{B}/Y^{obs}_B$ for the three types of $R$ matrices R-1st, R-2nd and R-3rd are displayed in the plane $\phi$ versus $\vartheta$ in figure~\ref{fig:YB_C11}, figure~\ref{fig:YB_C22} and figure~\ref{fig:YB_C23}, respectively. Here both the three lepton mixing angles and the mass-squared splittings are set to their best fit values~\cite{Capozzi:2013csa} and two representative values $\delta=0, -\pi/2$ are considered.
From Eqs.~\eqref{eq:epsilon_alpha} and \eqref{eq:Yb} we know that the final baryon asymmetry $Y_{B}$ is proportional to $M_{1}$. We shall take $M_1=5\times10^{11}$ GeV for illustration in this work, and the conclusions would not change qualitatively for other values of $M_1$. The neutrino mass spectrum is NO and IO respectively in the first row and second row of these plots, and we choose $\delta=0$ in the left column and $\delta=-\pi/2$ in the right column. Note that the period of $\vartheta$ for R-1st is $\pi$ and there are no phenomenologically viable points in the region of $|\vartheta|>0.6\pi$ for both R-2nd and R-3rd. For R-1st, we find that the experimentally measured value of the baryon asymmetry can be accommodated in the case of NO, while $Y_{B}$ is too small to account for its observed value for IO. The second case R-2nd can result in successful leptogenesis regardless of whether the neutrino mass spectrum is NO or IO. From figure~\ref{fig:YB_C23}, we see that the realistic baryon asymmetry can be generated in the case of R-3rd plus IO while $Y_{B}$ is determined to be smaller than its measured value for R-3rd plus NO.

 \begin{figure}[t!]
\centering
\begin{tabular}{c}
\includegraphics[width=0.42\linewidth]{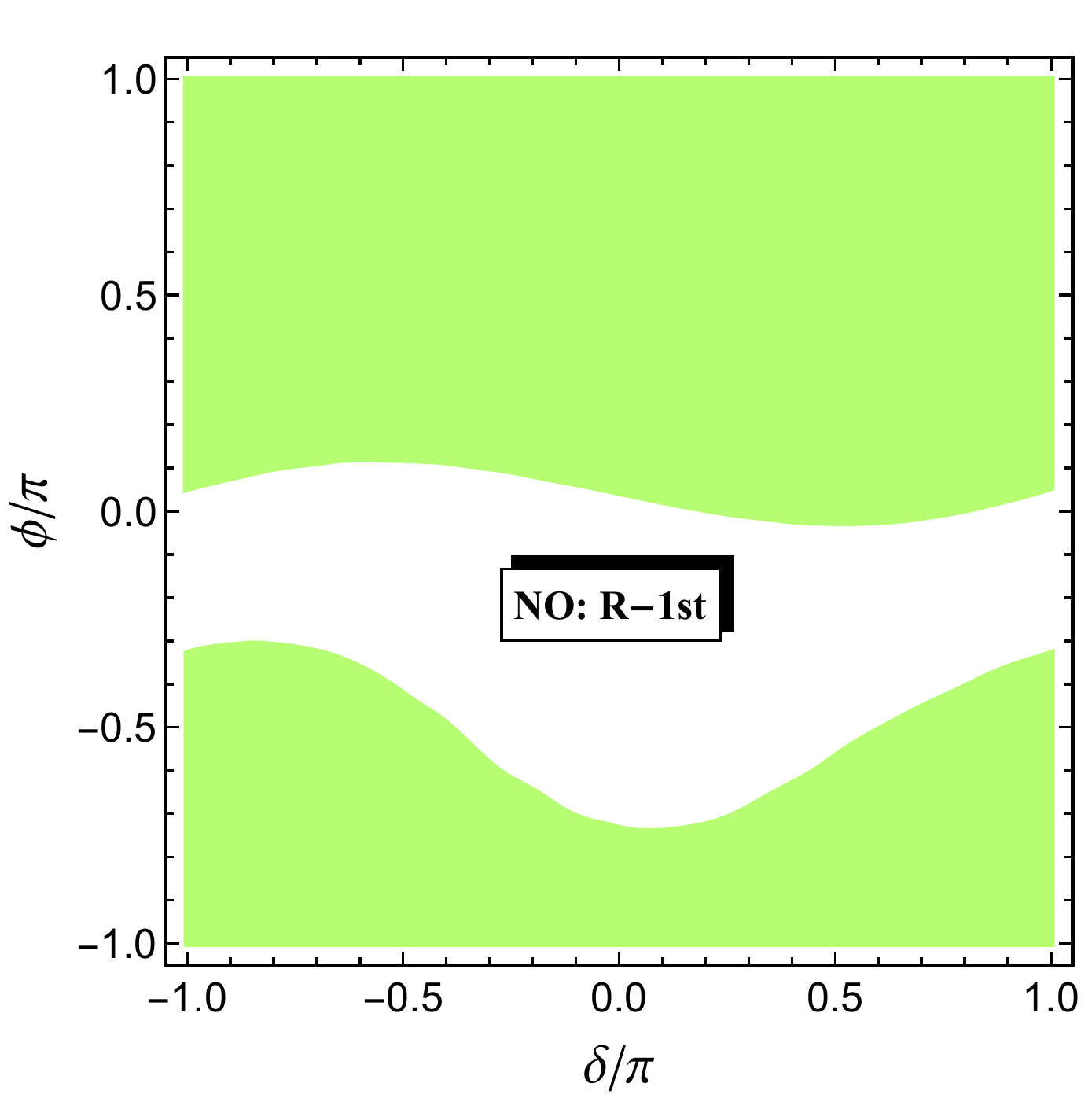}
\includegraphics[width=0.42\linewidth]{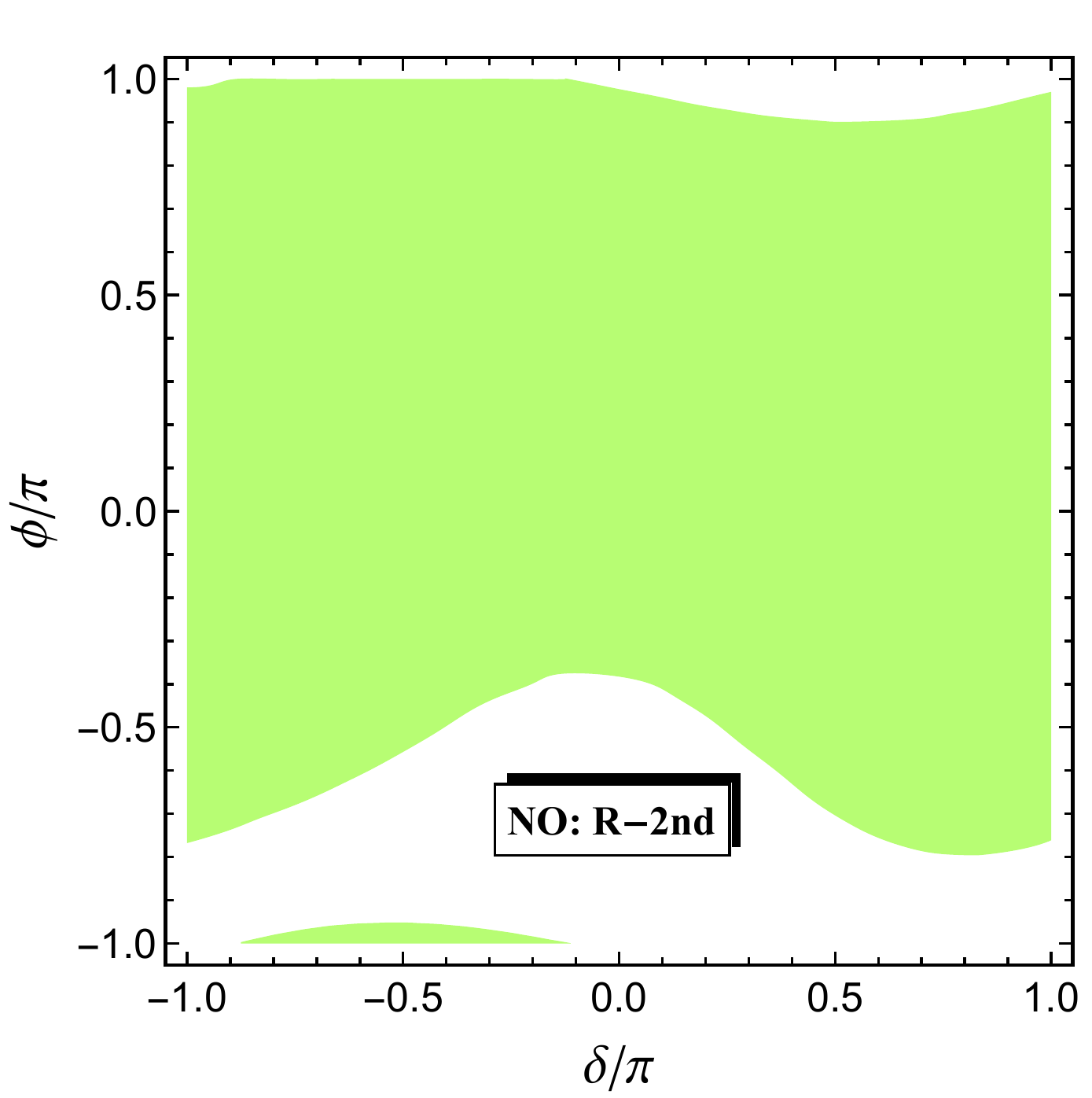}\\
\includegraphics[width=0.42\linewidth]{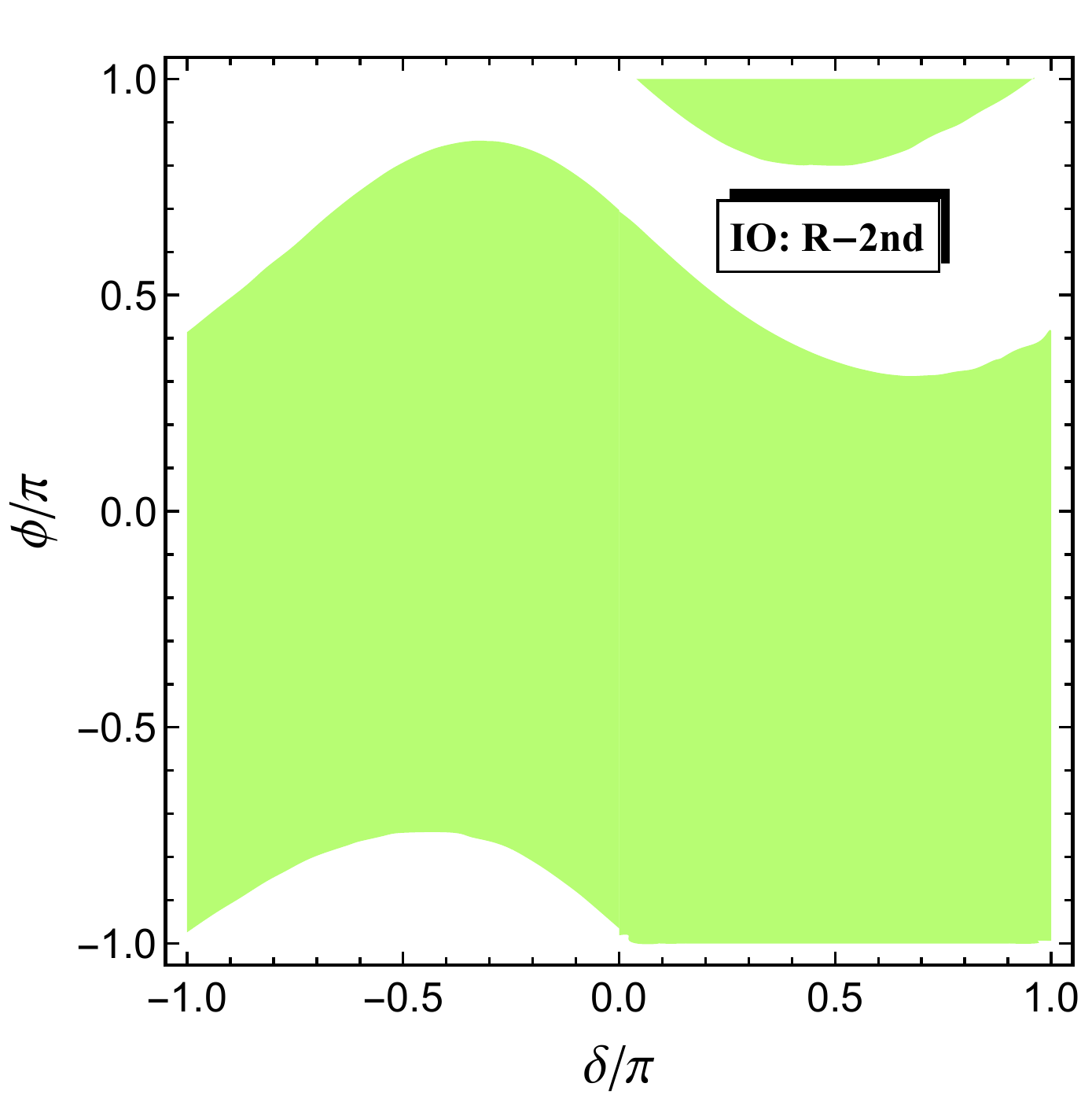}
\includegraphics[width=0.42\linewidth]{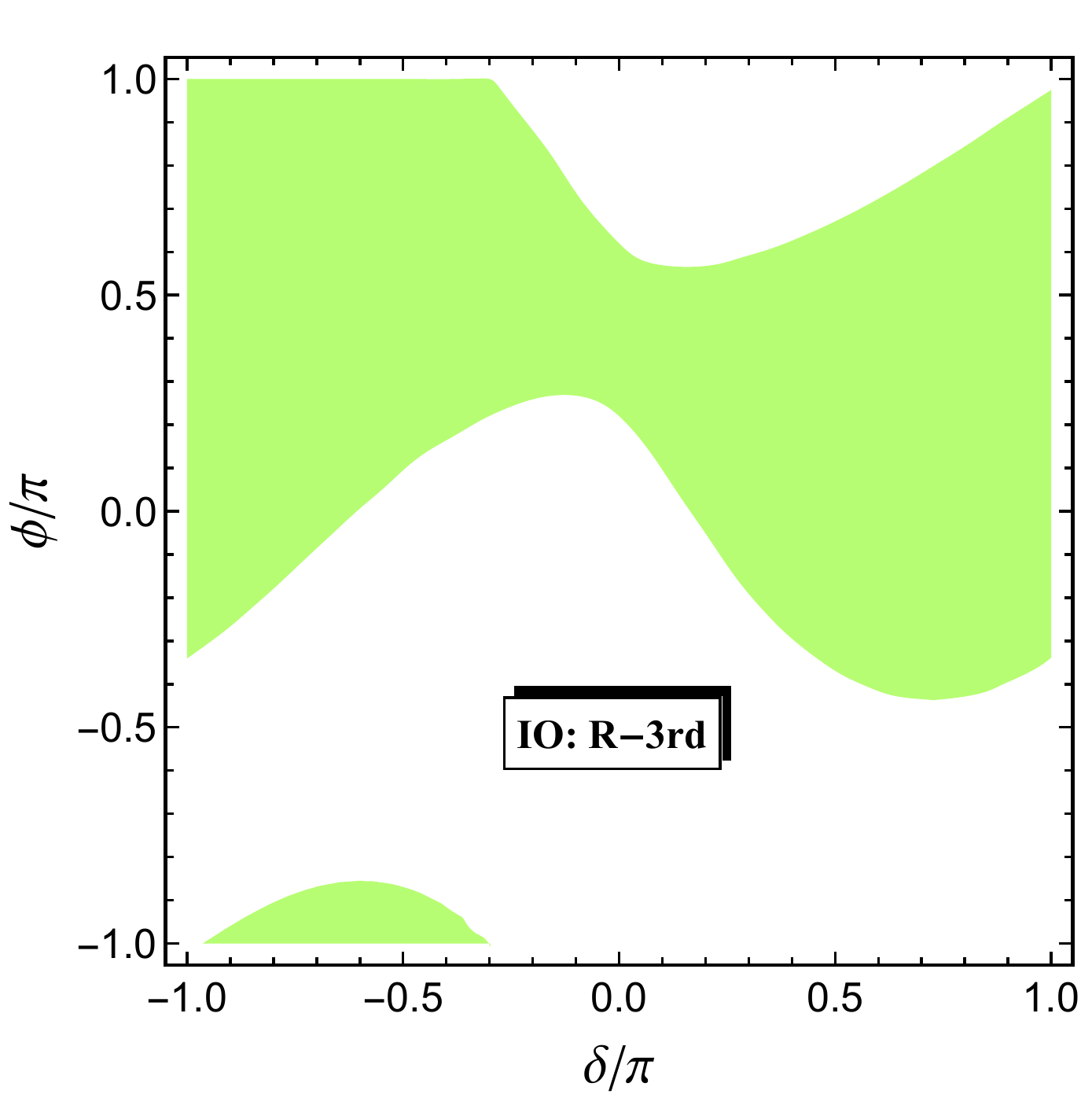}
\end{tabular}
\caption{\label{fig:YB_delta_phi}
The viable regions in the $\delta-\phi$ plane where the cosmological matter/antimatter asymmetry can be generated for certain values of $\vartheta$. For the cases of IO:R-1st and NO:R-3rd, the baryon asymmetry $Y_{B}$ is smaller than its observed value for any values of $\delta$, $\phi$ and $\vartheta$.
}
\end{figure}

We have chosen the representative values $\delta=0$ and $-\pi/2$ for illustration in figures~\ref{fig:YB_C11}--\ref{fig:YB_C23}. In view of the fact that the Dirac CP phase $\delta$ are not constrained at the $3\sigma$ level at present, we display the regions in $\delta-\phi$ plane in figure~\ref{fig:YB_delta_phi}, where successful leptogenesis ($Y_{B}/Y^{obs}_B=1$) can be realized for certain values of $\vartheta$. We notice that the observe baryon asymmetry can be generated in quite large regions of the $\delta-\phi$ plane. For the cases of IO:R-1st and NO:R-3rd with $M_1=5\times10^{11}$ GeV, the baryon asymmetry $Y_{B}$ is too smaller to be in accordance with experimental data. Eq.~\eqref{eq:epsilon_alpha} and Eq.~\eqref{eq:Yb} implies that $Y_{B}$ increases with $M_1$. The maximal value of $M_1$ is $10^{12}$ GeV in the flavored leptogenesis regime, accordingly we find that the maximum of $Y_{B}/Y^{obs}_B$ is $0.226$ and $0.968$ for IO:R-1st and NO:R-3rd respectively, when $\delta$, $\phi$ and $\vartheta$ are treated as free parameters. Therefore these two cases can not lead to successful leptogenesis in our setup even if $\delta$ and $M_1$ are not fixed to the above example values. From the figure~\ref{fig:YB_delta_phi}, we can easily see whether the minimal seesaw model plus a residual CP symmetry is capable of explaining the matter/antimatter asymmetry or not for each possible experimental outcome of $\delta$ and $\phi$.

The $0\nu\beta\beta$ decay process is an important probe for the Majorana nature of the neutrinos. If it was observed in future, neutrinos must be Majorana particles. The amplitude of the $0\nu\beta\beta$ decay is proportional to the effective Majorana neutrino mass $m_{ee}$ which is defined as~\cite{Olive:2016xmw}
\begin{equation}
m_{ee}=\left|\sum_{i}m_iU^{2}_{1i}\right|=\left|m_1c^2_{12}c^2_{13}+m_2s^2_{12}c^2_{13}e^{i\phi}+m_3s^2_{13}e^{-2i\delta}\right|\,.
\end{equation}
In the 2RNH model, the lightest neutrino is massless with $m_{1}=0$ for NO and $m_{3}=0$ for IO. Consequently the expression of $m_{ee}$ can be reduced to
\begin{equation}
\label{mee_2RHN}m_{ee}=\left\{\begin{array}{l}
\big|\sqrt{\delta m^2}\,s^2_{12}c^2_{13}e^{i(\phi+2\delta)}+\sqrt{\Delta m^2+\delta m^2/2}\,s^2_{13}\big|\,, \quad \text{for} \quad \text{NO}\,, \\[0.1in]
\big|\sqrt{-\Delta m^2-\delta m^2/2}\,c^2_{12}c^2_{13}+\sqrt{-\Delta m^2+\delta m^2/2}\,s^2_{12}c^2_{13}e^{i\phi}\big|\,, \quad \text{for} \quad \text{IO}\,,
\end{array}\right.
\end{equation}
where the light neutrino masses in Eq.~\eqref{eq:mass_relation} are used. We see that the effective mass $m_{ee}$ depends on the combination $\phi+2\delta$ for NO and on the phase $\phi$ for IO case. From the panels in the first row of figure~\ref{fig:YB_delta_phi}, we find that the phase $\phi+2\delta$ can take any value between $-\pi$ and $\pi$ when sufficient baryon asymmetries are generated for NO case. Similarly the panels in the second row of figure~\ref{fig:YB_delta_phi} imply that the phase $\phi$ can vary in the range of $[-\pi,\pi]$ if the observed baryon asymmetry is generated for IO. Thus the effective Majorana mass $m_{ee}$ reaches the maximal value when $\phi+2\delta=0$ ($\phi=0$) and the minimal value when $\phi+2\delta=\pi$ $(\phi=\pi)$ for NO (IO) spectrum. Therefore in the parameter space of successful leptogenesis the effective mass $m_{ee}$ varies in the interval
\begin{align}
\nonumber&0.000717\text{eV}\leq m_{ee}\leq0.00449\text{eV} \quad \text{for}\quad \text{NO}\,,\\
\label{eq:mee_general}&0.0130\text{eV}\leq m_{ee}\leq0.0478\text{eV} \quad \text{for}\quad \text{IO}\,.
\end{align}
The predictions of IO case can be tested in future $0\nu\beta\beta$ decay experiments.

\section{\label{sec:Lep_CP_flavor}Leptogenesis with two residual CP transformations or a cyclic residual flavor symmetry }

In this section, we shall proceed to discuss the predictions for leptogenesis in the case that two residual CP transformations or a cyclic residual flavor symmetry is preserved by the seesaw Lagrangian in the 2RHN model.

\subsection{\label{subsec:Lep_two_CP}Two residual CP transformations preserved}
Following the same method as section~\ref{sec:LepG_one_CP}, we investigate what we could learn if the parent CP symmetry at high energy scale is broken down to two residual CP transformations in the neutrino sectors. The lepton fields transform as
\begin{eqnarray}\label{eq:res_CP}
\nonumber&&\nu_{L}
\xmapsto{\text{CP}_{1}}
iX_{\nu1}\gamma_0C\bar{\nu}^{T}_{L}\,,\qquad N_{R}
\xmapsto{\text{CP}_{1}}i\widehat{X}_{N1}\gamma_{0}C\bar{N}^{T}_{R}\,,\qquad \\
&&\nu_{L}\xmapsto{\text{CP}_{2}}iX_{\nu2}\gamma_0C\bar{\nu}^{T}_{L}\,,\qquad N_{R}\xmapsto{\text{CP}_{2}}i\widehat{X}_{N2}\gamma_0C\bar{N}^{T}_{R}\,.
\end{eqnarray}
with $X_{\nu1}\neq X_{\nu2}$ and $\widehat{X}_{N1}\neq\widehat{X}_{N2}$. The invariance of $\lambda$ and $M$ under the action of the above CP transformations $X_{\nu i}$ and $\widehat{X}_{Ni}$ implies
\begin{subequations}
\begin{eqnarray}
\label{eq:cons1}&&\widehat{X}^{\dagger}_{N1}\lambda X_{\nu1}=\lambda^{\ast},~\quad \widehat{X}^{\dagger}_{N1}M\widehat{X}^{\ast}_{N1}=M^{\ast},\\
\label{eq:cons2}&&\widehat{X}^{\dagger}_{N2}\lambda X_{\nu2}=\lambda^{\ast}, ~\quad \widehat{X}^{\dagger}_{N2}M\widehat{X}^{\ast}_{N2}=M^{\ast}\,.
\end{eqnarray}
\label{eq:cons_rcp}
\end{subequations}
Notice that $-X_{\nu i}$, $-\widehat{X}_{Ni}$ leads to the same constraints as $X_{\nu i}$, $\widehat{X}_{Ni}$, hence they are identified as the same residual CP transformation. Because the RH neutrino fields $N_{1R}$ and $N_{2R}$ are assumed to be in the mass eigenstates, $\widehat{X}_{N1}$ and $\widehat{X}_{N2}$ must be diagonal with elements $+1$ or $-1$. i.e.,
\begin{equation}\label{eq:XR12}
\widehat{X}_{N1},\widehat{X}_{N2}=\text{diag}(\pm1,\pm1)\,,
\end{equation}
The light neutrino mass matrix $m_{\nu}$ is given by the seesaw relation. We can straightforwardly check that the residual CP transformations lead to the following two constraints on $m_{\nu}$,
\begin{equation}
\label{eq:mnu_invariant_2CP}X^{T}_{\nu1}m_{\nu}X_{\nu1}=m^{\ast}_{\nu},\qquad X^{T}_{\nu2}m_{\nu}X_{\nu2}=m^{\ast}_{\nu}\,.
\end{equation}
This is exactly the condition that $m_{\nu}$ is invariant under the residual CP transformations $X_{\nu1}$ and $X_{\nu2}$. From Eq.~\eqref{eq:mnu_invariant_2CP} we can derive that the unitary transformation $U_{\nu}$ which diagonalizes $m_{\nu}$ should satisfy
\begin{equation}\label{eq:constraint_resodual_CP}
U^{\dagger}_{\nu}X_{\nu1}U^{\ast}_{\nu}=\widehat{X}_{\nu1},~~\quad U^{\dagger}_{\nu}X_{\nu2}U^{\ast}_{\nu}=\widehat{X}_{\nu2}\,,
\end{equation}
with
\begin{equation}
\begin{array}{ll}
\widehat{X}_{\nu1}, \widehat{X}_{\nu2}=\text{diag}\left(e^{i\alpha_{1,2}}, \pm1, \pm1\right) ~~&~~ \text{for}~~\text{NO}\,,\\
\widehat{X}_{\nu1}, \widehat{X}_{\nu2}=\text{diag}\left(\pm1, \pm1,e^{i\alpha_{1,2}}\right) ~~&~~ \text{for}~~\text{IO}\,, \\
\end{array}
\end{equation}
where $\alpha_1$ and $\alpha_{2}$ are arbitrary real parameters. Eq.~\eqref{eq:constraint_resodual_CP} indicates that both residual CP transformations $X_{\nu1}$ and $X_{\nu2}$ must be symmetric unitary matrices. Using the symmetry properties of $\lambda$, $M$ and $U_{\nu}$ shown in Eqs.~(\ref{eq:cons1},\ref{eq:cons2},\ref{eq:mnu_invariant_2CP}),
we find that the $R$-matrix is subject to the following constraints
\begin{equation}\label{eq:constr_R2}
\widehat{X}_{N1}R^{*}\widehat{X}^{-1}_{\nu1}=R,~~\quad \widehat{X}_{N2}R^{*}\widehat{X}^{-1}_{\nu2}=R\,,
\end{equation}
which imply
\begin{equation}\label{eq:constr_R3}
R=\widehat{X}_{N1} \widehat{X}_{N2}R\widehat{X}_{\nu1} \widehat{X}^{-1}_{\nu2}\,.
\end{equation}
Because the residual CP transformations $X_{\nu1}$, $\hat{X}_{N1}$ are distinct from $X_{\nu2}$, $\hat{X}_{N2}$, the combinations $\widehat{X}_{N1}\widehat{X}_{N2}$ and $\widehat{X}_{\nu1} \widehat{X}^{-1}_{\nu2}$ should take the form\footnote{The same results for the $R$-matrix would be obtained in the case of $\widehat{X}_{N1}\widehat{X}_{N2}=-\text{diag}(1, -1)$, $\widehat{X}_{\nu1} \widehat{X}^{-1}_{\nu2}=P_{\nu}\text{diag}(e^{i(\alpha_1-\alpha_2)},1,-1)P^T_{\nu}$.}
\begin{equation}\label{eq:XR12}
\widehat{X}_{N1}\widehat{X}_{N2}=\text{diag}(1,-1), \qquad
\widehat{X}_{\nu1} \widehat{X}^{-1}_{\nu2}=P_{\nu}\text{diag}(e^{i(\alpha_1-\alpha_2)},1,-1)P^T_{\nu}\,,
\end{equation}
where $P_\nu$ is a permutation matrix with $P_\nu=P_{123},P_{132}$ for NO and $P_\nu=P_{231},P_{321}$ for IO. Here the six $3\times3$ permutation matrices are denoted as
\begin{equation}\label{eq:permutation_matrices}
\begin{array}{lll}
P_{123}=\begin{pmatrix}
1  &~ 0  ~&  0 \\
0  &~ 1  ~&  0\\
0  & ~0~  &  1
\end{pmatrix},~~&~~ P_{132}=\begin{pmatrix}
1  &  ~0~ &  0 \\
0  &  ~0~ &  1 \\
0  &  ~1~ &  0
\end{pmatrix},~~&~~ P_{213}=\begin{pmatrix}
0  &  ~1~  &  0 \\
1  &  ~0~  &  0 \\
0  &  ~0~  &  1
\end{pmatrix},\\
& & \\[-10pt]
P_{231}=\begin{pmatrix}
0   &  ~1~   &  0 \\
0   &  ~0~   &  1  \\
1   &  ~0~   &  0
\end{pmatrix},~~&~~ P_{312}=\begin{pmatrix}
0   &  ~0~  &   1  \\
1   &  ~0~  &   0 \\
0   &  ~1~  &  0
\end{pmatrix},~~&~~ P_{321}=\begin{pmatrix}
0    &   ~0~    &   1  \\
0    &   ~1~    &   0  \\
1    &   ~0~    &   0
\end{pmatrix}\,.
\end{array}
\end{equation}
Inserting Eq.~\eqref{eq:XR12} into Eq.~\eqref{eq:constr_R3} we obtain
\begin{equation}\label{eq:PN_R_PnuT}
RP_{\nu}=\text{diag}(1, -1)RP_{\nu}\text{diag}(e^{i(\alpha_1-\alpha_2)}, 1, -1)\,.
\end{equation}
Consequently the (13) and $(22)$ elements of the matrix $RP_{\nu}$ are vanishing. The explicit forms of the $R$-matrix for all possible values of $P_{\nu}$ are summarized in table~\ref{tab:orth_R}. It is easy to check that all the flavored CP symmetry $\epsilon_{\alpha}$ is vanishing, i.e.
\begin{equation}
\epsilon_{e}=\epsilon_{\mu}=\epsilon_{\tau}=0\,.
\end{equation}
As a result, a net baryon asymmetry can not be generated at leading order in this case, and moderate high order corrections are necessary in order to to make leptogenesis viable. We would like to emphasize that this result is quite general and it is independent of the explicit form of the residual CP transformations $X_{\nu i}$ and $\hat{X}_{N i}$.

\begin{table}[t!]
\addtolength{\tabcolsep}{-2pt}
\begin{center}
\begin{tabular}{|c|c|c|c|c|c|} \hline\hline
\texttt{Mass Ordering} & $P_{\nu}$ & $\hat{\theta} $ & $R$\\\hline
\multirow{3}{*}{NO} & $P_{123}$ & $0,\pi$   & $R=\begin{pmatrix}
0  ~&~ \pm1  ~&~ 0 \\
0  ~&~ 0 ~&~ \pm\xi
\end{pmatrix}$ \\ \cline{2-4}
����&��$P_{132}$ & $\pm\frac{\pi}{2}$   & $R=\begin{pmatrix}
0  ~&~ 0  ~&~ \pm\xi \\
0  ~&~ \mp1 ~&~ 0
\end{pmatrix}$ \\ \hline
\multirow{3}{*}{IO} &��$P_{231}$ & $0,\pi$   & $R=\begin{pmatrix}
 \pm1  ~&~ 0  ~&~ 0\\
 0 ~&~ \pm\xi ~&~ 0
\end{pmatrix}$ \\ \cline{2-4}
��& $P_{321}$ & $\pm\frac{\pi}{2}$   & $R=\begin{pmatrix}
 0  ~&~ \pm\xi ~&~ 0 \\
~ \mp1 ~&~ 0 ~&~ 0
\end{pmatrix}$
\\\hline\hline
\end{tabular}
\end{center}
\renewcommand{\arraystretch}{1.0}
\caption{The explicit form of $R$-matrix for different possible values $P_{\nu}$ , where $\xi$ is either $+1$ or $-1$. \label{tab:orth_R}}
\end{table}

\subsection{\label{subsec:lept_Flavor} A cyclic residual flavor symmetry preserved}

In this section we shall proceed to discuss the implications of the residual flavor symmetry (without residual CP) for leptogenesis. We assume that the flavor symmetry group is broken down to a cyclic $Z_n$ subgroup in the neutrino sector, where the subscript $n$ denotes the order of the cyclic group. Under the action of the generator of the residual flavor symmetry $Z_n$, the neutrinos fields transform as \begin{equation}\label{eq:res_Zn}
\nu_{L}\stackrel{Z_{n}}{\longmapsto} G_{\nu}\nu_{L}\,,~~\quad N_{R}\stackrel{Z_{n}}{\longmapsto} \widehat{G}_{N} N_{R}\,,
\end{equation}
where $G_{\nu}$ is a $3\times3$ unitary matrix with $G^n_{\nu}=1_{3\times3}$ and $\widehat{G}_{N}=\text{diag}\left(\pm1, \pm1\right)$ in our working basis. For this residual symmetry to hold, the Yukawa coupling $\lambda$ and the RH neutrino mass matrix $M\equiv\text{diag}\left(M_1, M_2\right)$ have to fulfill
\begin{equation}\label{eq:cons_flavour}
\widehat{G}^{\dagger}_{N}\lambda G_{\nu}=\lambda\,,~~\quad \widehat{G}^{\dagger}_{N}M\widehat{G}^{\ast}_{N}=M\,.
\end{equation}
Subsequently we can check that the light neutrino mass matrix is invariant under the residual flavor symmetry
\begin{equation}
\label{eq:mnu_flavor_constraint}G_{\nu}^T m_{\nu}G_{\nu}=m_{\nu}\,.
\end{equation}
From this condition we find that the neutrino diagonalization matrix $U_{\nu}$ can diagonalize the residual flavor symmetry transformation $G_{\nu}$ as well,
\begin{equation}\label{eq:U_flavour_cons}
U^{\dagger}_{\nu}G_{\nu}U_{\nu}=\widehat{G}_{\nu},~~\text{with}~~
\widehat{G}_{\nu}=\left\{\begin{array}{lll}
\text{diag}\left(e^{i\alpha}, \pm1, \pm1\right) ~~& \text{for} ~~& \text{NO} \\
\text{diag}\left(\pm1, \pm1, e^{i\alpha}\right) ~~& \text{for} ~~& \text{IO}
\end{array}\right.\,,
\end{equation}
where $\alpha=2\pi k/n$ with $k$ coprime to $n$ is a rational multiple of $\pi$. Notice that the maximal invariance group of the neutrino mass matrix is $U(1)\times Z_2\times Z_2$ not a Klein group $Z_2\times Z_2$ because one light neutrino mass is zero in this case. From Eq.~\eqref{eq:cons_flavour} and Eq.~\eqref{eq:U_flavour_cons}, we can determine that the residual flavor symmetry gives rise to the following constraint on the $R$-matrix,
\begin{equation}\label{eq:R_cons_flavour}R=\widehat{G}_{N}R\widehat{G}_{\nu}\,.
\end{equation}
\begin{table}[t!]
\addtolength{\tabcolsep}{-2pt}
\begin{center}
\begin{tabular}{|c|c|c|c|c|c|} \hline\hline
$\widehat{G}_{N} $ & $ \widehat{G}_{\nu} $ & $R$ (NO)  & $R$ (IO)  \\ \hline
$\text{diag}(1,-1)$   & $\mathcal{D}(1,1)$   & \xmark & \xmark \\ \hline
$\text{diag}(1,-1)$   & $\mathcal{D}(1,-1)$ & $\begin{pmatrix}
0  ~&~ \pm1  ~&~ 0 \\
0  ~&~ 0 ~&~ \pm\xi
\end{pmatrix}$  & $\begin{pmatrix}
 \pm1  ~&~ 0 ~&~ 0  \\
 0 ~&~ \pm\xi ~&~ 0
\end{pmatrix}$ \\ \hline
$\text{diag}(1, -1)$   & $\mathcal{D}(-1,1)$ & $\begin{pmatrix}
0  ~&~  0 ~&~  \pm\xi \\
0  ~&~  \mp1   ~&~ 0
\end{pmatrix}$  & $\begin{pmatrix}
 0 ~&~  \pm\xi ~&~ 0  \\
  \mp1   ~&~ 0 ~&~ 0
\end{pmatrix}$   \\  \hline
$\text{diag}(1, -1)$   & $\mathcal{D}(-1,-1)$ & \xmark & \xmark  \\  \hline
$\text{diag}(-1,1)$   & $\mathcal{D}(1,1)$   & \xmark  & \xmark \\ \hline
$\text{diag}(-1, 1)$   & $\mathcal{D}(1,-1)$ & $\begin{pmatrix}
0  ~&~  0 ~&~  \pm\xi \\
0  ~&~  \mp1   ~&~ 0
\end{pmatrix}$   & $\begin{pmatrix}
  0 ~&~  \pm\xi ~&~ 0  \\
 \mp1   ~&~ 0 ~&~ 0
\end{pmatrix}$   \\  \hline
$\text{diag}(-1,1)$   & $\mathcal{D}(-1,1)$ & $\begin{pmatrix}
0  ~&~ \pm1  ~&~ 0 \\
0  ~&~ 0 ~&~ \pm\xi
\end{pmatrix}$  & $\begin{pmatrix}
 \pm1  ~&~ 0 ~&~ 0  \\
 0 ~&~ \pm\xi~&~ 0
\end{pmatrix}$  \\  \hline
 $\text{diag}(1, 1)$   & $\mathcal{D}(-1,-1)$ & \xmark  & \xmark\\ \hline
$-\text{diag}(1,1)$ & $\mathcal{D}(1,1)$   & \xmark  & \xmark  \\  \hline
$-\text{diag}(1,1)$ & $\mathcal{D}(1,-1)$ & \xmark  & \xmark \\  \hline
$-\text{diag}(1,1)$ & $\mathcal{D}(-1,1)$ & \xmark  & \xmark \\  \hline\hline
\end{tabular}
\end{center}
\renewcommand{\arraystretch}{1.0}
\caption{\label{tab:orth_R_v3}The explicit form of the $R$-matrix for different possible values of $\widehat{G}_{N}$ and $\widehat{G}_{\nu}$. The notation ``\xmark'' means that the solution for $R$-matrix does not exist, and $\mathcal{D}(x,y)$ with $x,y=\pm1$ refers to $\text{diag}(e^{i\alpha},x,y)$ and $\text{diag}(x,y,e^{i\alpha})$ for NO and IO respectively. Note that the residual flavor symmetry gives no constraint on the $R$-matrix for $\widehat{G}_{N}=-\text{diag}(1,1)$ and $\widehat{G}_{\nu}=\mathcal{D}(-1,-1)$. }
\end{table}
The explicit forms of the $R$-matrix for all possible values of $\widehat{G}_{\nu}$ and $\widehat{G}_{N}$ are listed in table~\ref{tab:orth_R_v3}. We see that there is only one nonzero element in each row of the $R$-matrix, consequently all the flavored CP asymmetries are zero
\begin{equation}
\epsilon_{e}=\epsilon_{\mu}=\epsilon_{\tau}=0\,.
\end{equation}
Hence the baryon asymmetry $Y_{B}$ would be generally vanishing in the 2RHN model with a remnant $Z_{n}$ flavor symmetry in the neutrino sector. In a concrete model, one could take into account the non-leading corrections arising from loop effects and higher dimensional operators to explain the correct size of matter/antimatter asymmetry~\cite{Jenkins:2008rb_leptogenesis}.

\section{\label{sec:example}Examples in $\Delta(6n^2)$ flavor symmetry and CP}

In section~\ref{sec:LepG_one_CP}, we have presented the general results for leptogenesis in the scenario that one residual CP transformation is preserved in the neutrino sector. In order to show concrete examples, we shall study the case that the single residual CP transformation arises from the breaking of the generalized CP symmetry compatible with the $\Delta(6n^2)$ flavor group.

$\Delta(6n^2)$ as flavor symmetry group and the resulting phenomenological consequence for lepton flavor mixing have been discussed in the literature~\cite{King:2013vna,King:2014rwa,Hagedorn:2014wha,Ding:2014ora}. In the present work, we shall adopt the conventions and notations of Ref.~\cite{Ding:2014ora} for the $\Delta(6n^2)$ group. The $\Delta(6n^2)$ group is isomorphic to $(Z_n\times Z_n)\rtimes S_3$ where the index $n$ is a generic integer. The $\Delta(6n^2)$ group can be generated by four generators $a$, $b$, $c$ and $d$ which obey the following relations~\cite{Ding:2014ora,Escobar:2008vc}:
\begin{equation} \label{eq:relations_Delta}
\begin{array}{l}
a^3=b^2=(ab)^2=c^{n}=d^{n}=1, \qquad   cd=dc, \qquad aca^{-1}=c^{-1}d^{-1}, \\
 ada^{-1}=c\,, \qquad  bcb^{-1}=d^{-1}, \quad bdb^{-1}=c^{-1}\,.
\end{array}
\end{equation}
The $\Delta(6n^2)$ group has $6n^2$ elements which can be expressed as
\begin{equation}
g=a^{\alpha}b^{\beta}c^{\gamma}d^{\delta}\,,~~ \quad \alpha=0,1,2,\quad \beta=0,1,\quad \gamma,\delta=0,1,\ldots, n-1\,.
\end{equation}
The group $\Delta(6n^2)$ has one-dimensional, two-dimensional, three-dimensional and six-dimensional irreducible representations~\cite{Ding:2014ora,Escobar:2008vc}. It bas been shown that $\Delta(6n^2)$ has $2(n-1)$ three-dimensional irreducible representations denoted by $\mathbf{3}_{k,l}$ in which the explicit form of the four generators can be chosen as
\begin{equation}
\label{eq:rep_3d_Delta6n2}\mathbf{3}_{k, l}~:~ a=\begin{pmatrix}0 & ~1~ &0 \\ 0&~0~&1 \\
   1&~0~&0\end{pmatrix},~~
   b=(-1)^{k}\begin{pmatrix} 0 &~0~ &1 \\ 0&~1~&0 \\
   1&~0~&0\end{pmatrix},~~
   c=\begin{pmatrix} \eta^{l}&~0~ &0 \\ 0&~\eta^{-l}~&0 \\
   0&~0~&1\end{pmatrix},~~
   d=\begin{pmatrix}1 &~0~ &0 \\ 0&~\eta^{l}~&0 \\
   0&~0~&\eta^{-l}\end{pmatrix}\,,
\end{equation}
where $\eta\equiv e^{2\pi i/n}$, $k=1, 2$ and $l=1,2,\ldots, n-1$. In the following, without loss of generality we shall embed the three generations of left-handed lepton doublets into the faithful triplet $\mathbf{3}_{1,1}$ which is denoted by $\mathbf{3}$ for simplicity, while the two right-handed neutrinos are assumed to transform as a doublet of $\Delta(6n^2)$. As has been shown in Ref.~\cite{Ding:2014ora}, the most general CP transformation consistent with the $\Delta(6n^2)$ flavor symmetry is of the same form as the flavor symmetry transformation in the basis of Eq.~\eqref{eq:rep_3d_Delta6n2}, i.e.
\begin{equation}
X_{\bf r}=\rho_{\bf r}(g), \qquad g\in \Delta(6n^2)\,,
\end{equation}
where $\rho_{\mathbf{r}}(g)$ denotes the representation matrix of the element $g$ in the irreducible representation $\mathbf{r}$ of the $\Delta(6n^2)$ group. Moreover, we assume that the $\Delta(6n^2)$ flavor symmetry is broken down to an abelian subgroup $G_{l}$ in the charged lepton sector and $G_{l}$ is capable of distinguishing among the three generations of the charged leptons. As a result, the charged lepton mass matrix is invariant under the action of the generator $g_{l}$ of $G_{l}$,
\begin{equation}
\label{eq:cons_char_lep}\rho^{\dagger}_{\mathbf{3}}(g_l)m^{\dagger}_lm_{l}\rho_{\mathbf{3}}(g_l)=m^{\dagger}_lm_{l}\,,
\end{equation}
where the charged lepton mass matrix $m_l$ is given in the right-left basis. The matrix $\rho_{\mathbf{3}}(g_l)$ can be diagonalized by a unitary transformation $U_{l}$,
\begin{table}[t!]
\centering
\begin{tabular}{|c|c|c|c|}
\hline \hline
 &  &     \\ [-0.15in]
 $G_{l}$ &  $U_{l}$  &  \texttt{Constraints}  \\

  &   &      \\ [-0.16in]\hline
 &   &       \\ [-0.15in]

 &   &  $s+t\neq0$ $\text{mod}(n)$  \\[-0.16in]

$\langle c^{s}d^{t}\rangle$  &   $\begin{pmatrix}
 1  &~  0   &~  0  \\
 0  &~  1   &~  0  \\
 0  &~  0   &~  1
\end{pmatrix}$  &  $s-2t\neq0$ $\text{mod}(n)$   \\[-0.16in]

  &    &   $t-2s\neq0$ $\text{mod}(n)$   \\

&   &      \\ [-0.14in]\hline

 &   &     \\ [-0.14in]

$\langle bc^{s}d^{t}\rangle$  &   $\frac{1}{\sqrt{2}}\left(\begin{array}{ccc}
e^{-i\pi\frac{s+t}{2n}}  &  0~   &~ e^{-i\pi\frac{s+t}{2n}}\\
0                        &  \sqrt{2}~  &~  0  \\
-e^{i\pi\frac{s+t}{2n}}  &  0~   &~ e^{i\pi\frac{s+t}{2n}}
\end{array}
\right)$  &  $s-t\neq0,\frac{n}{3},\frac{2n}{3}$ $\text{mod}(n)$    \\[-0.14in]
&   &      \\ \hline
&   &   \\[-0.14in]

$\langle ac^{s}d^{t}\rangle$  &   $\frac{1}{\sqrt{3}}\left(\begin{array}{ccc}
e^{-2i\pi\frac{s}{n}}   ~&~   \omega^2e^{-2i\pi\frac{s}{n}} ~&~  \omega
e^{-2i\pi\frac{s}{n}} \\
e^{-2i\pi\frac{t}{n}}  ~&~  \omega e^{-2i\pi\frac{t}{n}} ~&~
\omega^2e^{-2i\pi\frac{t}{n}} \\
1    ~&~   1    ~&~  1
\end{array}
\right)$  &  ---  \\[-0.14in]

&   &      \\ \hline
 &   &       \\ [-0.16in]

  &   &    \\[-0.15in]

$\langle a^2c^{s}d^{t}\rangle$  &   $\frac{1}{\sqrt{3}}\left(\begin{array}{ccc}
e^{-2i\pi\frac{t}{n}}  ~&~  \omega^2e^{-2i\pi\frac{t}{n}}  ~&~  \omega
e^{-2i\pi\frac{t}{n}} \\
e^{2i\pi\frac{s-t}{n}}   ~&~  \omega e^{2i\pi\frac{s-t}{n}} ~&~
\omega^2e^{2i\pi\frac{s-t}{n}} \\
1   ~&~  1  ~&~  1
\end{array}
\right)$  &  ---  \\[-0.14in]

&   &      \\ \hline
 &   &       \\ [-0.15in]

$\langle abc^{s}d^{t}\rangle$  &   $\frac{1}{\sqrt{2}}\left(\begin{array}{ccc}
e^{i\pi\frac{t-2s}{2n}}  ~&~  e^{i\pi\frac{t-2s}{2n}}   ~&~    0  \\
-e^{-i\pi\frac{t-2s}{2n}}  ~&~ e^{-i\pi\frac{t-2s}{2n}}  ~&~  0  \\
0   ~&~  0   ~&~  \sqrt{2}
\end{array}
\right)$  &  $t\neq0,\frac{n}{3},\frac{2n}{3}$  \\

&   &      \\ [-0.14in]\hline
 &   &       \\ [-0.15in]

$\langle a^2bc^{s}d^{t}\rangle$  &   $\frac{1}{\sqrt{2}}\left(\begin{array}{ccc}
\sqrt{2}  ~&~  0   ~&~   0  \\
0   ~&~  e^{i\pi\frac{s-2t}{2n}}   ~&~   e^{i\pi\frac{s-2t}{2n}} \\
0   ~&~  -e^{-i\pi\frac{s-2t}{2n}}  ~&~  e^{-i\pi\frac{s-2t}{2n}}
\end{array}
\right)$  &  $s\neq0,\frac{n}{3},\frac{2n}{3}$   \\
&   &      \\ [-0.14in]
\hline\hline
\end{tabular}
\caption{\label{tab:cle_diagonal_matrix} The unitary transformation $U_{l}$ for the possible remnant subgroup $G_{l}$. Here the notation $\langle g\rangle$ denotes a group generated by the element $g$. The allowed values of the parameters $s$ and $t$ are $s,t=0, 1, \ldots , n-1$ and $\omega=e^{2\pi i/3}$ is the cube root of unit. Note that the identity $\left(ac^{t}d^{t-s}\right)^2=a^2c^sd^t$ is fulfilled, consequently the unitary matrix $U_{l}$ for $G_{l}=\langle a^2c^{s}d^{t}\rangle$ can be obtained from that corresponding one of $G_{l}=\langle ac^{s}d^{t}\rangle$ through the replacement $s\rightarrow t$ and $t\rightarrow t-s$. The constraints on the parameters $s$ and $t$ are to eliminate the degeneracy among the eigenvalues of the generator of $G_{l}$, and they can be completely relaxed by extending $G_{l}$ to be the direct product of several cyclic groups~\cite{Yao:2015dwa}. }
\end{table}

\begin{equation}
U^{\dagger}_{l}\rho_{\mathbf{3}}(g_{l})U_{l}=\rho^{\text{diag}}_{\mathbf{3}}(g_{l})\,.
\end{equation}
Then Eq.~\eqref{eq:cons_char_lep} implies that $U_{l}$ also diagonalizes the charged lepton mass matrix $m^{\dagger}_lm_{l}$. Notice that $U_l$ is uniquely determined up to permutations and phases of their column vectors. All possible residual subgroup $G_{l}$ and the corresponding diagonalization matrices $U_{l}$ are summarized in table~\ref{tab:cle_diagonal_matrix}, where $G_{l}$ is assumed to be generated by a single generator. If we further take into account the case that $G_{l}$ is a product of several cyclic groups, the constraints on the parameters $s$ and $t$ in table~\ref{tab:cle_diagonal_matrix} would be removed, yet no new additional form of $U_{l}$ is generated~\cite{Yao:2015dwa}. In the neutrino sector, a single remnant CP transformation $X_{\nu}$ is preserved by the neutrino mass matrix such that the neutrino mixing matrix $U_{\nu}$ is of the form of Eq.~\eqref{eq:Unu_one_CP}, as shown in section~\ref{sec:LepG_one_CP}. Hence the lepton mixing matrix is determined to be given by
\begin{equation}
U=P_{l}U^{\dagger}_{l}\Sigma_{\nu}O_{3\times3}\hat{X}^{-\frac{1}{2}}_{\nu}\,,
\end{equation}
where $P_{l}$ is a generic $3\times3$ permutation matrix since the charged lepton masses can not be predicted in this approach.
One can straightforwardly check that two pairs of subgroups $\left\{G_{l}, X_{\nu}\right\}$ and $\left\{G^{\prime}_{l}, X^{\prime}_{\nu}\right\}$ would yield the same results for the PMNS mixing matrix~\cite{Lu:2016jit}, if they are related by a similarity transformation $\Omega$,
\begin{equation}
\rho_{\mathbf{3}}(g^\prime_l)=\Omega\rho_{\bf 3}(g_l)\Omega^\dagger, \quad  X^{\prime}_{\nu}=\Omega X_{\nu}\Omega^T\,,
\end{equation}
where $g_{l}$ and $g'_{l}$ denote the generator of $G_{l}$ and $G'_{l}$ respectively. Moreover generally, we denote the mixing matrices predicted by two generic residual symmetries $\left\{G_{l}, X_{\nu}\right\}$ and $\left\{G^{\prime}_{l}, X^{\prime}_{\nu}\right\}$ as
\begin{equation}
U=P_{l}U^{\dagger}_{l}\Sigma_{\nu}O_{3\times3}\hat{X}^{-\frac{1}{2}}_{\nu},\quad U'=P'_{l}U'^{\dagger}_{l}\Sigma'_{\nu}O'_{3\times3}\hat{X}'^{-\frac{1}{2}}_{\nu}\,,
\end{equation}
The condition under which $U$ and $U'$ essentially lead to the same mixing pattern is found to be~\cite{Lu:2016jit}
\begin{equation}
\label{eq:equiv_condition}\Sigma\Sigma^{T}=Q_{L}P_{L}\Sigma'\Sigma'^{T}P^{T}_{L}Q_{L}\,,
\end{equation}
where $\Sigma\equiv U^{\dagger}_{l}\Sigma_{\nu}$, $\Sigma'\equiv U'^{\dagger}_{l}\Sigma'_{\nu}$, $P_{L}\equiv P^{T}_{l}P'_{l}$, and $Q_{L}$ is a diagonal phase matrix. As stated above, we assume that the concerned $\Delta(6n^2)$ flavor group and CP symmetry are broken to an abelian subgroup in the charged lepton sector and to a single remnant CP transformation $X_{\nu}$ in the neutrino sector. Thus $X_{\nu}$ has to be a symmetric unitary matrix and it can be
\begin{equation}\label{eq:X_nu}
X_{\nu}=\rho_{\mathbf{3}}(c^{x}d^{y}), \quad \rho_{\mathbf{3}}(bc^{x}d^{-x}), \quad \rho_{\mathbf{3}}(abc^{x}d^{2x}), \quad \rho_{\mathbf{3}}(a^2bc^{2x}d^{x}), \quad x,y=0,1,\ldots,n-1\,,
\end{equation}
which are related with each other by similarity transformation as follows
\begin{eqnarray}
\nonumber&& \rho_{\bf 3}(b)\rho_{\bf 3}(bc^{x}d^{-x})\rho^{T}_{\bf 3}(b)=\rho_{\bf 3}(bc^{x}d^{-x}),\\
\nonumber && \rho_{\bf 3}(a^{2})\rho_{\bf 3}(bc^{x}d^{-x})\rho^{T}_{\bf 3}(a^{2})=\rho_{\bf 3}(abc^{x}d^{2x})\,,\\
\label{eq:con_rel_CP} && \rho_{\bf 3}(ad^{2x})\rho_{\bf 3}(bc^{x}d^{-x})\rho^{T}_{\bf 3}(ad^{2x})=\rho_{\bf 3}(a^2bc^{2x}d^{x}) \,.
\end{eqnarray}
Hence it is sufficient to consider the choices of $X_{\nu}=\rho_{\bf 3}(c^{x}d^{y})$ and  $X_{\nu}=\rho_{\bf 3}(bc^{x}d^{-x})$ with $x,y=0,1,\ldots,n-1$. The corresponding Takagi factorization matrix can be read out as,
\begin{eqnarray}
\label{eq:Unu_1}&& X_{\nu}=\rho_{\bf 3}(c^{x}d^{y}),\quad \Sigma_{\nu}=\text{diag}(e^{\frac{x\pi i}{n}},e^{\frac{(y-x)\pi i}{n}},e^{-\frac{y\pi i}{n}})\,,\\
\label{eq:Unu_2}&&X_{\nu}=\rho_{\bf 3}(bc^{x}d^{-x}),\quad \Sigma_{\nu}=\begin{pmatrix}
 0 &~ -ie^{\frac{x\pi i}{n}} ~& e^{\frac{x\pi i}{n}} \\
 \sqrt{2}e^{-\frac{2x\pi i}{n}} &~ 0 ~& 0 \\
 0 &~ ie^{\frac{x\pi i}{n}} ~& e^{\frac{x\pi i}{n}} \\
\end{pmatrix}\,.
\end{eqnarray}
Furthermore taking into account the following conjugate relations
\begin{eqnarray}
\nonumber&&b(abc^sd^t)b^{-1}=a^2bc^{-t}d^{-s},\\
\nonumber&&\rho_{\mathbf{3}}(b)\rho_{\mathbf{3}}(c^{x}d^{y})\rho^{T}_{\mathbf{3}}(b)=\rho_{\mathbf{3}}(c^{-y}d^{-x}),\\
&&\rho_{\mathbf{3}}(b)\rho_{\mathbf{3}}(bc^{x}d^{-x})\rho^{T}_{\mathbf{3}}(b)=\rho_{\mathbf{3}}(bc^{x}d^{-x})\,,
\end{eqnarray}
we only need to consider eight possible remnant symmetries constituted by $G_{l}=\langle c^{s}d^{t}\rangle$, $\langle bc^{s}d^{t}\rangle$, $\langle ac^{s}d^{t}\rangle$, $\langle abc^{s}d^{t}\rangle$ and $X_{\nu}=\rho_{\bf 3}(c^{x}d^{y})$, $X_{\nu}=\rho_{\bf 3}(bc^{x}d^{-x})$. In this section, we shall investigate the predictions for lepton flavor mixing and matter-antimatter asymmetry via leptogenesis for each possible case. The explicit form of the lepton mixing matrix and the expressions of the mixing parameters and rephasing bilinear invariants are given in Appendix~\ref{sec:D6n2_example}.

\begin{figure}[t!]
\centering
\begin{tabular}{c}
\includegraphics[width=1\linewidth]{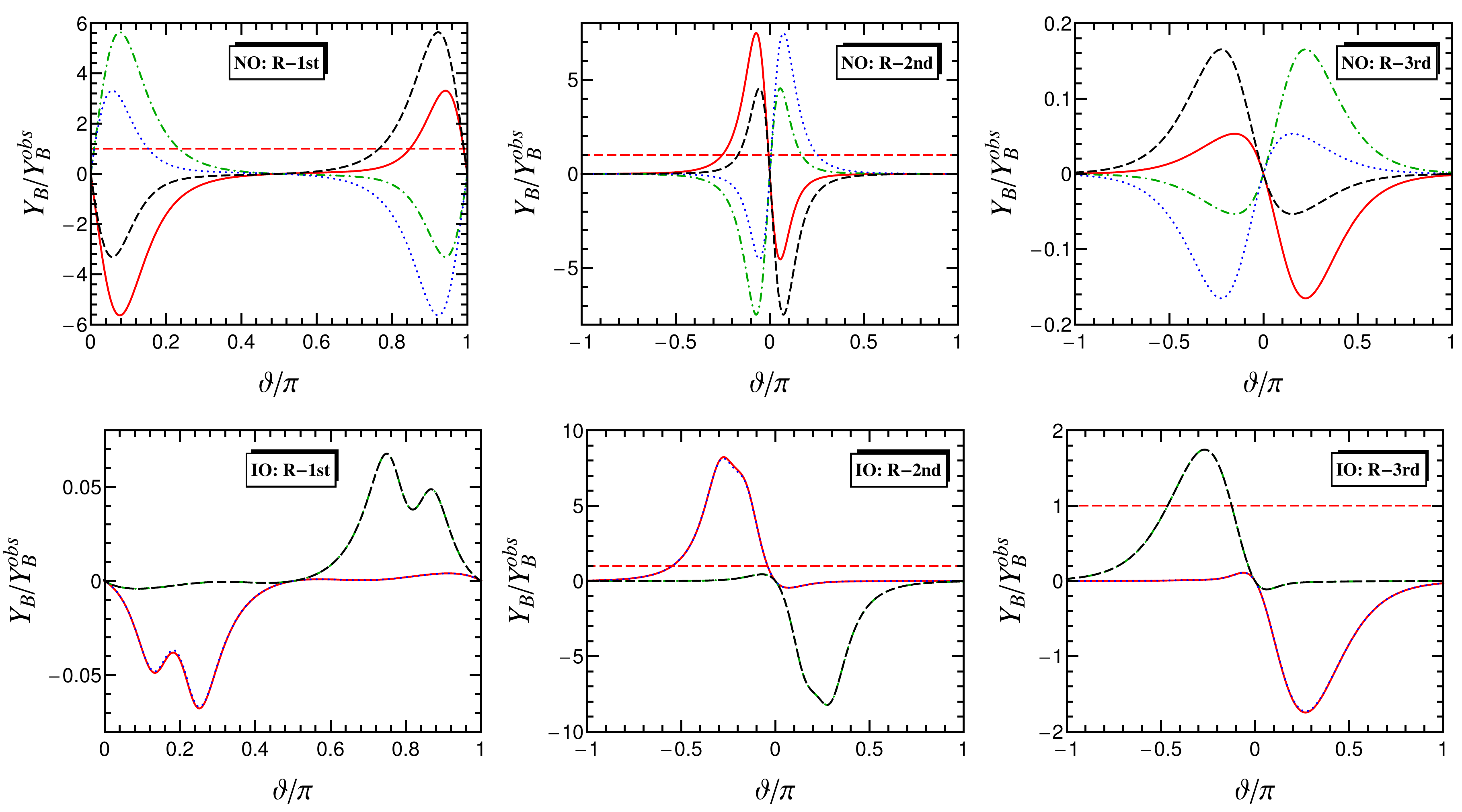}
\end{tabular}
\caption{\label{fig:YB_II}$Y_B/Y_B^{obs}$ as a function of the parameter $\vartheta$ in case II, where we choose the RH neutrino mass $M_1=5\times10^{11}\,\mathrm{GeV}$. The red solid, green dash-dotted, blue dotted and black dashed lines correspond to the four best fitting points shown in table~\ref{tab:best_fit} one by one. The horizontal red dashed line represents the experimental measured value $Y^{obs}_{B}$. The neutrino mass spectrum is NO and IO in the first row and the second row respectively. The panels in the left, middle and right columns are for the three admissible forms of the $R$-matrix such as R-1st, R-2nd and R-3rd respectively.}
\end{figure}

\begin{figure}[t!]
\centering
\begin{tabular}{c}
\includegraphics[width=1\linewidth]{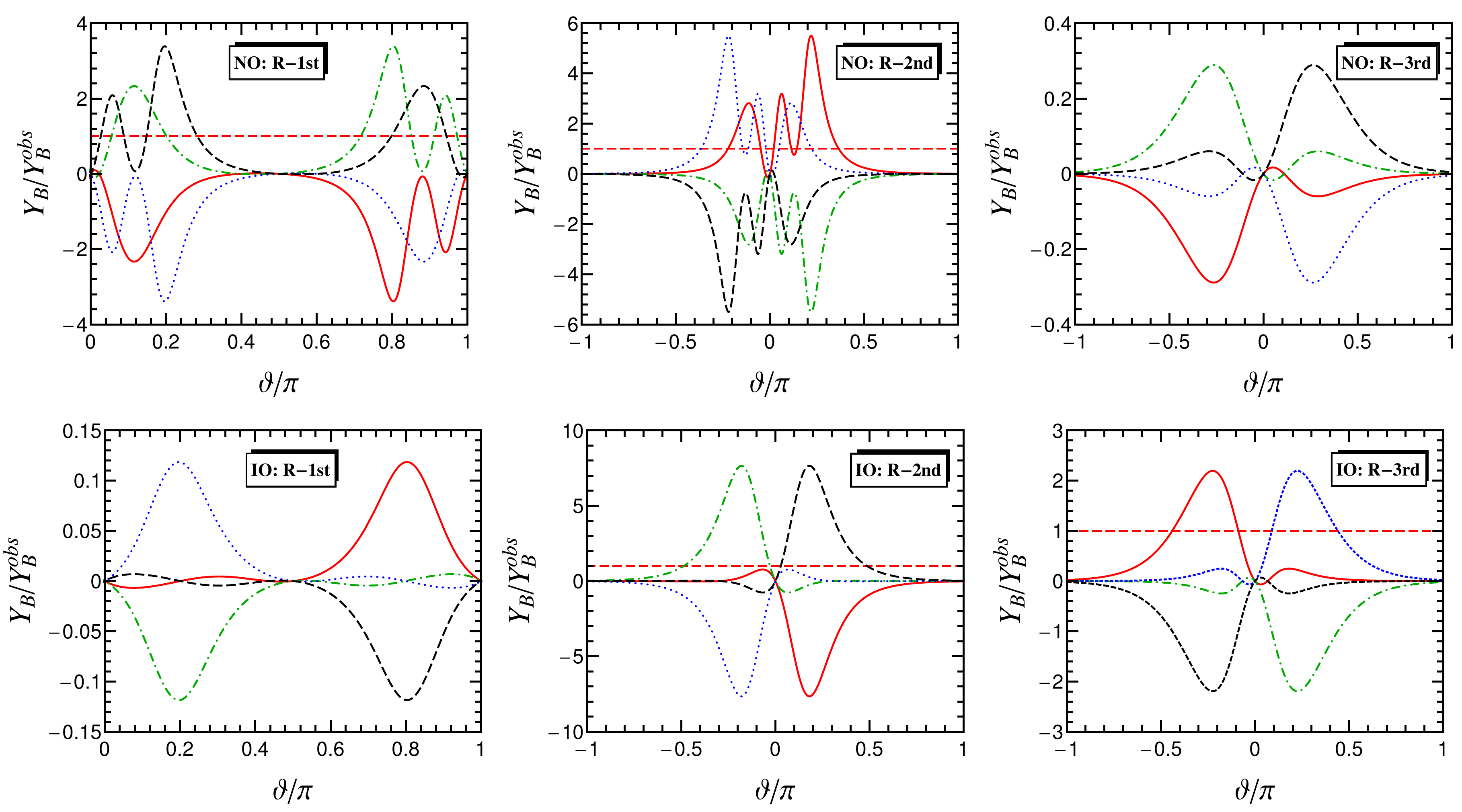}
\end{tabular}
\caption{\label{fig:YB_III_P123}$Y_B/Y_B^{obs}$ as a function of the parameter $\vartheta$ for the mixing pattern $U_{III,1}$ with $\varrho_{1}=\pi/6$, where we choose the RH neutrino mass $M_1=5\times10^{11}\,\mathrm{GeV}$. The red solid, green dash-dotted, blue dotted and black dashed lines correspond to the four best fitting points shown in table~\ref{tab:best_fit} one by one. The horizontal red dashed line represents the experimental measured value $Y^{obs}_{B}$. The neutrino mass spectrum is NO and IO in the first row and the second row respectively. The panels in the left, middle and right columns are for the three admissible forms of the $R$-matrix such as R-1st, R-2nd and R-3rd respectively.}
\end{figure}

For the first case, the lepton mixing matrix is given by Eq.~\eqref{eq:PMNS_I},  both Dirac and Majorana CP violation phases are trivial. The CP asymmetries $\epsilon_{\alpha}$ are found to be vanishing, therefore  non-zero baryon asymmetry can not be generated although the experimental data on lepton mixing angles can be accommodated. Freely varying the parameters $\theta_{1,2,3}$ and requiring the three mixing angles in the experimentally preferred $3\sigma$ ranges~\cite{Capozzi:2013csa}, we find the effective Majorana neutrino mass $m_{ee}$ takes values in the following intervals,
\begin{align}
\nonumber& \text{NO} : 0.000717\text{eV}\leq m_{ee}\leq0.00219\text{eV}  \quad \text{and}  \quad 0.00308\text{eV}\leq m_{ee}\leq0.00449\text{eV}\,,\\
\label{eq:mee_caseI}&\text{IO} : 0.0130\text{eV}\leq m_{ee}\leq0.0227\text{eV}  \quad \text{and}\quad  0.0471\text{eV}\leq m_{ee}\leq0.0478\text{eV} \,.
\end{align}
Here the two different regimes for both NO and IO arise from the CP parity matrix $\widehat{X}_{\nu}$. In other words, the CP parities of the two massive light neutrinos can be identical or opposite, and accordingly two distinct values of $m_{ee}$ are obtained.

The second kind of residual symmetry gives rise to the lepton mixing pattern of Eq.~\eqref{eq:PMNS_II}. Three independent mixing patterns can be obtained from the six row permutations, yet only the mixing matrix $U_{II,3}$ is viable. Eq.~\eqref{eq:mixing_para_caseII} indicates that both Dirac CP phase $\delta$ and the atmospheric mixing angle $\theta_{23}$ are maximal while the values of $\theta_{12}$ and $\theta_{13}$ are not constrained for the mixing matrix $U_{II,3}$. The best fitting values $(\sin^2\theta_{13})^{\text{bf}}=0.0234$ ($(\sin^2\theta_{13})^{\text{bf}}=0.0240$) and $(\sin^2\theta_{12})^{\text{bf}}=0.308$~\cite{Capozzi:2013csa} for NO (IO) can be reproduced for certain values of the parameters $\theta_{1,2,3}$, as shown in table~\ref{tab:best_fit}. Since the lepton mixing angles in Eq.~\eqref{eq:mixing_para_caseII} are invariant under the transformations  $(\theta_{2}, \theta_{3})\rightarrow (\pi-\theta_{2}, \theta_{3})$,  $(\theta_{2}, \theta_{3})\rightarrow(\theta_{2}, \pi-\theta_{3})$ and $(\theta_{2}, \theta_{3})\rightarrow(\pi-\theta_{2}, \pi-\theta_{3})$, four best fitting values for $\theta_{2,3}$ can be found. Recently T2K and NOvA have reported a slight preference for $\delta$ close to $3\pi/2$ while maximal $\theta_{23}$ is favored by T2K and disfavored by NOvA~\cite{Abe:2015awa,T2K_delta_CP,Abe:2017uxa,NovA_delta_CP,Adamson:2017gxd}. T2K and NOvA are expected to be able to exclude maximal $\theta_{23}$ at 90\% confidence level after their full period of data taking. These two experiments can also contribute to the measurement of the Dirac phase $\delta$, if running in both the neutrino and the anti-neutrino modes. They can possibly exclude certain ranges of $\delta$ especially the values around $\delta=\pm\pi/2$, depending on $\theta_{23}$ and the neutrino mass hierarchy. Future long-baseline experiments DUNE~\cite{Acciarri:2016crz}, T2HK~\cite{Kearns:2013lea} and T2HKK~\cite{Abe:2016ero} will allow for
a measurement of the Dirac phase and atmospheric mixng angle with significantly improved sensitivities and thus can fully test the maximal-maximal predictions. Note that the next generation neutrino experiments~\cite{Acciarri:2016crz,Kearns:2013lea} are capable of testing the predictions for maximal $\delta$ and $\theta_{23}$. Furthermore, we plot the numerical results of the baryon asymmetry $Y_{B}$ with respect to the free parameter $\vartheta$ in figure~\ref{fig:YB_II}, where the parameters $\theta_{1,2,3}$ are set to their best fit values. Obviously the observed mater-antimatter asymmetry in the universe can be obtained for particular values of $\vartheta$ except the cases of NO:R-3rd
and IO:R-1st. This conclusion is consistent with the general results of section~\ref{sec:LepG_one_CP}. In addition, the allowed regions of the effective Majorana mass $m_{ee}$ are found to be the same as case I, and they are given in Eq.~\eqref{eq:mee_caseI}. The reason is because $m_{ee}$ is independent of $\theta_{23}$, as shown in Eq.~\eqref{mee_2RHN}. We also present the value of $m_{ee}$ at the best fitting points $\theta^{\rm{bf}}_{1,2,3}$ in table~\ref{tab:best_fit}.

\begin{figure}[t!]
\centering
\begin{tabular}{c}
\includegraphics[width=1\linewidth]{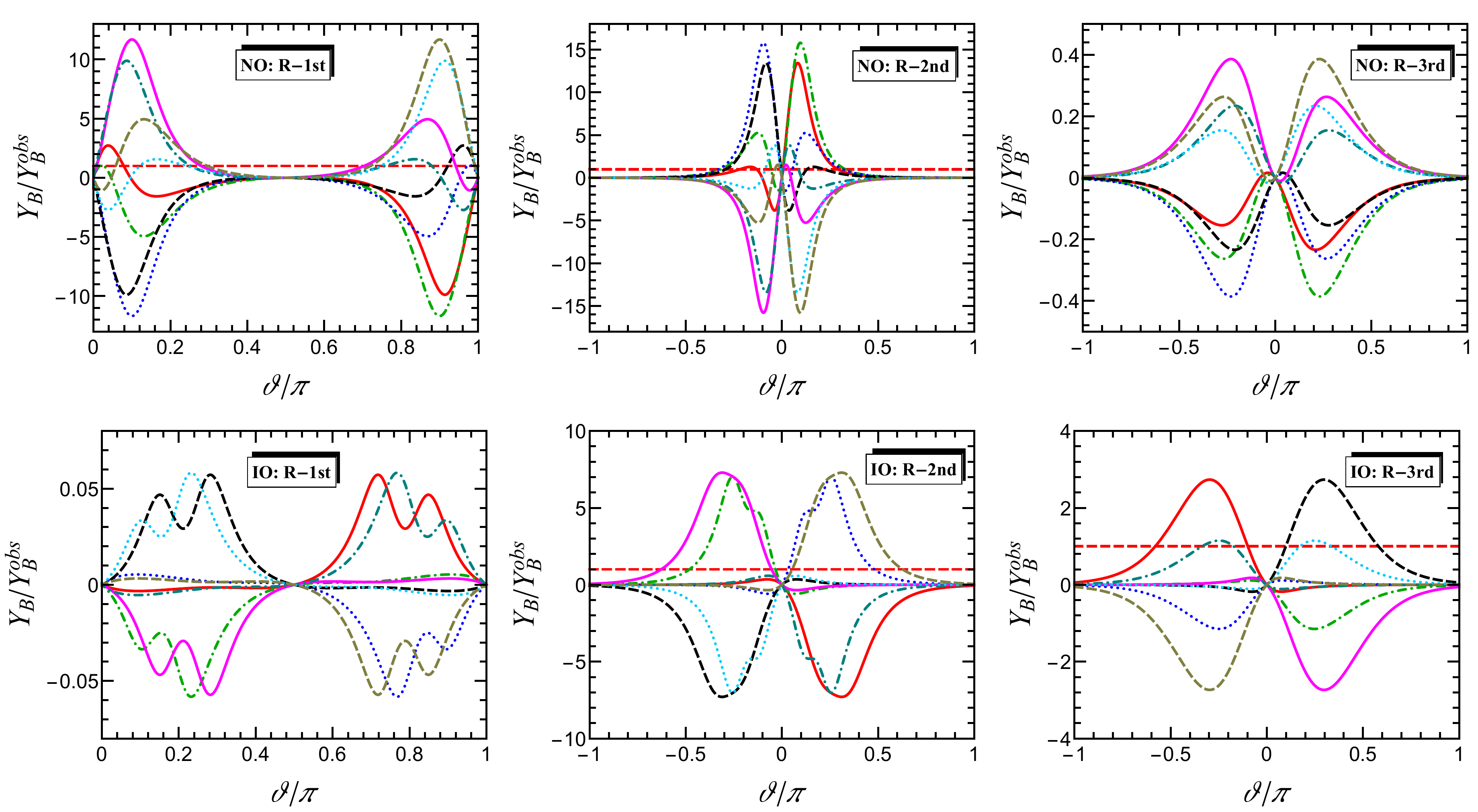}
\end{tabular}
\caption{\label{fig:YB_III_P213}$Y_B/Y_B^{obs}$ as a function of the parameter $\vartheta$ for the mixing pattern $U_{III,3}$ with $\varrho_{1}=\pi/3$, where we choose the RH neutrino mass $M_1=5\times10^{11}\,\mathrm{GeV}$. The red solid, green dash-dotted, blue dotted, black dashed, pink solid, cyan dash-dotted, dark green dotted and brown dashed lines correspond to the eight best fitting points shown in table~\ref{tab:best_fit} one by one. The horizontal red dashed line represents the experimental measured value $Y^{obs}_{B}$. The neutrino mass spectrum is NO and IO in the first row and the second row respectively. The panels in the left, middle and right columns are for the three admissible forms of the $R$-matrix such as R-1st, R-2nd and R-3rd respectively. }
\end{figure}

For the case III, three independent lepton mixing patterns can be obtained as shown in Eq.~\eqref{eq:mix_III_ind}. The predictions of the mixing parameters for mixing matrix $U_{III,1}$ are given in Eq.~\eqref{eq:mixing_para_caseIII1}. The parameter value of $\varrho_{1}=0$ is always admissible, and the resulting lepton mixing matrix is the same as $U_{I}$ if the possible shifts in $\theta_{1,2,3}$ are taken into account. Consequently the lepton mixing angles in the experimentally preferred range can be achieved for appropriate choices of the parameters $\theta_{1,2,3}$. However, both Dirac phase $\delta$ and Majorana phase $\phi$ would be determined to be trivial, such that successful leptogenesis can not be achieved. The smallest value of the index $n$ which is capable of accommodating the experimental data and nontrivial CP violating phases is $n=6$ with $\varrho_{1}=\pi/6$ up to the symmetry transformations shown in Eq.~\eqref{eq:symmetry_caseIII_1st}. Please see table~\ref{tab:best_fit} for the corresponding results of the $\chi^2$ analysis. We notice that the atmospheric mixing angle deviates from maximal mixing with $\sin^2\theta_{23}=0.577$ (0.641) for NO (IO) at the best fit point where the $\chi^2$ function reaches a global minimum, and the Dirac CP phase is approximately maximal with $|\sin\delta|=0.985$ (0.983). This result is consistent with the weak evidence of maximal Dirac CP violation reported by T2K~\cite{Abe:2015awa,T2K_delta_CP,Abe:2017uxa} and NO$\nu$A~\cite{NovA_delta_CP,Adamson:2017gxd} and global data fitting~\cite{Capozzi:2013csa,Forero:2014bxa,Gonzalez-Garcia:2014bfa,Capozzi:2016rtj,Esteban:2016qun}, and it can be tested in forthcoming neutrino oscillation experiments~\cite{Acciarri:2016crz,Kearns:2013lea,Abe:2016ero}. The numerical results of $Y_{B}$ versus $\vartheta$ for $\varrho_{1}=\pi/6$ are shown in figure~\ref{fig:YB_III_P123}. We see that the correct value of the baryon asymmetry can be obtained for particular values of $\vartheta$ except in the case of R-3rd with NO spectrum and R-1st with IO. Moreover, we find the effective mass $m_{ee}$ varies in the intervals,
\begin{align}
\nonumber&\text{NO} :~0.000723\text{eV}\leq m_{ee}\leq0.00449 \text{eV} \,,\\
\label{eq:mee_caseIII_pi6}&\text{IO} : ~ 0.0223\text{eV}\leq m_{ee}\leq0.0297\text{eV}  \quad \text{and} \quad  0.0430\text{eV}\leq m_{ee}\leq0.0436\text{eV}\,.
\end{align}

\begin{figure}[t!]
\centering
\begin{tabular}{c}
\includegraphics[width=1\linewidth]{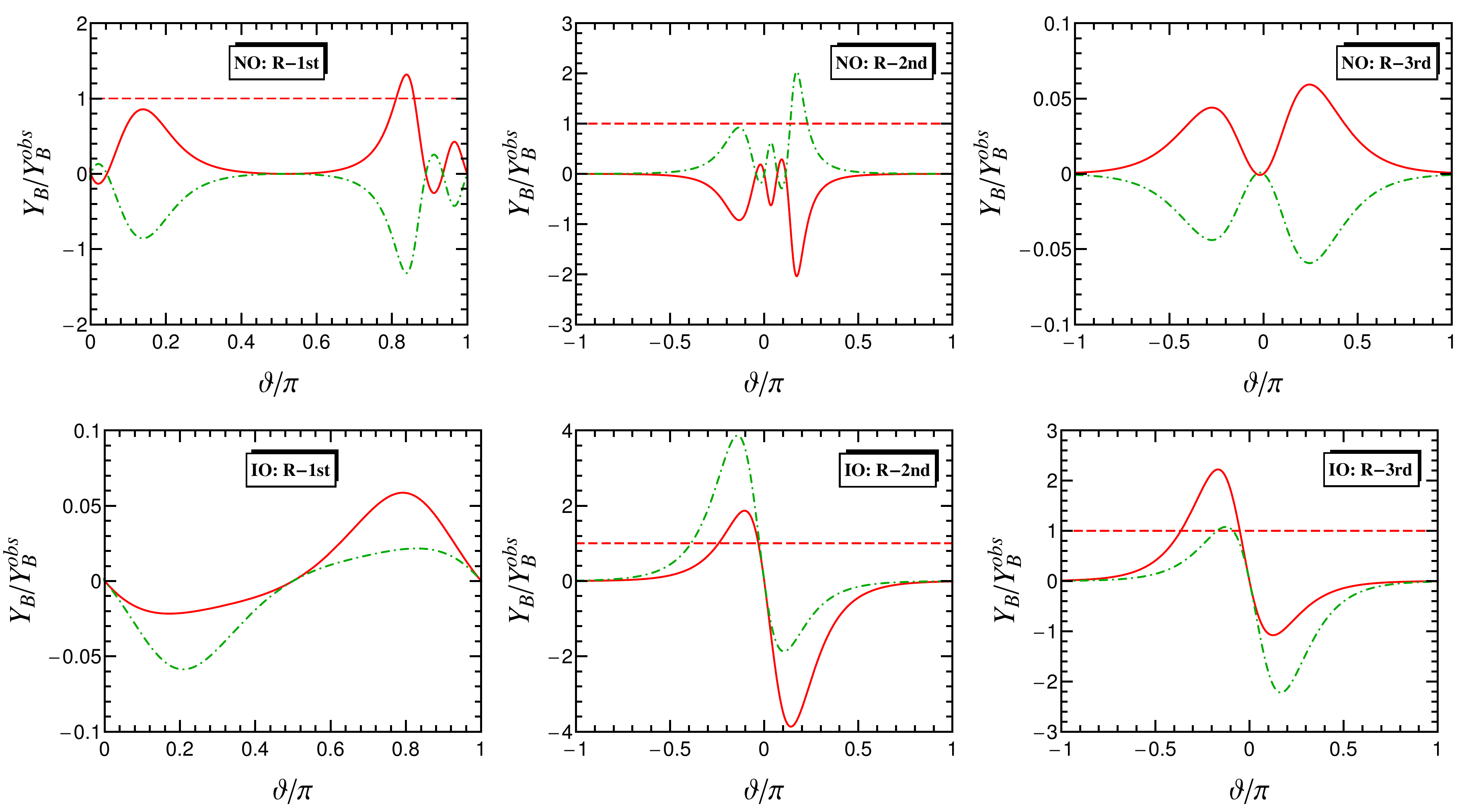}
\end{tabular}
\caption{\label{fig:YB_V_case1}$Y_B/Y_B^{obs}$ as a function of the parameter $\vartheta$ for the mixing pattern $U_{V,1}$ with $\varrho_{3}=0$ and $\varrho_{4}=\pi/2$, where we choose the RH neutrino mass $M_1=5\times10^{11}\,\mathrm{GeV}$. The red solid and green dash-dotted lines correspond to the two best fitting points shown in table~\ref{tab:best_fit} one by one. The horizontal red dashed line represents the experimental measured value $Y^{obs}_{B}$. The neutrino mass spectrum is NO and IO in the first row and the second row respectively. The panels in the left, middle and right columns are for the three admissible forms of the $R$-matrix such as R-1st, R-2nd and R-3rd respectively. }
\end{figure}

\begin{figure}[t!]
\centering
\begin{tabular}{c}
\includegraphics[width=1\linewidth]{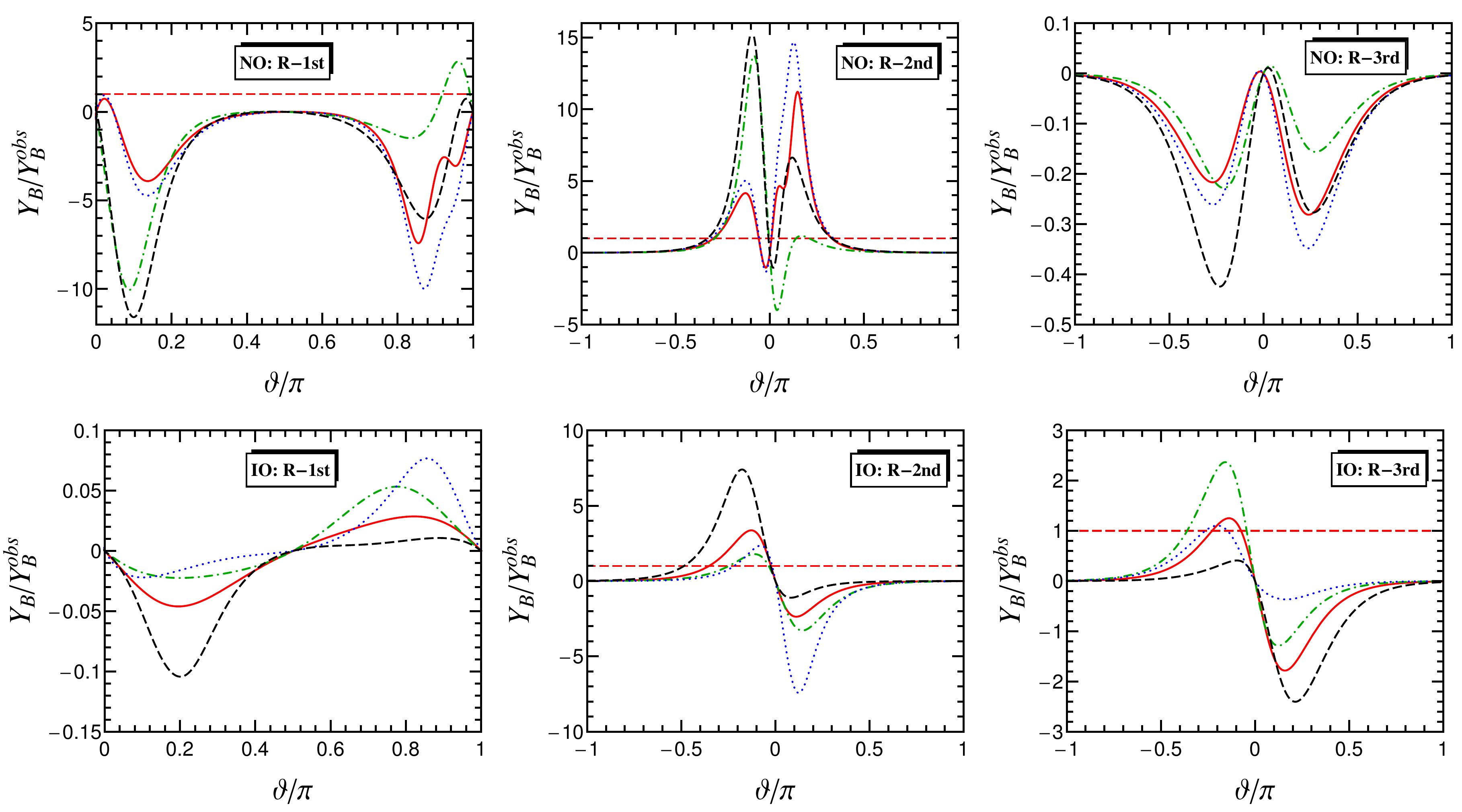}
\end{tabular}
\caption{\label{fig:YB_V_case2}$Y_B/Y_B^{obs}$ as a function of the parameter $\vartheta$ for the mixing pattern $U_{V,1}$ with $\varrho_{3}=0$ and $\varrho_{4}=\pi/3$, where we choose the RH neutrino mass $M_1=5\times10^{11}\,\mathrm{GeV}$. The red solid, green dash-dotted, blue dotted and black dashed lines correspond to the four best fitting points shown in table~\ref{tab:best_fit} one by one. The horizontal red dashed line represents the experimental measured value $Y^{obs}_{B}$. The neutrino mass spectrum is NO and IO in the first row and the second row respectively. The panels in the left, middle and right columns are for the three admissible forms of the $R$-matrix such as R-1st, R-2nd and R-3rd respectively. }
\end{figure}

The mixing parameters for $U_{III,2}$ are given by Eq.~\eqref{eq:mixing_para_caseIII2}. As shown in table~\ref{tab:best_fit}, agreement with the experimental data can be achieved for both $\varrho_1=0$ and $\varrho_1=\pi/6$. The CP asymmetries fulfill $\epsilon_2=\epsilon_{\tau}=0$ in this case, therefore the baryon asymmetry $Y_{B}$ is predicted to be zero. As regards the third mixing pattern $U_{III,3}$, the expressions of mixing parameters and the rephase invariants are shown in Eq.~\eqref{eq:mixing_para_caseIII3} and Eq.~\eqref{eq:Ia_III3}, respectively. For the smallest group index $n=2$, the parameter $\varrho_1$ can be either $0$ or $\pi/2$. We find that the experimental data on lepton mixing angles can be accommodated well for both $\varrho_{1}=0$ and $\varrho_{1}=\pi/2$. The mixing pattern $U_{III, 3}$ with $\varrho_1=0$ is equivalent to $U_I$ in Eq.~\eqref{eq:PMNS_I}, the Dirac as well as Majorana CP phases are trivial, and consequently a nonzero baryon asymmetry can not be generated. The mixing matrix $U_{III, 3}$ for $\varrho_1=\pi/2$ is related to $U_{II, 3}$ as follow,
\begin{equation}
U_{III,3}(\varrho_1=\pi/2,\theta_{1},\theta_{2},\theta_{3})=U_{II,3}(\theta^{\prime}_{1},\theta^{\prime}_{2},\theta^{\prime}_{3})\,,
\end{equation}
where $\theta'_{1,2,3}$ are defined through $O_{3\times3}(\theta^{\prime}_{1},\theta^{\prime}_{2},\theta^{\prime}_{3})=P_{231}O_{3\times3}(\theta_{1},\theta_{2},\theta_{3})$. Hence $U_{III, 3}$ with $\varrho_1=\pi/2$ and $U_{II, 3}$ lead to the same predictions for lepton mixing parameters and $Y_{B}$. Furthermore, new mixing pattern can be obtained from the $\Delta(6\cdot3^2)=\Delta(54)$ group for $\varrho_1=\pi/3$. Note that $\varrho_1=2\pi/3$ leads to the same mixing matrix as $\varrho_1=\pi/3$ after the shift of $\theta_{1, 2, 3}$ is considered. As shown in table~\ref{tab:best_fit}, the best fit values~\cite{Capozzi:2013csa} of the three mixing angles can be achieved for certain values of the parameters $\theta_{1,2,3}$. The corresponding predictions for $Y_{B}$ as a function of $\vartheta$ are plotted in figure~\ref{fig:YB_III_P213}. The observed matter-antimatter asymmetry could be reproduced except for the cases of NO:R-3rd and IO:R-1st.
Furthermore, we obtain the effective mass $m_{ee}$ of the $0\nu\beta\beta$ decay is
\begin{align}
\nonumber&\text{NO} :~0.000717\text{eV}\leq m_{ee}\leq0.00219\text{eV} \quad \text{and} \quad 0.00308\text{eV}\leq m_{ee}\leq0.00449\text{eV} \,,\\
\label{eq:mee_caseIIIpi3}&\text{IO} :~ 0.0130\text{eV}\leq m_{ee}\leq0.0227\text{eV}  \quad~ \text{and} \quad 0.0471\text{eV}\leq m_{ee}\leq0.0478\text{eV} \,.
\end{align}
For the case IV, the resulting lepton mixing matrix $U_{IV}$ is equivalent to $U_I$. Hence we get the same predictions for lepton flavor mixing, $0\nu\beta\beta$ decay and leptogenesis as those of case I.

\begin{figure}[t!]
\centering
\begin{tabular}{c}
\includegraphics[width=1\linewidth]{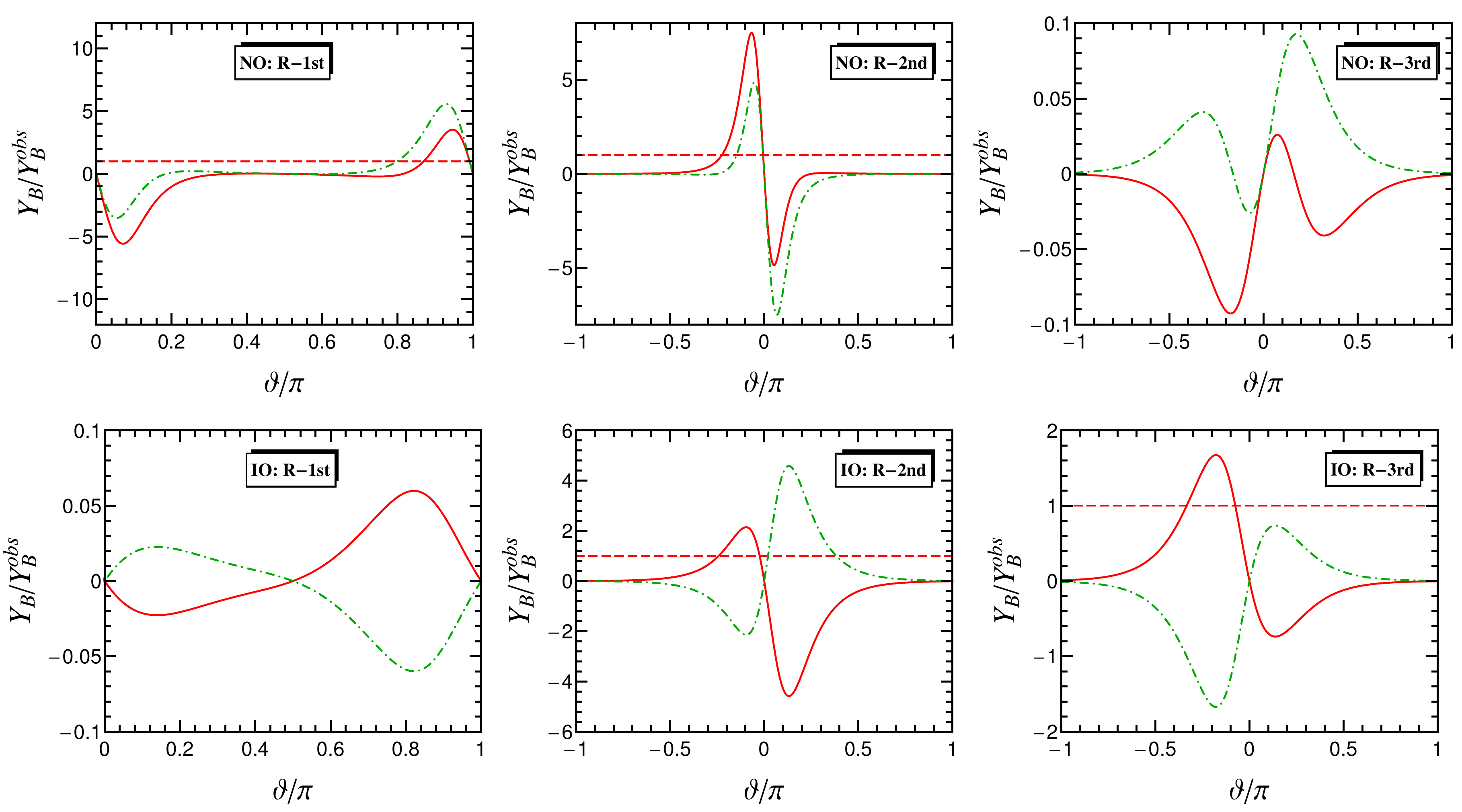}
\end{tabular}
\caption{\label{fig:YB_VI}$Y_B/Y_B^{obs}$ as a function of the parameter $\vartheta$ for the mixing pattern $U_{VI}$ with $\varrho_{5}=\pi/2$, where we choose the RH neutrino mass $M_1=5\times10^{11}\,\mathrm{GeV}$. The red solid and green dash-dotted lines correspond to the two best fitting points shown in table~\ref{tab:best_fit} one by one. The horizontal red dashed line represents the experimental measured value $Y^{obs}_{B}$. The neutrino mass spectrum is NO and IO in the first row and the second row respectively. The panels in the left, middle and right columns are for the three admissible forms of the $R$-matrix such as R-1st, R-2nd and R-3rd respectively. }
\end{figure}

Similarly three mixing matrices can be obtained in the case V. For the mixing matrix $U_{V,1}$, the results of the mixing parameters are given by Eq.~\eqref{eq:mixing_para_caseV1}, the CP invariants $I^{\alpha}_{\text{NO}}$ and $I^{\alpha}_{\text{IO}}$ are shown in Eq.~\eqref{eq:Ia_V}. For the smallest $\Delta(6n^2)$ group with $n=2$, the values of $\varrho_{3}$ and $\varrho_{4}$ can be $0$ and $\pi/2$ in the fundamental interval. Utilizing the equivalence condition of Eq.~\eqref{eq:equiv_condition}, we find two independent mixing patterns with $\varrho_3=\varrho_{4}=0$ and $\varrho_{3}=0,\varrho_{4}=\pi/2$. Moreover, the mixing matrix $U_{V,1}$ for $\varrho_{3}=\varrho_{4}=0$ is the same as $U_{II,3}$ if the possible shifts of $\theta_{1,2,3}$ is considered. In the case of $\varrho_{3}=0$ and $\varrho_{4}=\pi/2$, the results of the $\chi^2$ analysis are summarized in table~\ref{tab:best_fit}, and the predictions for $Y_{B}$ are plotted in figure~\ref{fig:YB_V_case1}. The effective Majorana neutrino mass $m_{ee}$ is determined to take values in the intervals
\begin{align}
\nonumber& \text{NO} : ~0.00143\text{eV}\leq m_{ee}\leq0.00449\text{eV}\,,\\
\label{eq:mee_caseVI1pi2}&\text{IO} : ~ 0.0144\text{eV}\leq m_{ee}\leq0.0161\text{eV} \quad \text{and} \quad  0.0464\text{eV}\leq m_{ee}\leq0.0478\text{eV} \,.
\end{align}
For the flavor group $\Delta(6\cdot3^2)=\Delta(54)$, the possible values of $\varrho_{3}$ and $\varrho_{4}$ are $0$, $\pi/3$ and $2\pi/3$. We can obtain three  phenomenologically viable mixing patterns corresponding to $(\varrho_{3}, \varrho_{4})=(0, 0)$, $(0, \pi/3)$, $(0, 2\pi/3)$. Note that $U_{V,1}$ for $(\varrho_{3},\varrho_{4})=(0, 2\pi/3)$ is equivalent to the complex conjugate of $U_{V,1}$ with $(\varrho_{3},\varrho_{4})=(0, \pi/3)$. The best fit values of the three lepton mixing angles can be reproduced for particular values of $\theta_{1, 2, 3}$ in the case of $(\varrho_{3},\varrho_{4})=(0, \pi/3)$, the resulting predictions for CP violation phases are listed in table~\ref{tab:best_fit}, and the variation of $Y_B$ with respect to $\vartheta$ is plotted in figure~\ref{fig:YB_V_case2}. In addition, we find the effective mass $m_{ee}$ is
\begin{align}
\nonumber&\text{NO} : ~0.000717\text{eV}\leq m_{ee}\leq0.00449\text{eV}\,,\\
\label{eq:mee_caseV1pi3}&\text{IO} : ~ 0.0264\text{eV}\leq m_{ee}\leq0.0285\text{eV} \quad \text{and}  \quad  0.0399\text{eV}\leq m_{ee}\leq0.0455\text{eV}\,.
\end{align}
The other two mixing matrices $U_{V,2}$ and $U_{V,3}$ can not accommodate the experimental data on the lepton mixing angles for $n=2$, and they give rise to the same mixing patterns as $U_{V,1}$ in the case of $n=3$.

\begin{figure}[t!]
\centering
\begin{tabular}{c}
\includegraphics[width=1\linewidth]{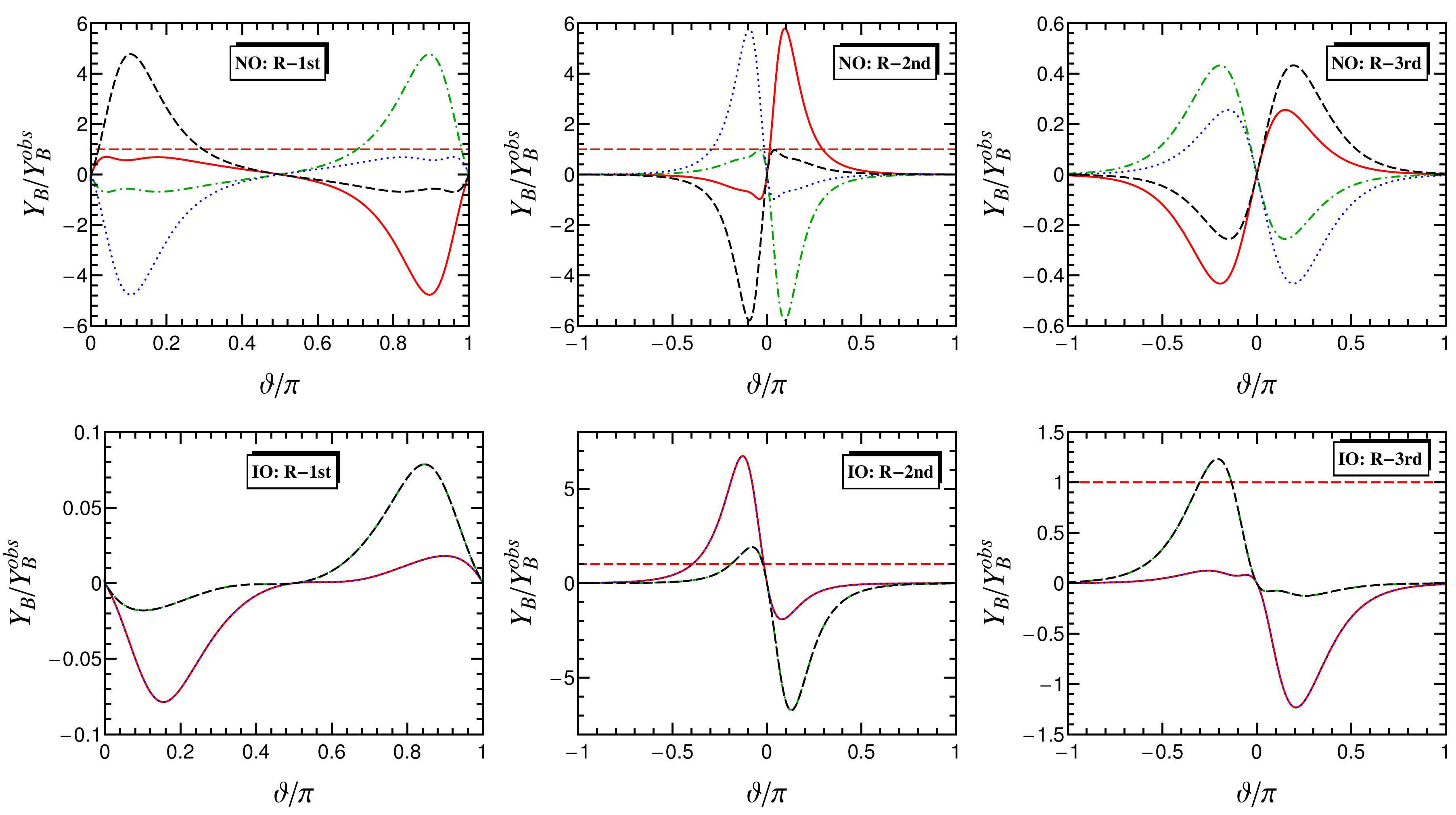}
\end{tabular}
\caption{\label{fig:YB_VIII_P123}$Y_B/Y_B^{obs}$ as a function of the parameter $\vartheta$ for the mixing pattern $U_{VIII,1}$, where we choose the RH neutrino mass $M_1=5\times10^{11}\,\mathrm{GeV}$. The red solid, green dash-dotted, blue dotted and black dashed lines correspond to the four best fitting points shown in table~\ref{tab:best_fit} one by one. The horizontal red dashed line represents the experimental measured value $Y^{obs}_{B}$. The neutrino mass spectrum is NO and IO in the first row and the second row respectively. The panels in the left, middle and right columns are for the three admissible forms of the $R$-matrix such as R-1st, R-2nd and R-3rd respectively. }
\end{figure}

The symmetry breaking pattern VI leads to only one independent mixing matrix which is given in  Eq.~\eqref{eq:PMNS_VI}. The predictions for the mixing parameters and the rephasing invariants are reported in Eqs.~\eqref{eq:mixing_para_caseVI} and \eqref{eq:Ia_VI}, respectively. For the $\Delta(6\cdot2^2)\cong S_4$ flavor group, the value of $\varrho_{5}$ is either $0$ or $\pi/2$ in the fundamental region. Notice that $U_{VI}$ for $\varrho_5=0$ is equivalent to $U_{I}$. Hence the three lepton mixing angles are not subject to any constraint, and the Dirac as well as Majorana CP phases are trivial. The results of the $\chi^2$ analysis for $\varrho_5=\pi/2$ are collected in table~\ref{tab:best_fit}. We display the variation of $Y_B$ with respect to $\vartheta$ in figure~\ref{fig:YB_VI}. The observed baryon asymmetry can be generated for certain values of $\vartheta$ except in the case of NO:R-3rd and IO:R-1st. Moreover, we find the effective Majorana neutrino mass $m_{ee}$ is in the intervals
\begin{align}
\nonumber & \text{NO} : ~0.00117\text{eV}\leq m_{ee}\leq0.00449\text{eV}\,,\\
\label{eq:mee_caseVI} &\text{IO} :~ 0.0144\text{eV}\leq m_{ee}\leq0.0174\text{eV} \quad \text{and}  \quad 0.0458\text{eV}\leq m_{ee}\leq0.0478\text{eV} \,.
\end{align}
As shown in Appendix~\ref{sec:D6n2_example}, case VII leads to the same predictions for lepton mixing, neutrinoless double decay and matter/antimatter asymmetry via leptogenesis as case III.

\begin{figure}[t!]
\centering
\begin{tabular}{c}
\includegraphics[width=1\linewidth]{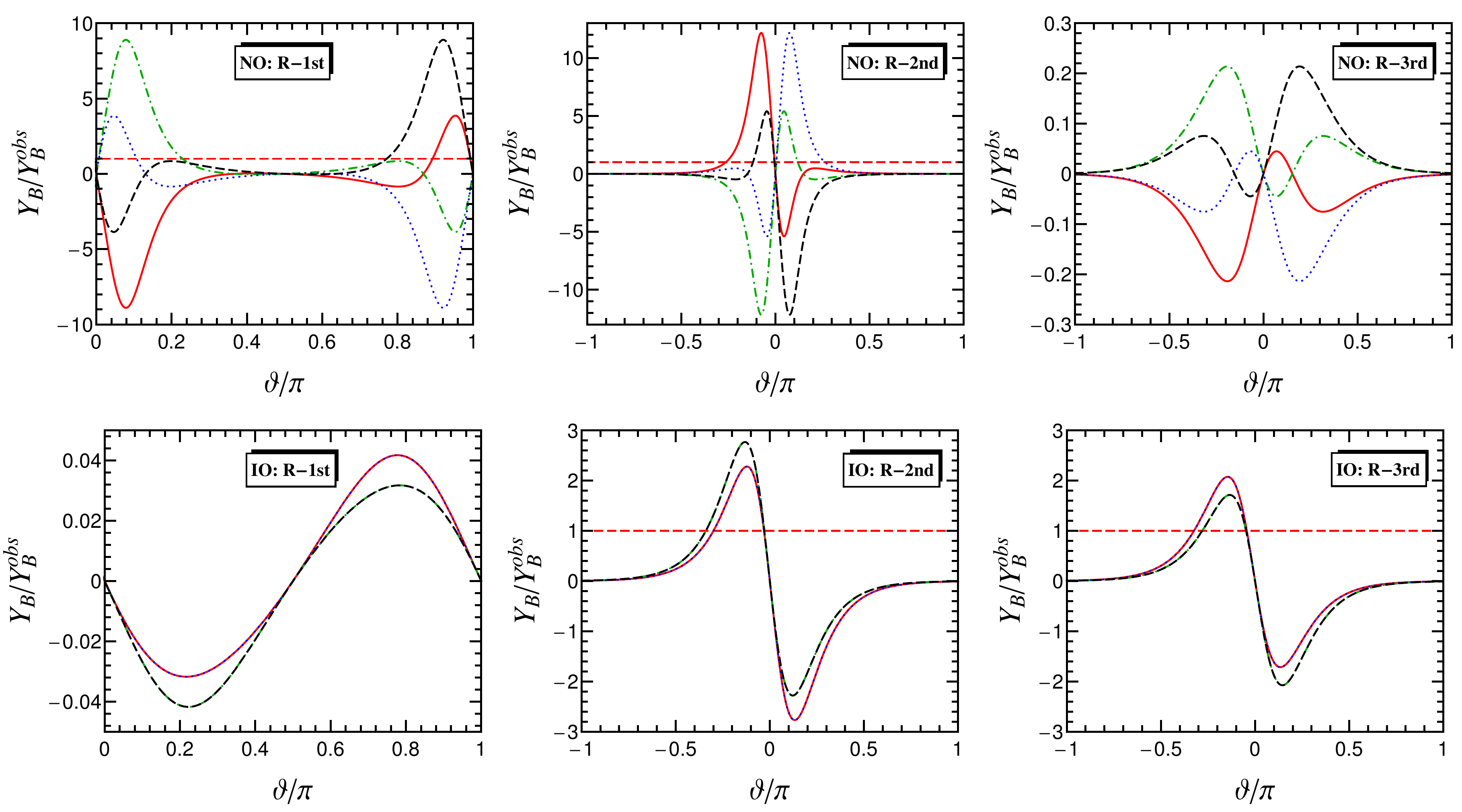}
\end{tabular}
\caption{\label{fig:YB_VIII_P132}$Y_B/Y_B^{obs}$ as a function of the parameter $\vartheta$ for the mixing pattern $U_{VIII,2}$, where we choose the RH neutrino mass $M_1=5\times10^{11}\,\mathrm{GeV}$. The red solid, green dash-dotted, blue dotted and black dashed lines correspond to the four best fitting points shown in table~\ref{tab:best_fit} one by one. The horizontal red dashed line represents the experimental measured value $Y^{obs}_{B}$. The neutrino mass spectrum is NO and IO in the first row and the second row respectively. The panels in the left, middle and right columns are for the three admissible forms of the $R$-matrix such as R-1st, R-2nd and R-3rd respectively. }
\end{figure}

In the end of this section, we proceed to discuss the last case VIII. After considering all possible row permutations, we can obtain three independent mixing matrices which are given by Eq.~\eqref{eq:all_PMNS_VIII}. The mixing parameters and CP invariants for the mixing pattern $U_{VIII,1}$ are summarized in Eqs.~\eqref{eq:mixing_para_caseVIII1} and \eqref{eq:Ia_VIII}, respectively. Our numerical results for this case are summarized in table~\ref{tab:best_fit}, and the variation of $Y_{B}$ as a function of $\vartheta$ is shown in figure~\ref{fig:YB_VIII_P123}.
From the expressions of the CP asymmetry $\epsilon_{\alpha}$ and the washout mass $\tilde{m}_{\alpha}$, we can see that the final baryon asymmetry $Y_{B}$ has the following symmetry properties:
\begin{eqnarray}
&&\nonumber Y_{B}(\vartheta,\theta_{1}=\pi,\theta_{2},\theta_{3})=-Y_{B}(\vartheta,\theta_{1}=\pi,\theta_{2},\pi-\theta_{3})\\
\label{eq:Yb_sym_NO_VIII}&&~=Y_{B}(-\vartheta,\theta_{1}=0,\pi-\theta_{2},\theta_{3})=-Y_{B}(-\vartheta,\theta_{1}=0,\pi-\theta_{2},\pi-\theta_{3}), ~~\text{for}~~ \text{NO}\,,  \\[0.04in]
&&\nonumber Y_{B}(\vartheta,\theta_{1}=\pi,\theta_{2},\theta_{3})=-Y_{B}(-\vartheta,\theta_{1}=\pi,\theta_{2},\pi-\theta_{3})\\
\label{eq:Yb_sym_IO_VIII}&&~=Y_{B}(\vartheta,\theta_{1}=0,\pi-\theta_{2},\theta_{3})=-Y_{B}(-\vartheta,\theta_{1}=0,\pi-\theta_{2},\pi-\theta_{3}), ~~\text{for}~~ \text{IO}\,.
\end{eqnarray}
Thus the coincidence of two pair of curves in IO case can be easily understood from Eq.~\eqref{eq:Yb_sym_IO_VIII}.
The second mixing matrix $U_{VIII, 2}$ is related to $U_{VIII, 1}$ through the exchange of the second and third rows. Therefore they lead to the same reactor and solar mixing angles, while the atmospheric angle changes from $\theta_{23}$ to $\pi/2-\theta_{23}$ and the Dirac phase change from $\delta$ to $\pi+\delta$. The corresponding results of the $\chi^2$ analysis are listed in table~\ref{tab:best_fit}, and the predictions for the matter-antimatter asymmetry $Y_{B}$ are displayed in figure~\ref{fig:YB_VIII_P132}. $U_{VIII, 1}$ and $U_{VIII, 2}$ give the same prediction for the effective mass $m_{ee}$ as follow,
\begin{align}
\nonumber& \text{NO} :~0.000717\text{eV}\leq m_{ee}\leq0.00449\text{eV}\,,\\
\label{eq:mee_caseVIII}& \text{IO} :~0.0226\text{eV}\leq m_{ee}\leq0.0256\text{eV} \quad \text{and} \quad 0.0417\text{eV}\leq m_{ee}\leq0.0475\text{eV} \,.
\end{align}
Lastly, the third mixing matrix $U_{VIII, 3}$ can not describe the measured values of $\theta_{13}$ and $\theta_{12}$ simultaneously because of the sum rule shown in Eq.~\eqref{eq:caseVIII_sum}.

The above predicted lepton mixing patterns can be tested in various ways. The upcoming reactor neutrino oscillation experiments such as JUNO~\cite{An:2015jdp} will be able to make very precise sub-percent measurements of the solar mixing angle $\theta_{12}$. The current experiments T2K and NOvA have the potential to exclude maximal $\theta_{23}$ and maximal Dirac phase $\delta$. The next generation of long-baseline experiments DUNE~\cite{Acciarri:2016crz}, T2HK~\cite{Kearns:2013lea} and T2HKK~\cite{Abe:2016ero} will be able to place important constraints
on the parameters $\theta_{23}$ and $\delta$, in particular the sensitivity to the CP phase $\delta$ would be improved significantly. In short, future neutrino facilities would be able to improve our knowledge of the mixing parameters in a number of ways. This could allow many of the presented mixing patterns to be excluded. Moreover, forthcoming $0\nu\beta\beta$ experiments are expected to probe the full region of parameter space associated with IO neutrino mass spectrum. Thus all our models for IO mass spectrum can be tested independently of oscillation physics.

\begin{table}[!htbp]
\centering
\scriptsize
\renewcommand{\tabcolsep}{1.55mm}
\begin{tabular}{|c|c|c|c|c|c|c|c|c|c|c|c|c|}
\hline \hline
&  $\varrho_{i}$ & & $\theta^{\text{bf}}_{1}/\pi$ & $\theta^{\text{bf}}_{2}/\pi$  & $\theta^{\text{bf}}_{3}/\pi$ & $\chi^2_{\text{min}}$   & $\sin^2\theta_{13}$ & $\sin^2\theta_{12}$ & $\sin^2\theta_{23}$  & $|\sin\delta|$ & $|\sin\phi|$ & $m_{ee}/$eV \\ \hline

\multirow{7}{*}{$U_{II,3}$}& \multirow{7}{*}{$-$} &   &     &  \multirow{2}{*}{$0.049$} & $0.187$ &   &   & \multirow{7}{*}{$0.308$} & \multirow{7}{*}{$0.5$}  & \multirow{7}{*}{$1$} & \multirow{7}{*}{$0$}  &   \\[-0.02in] \cline{6-6}
&  & \multirow{2}{*}{NO} &  \multirow{2}{*}{$-$} &   & $0.813$ & \multirow{2}{*}{$3.645$} & \multirow{2}{*}{$0.0234$}  &&&& & \multirow{2}{*}{$0.00377$ or $0.00145$} \\ [-0.02in] \cline{5-6}
&  & & & \multirow{2}{*}{$0.951$}  & $0.187$ &  &  &&&& &  \\ [-0.02in] \cline{6-6}
&  & & &   & $0.813$ &  &  &&&&&  \\  [-0.02in] \cline{3-8} \cline{13-13}

& &  &   & \multirow{2}{*}{$0.050$} & $0.187$ &  &  &    & &  & & \\ \cline{6-6}
&  & \multirow{2}{*}{IO} & \multirow{2}{*}{$-$} &   & $0.813$ & \multirow{2}{*}{$0.105$} & \multirow{2}{*}{$0.024$}  &&&&& \multirow{2}{*}{$0.0475$ or $0.0179$} \\ [-0.02in] \cline{5-6}
&  & & & \multirow{2}{*}{$0.950$}  & $0.187$ &  &  &&&&& \\ [-0.02in] \cline{6-6}
&  & & &   & $0.813$ &  &  &&&&& \\  [-0.02in] \hline

\multirow{7}{*}{$U_{III,1}$}& \multirow{7}{*}{$\varrho_{1}=\frac{\pi}{6}$} &  &     \multirow{2}{*}{$0.322$} &  \multirow{2}{*}{$0.155$} & $0.614$ &  &  &  \multirow{7}{*}{$0.308$} & && & \\ [-0.02in] \cline{6-6}
&  &   \multirow{2}{*}{NO} & &   & $0.977$ & \multirow{2}{*}{$27.205$} & \multirow{2}{*}{$0.0295$} &&\multirow{2}{*}{$0.577$}   & \multirow{2}{*}{$0.985$} & \multirow{2}{*}{$0.253$}& \multirow{2}{*}{$0.00162$ or $0.00389$} \\ [-0.02in] \cline{4-6}
&  & & \multirow{2}{*}{$0.678$} & \multirow{2}{*}{$0.845$}  & $0.386$ &  &  &&&&& \\ [-0.02in] \cline{6-6}
&  & & &   & $0.023$ &  &  &&&&& \\ [-0.02in] \cline{3-8}  \cline{10-13}

& &  &  \multirow{2}{*}{$0.340$} & \multirow{2}{*}{$0.143$} & $0.606$ & &&&&&& \\ [-0.02in] \cline{6-6}
&  & \multirow{2}{*}{IO} & &   & $0.968$ &  \multirow{2}{*}{$2.143$} & \multirow{2}{*}{$0.0251$} &    & \multirow{2}{*}{$0.641$} & \multirow{2}{*}{$0.983$} & \multirow{2}{*}{$0.789$} & \multirow{2}{*}{$0.0433$ or $0.0263$} \\ [-0.02in] \cline{4-6}
&  & & \multirow{2}{*}{$0.660$} & \multirow{2}{*}{$0.857$}  & $0.394$ &  &  &&&&& \\ [-0.02in] \cline{6-6}
&  & & &   & $0.032$ &  &  &&&&& \\  [-0.02in] \hline

\multirow{7}{*}{$U_{III,2}$}& \multirow{7}{*}{$\varrho_{1}=\frac{\pi}{6}$} &  &     \multirow{2}{*}{$0.329$} &  \multirow{2}{*}{$0.150$} & $0.611$ &  &  &  \multirow{7}{*}{$0.308$} &     &  &&  \\ [-0.02in] \cline{6-6}
&  &  \multirow{2}{*}{NO} & &   & $0.974$ &  \multirow{2}{*}{$7.674$} &  \multirow{2}{*}{$0.0278$} &  &  \multirow{2}{*}{$0.398$}   & \multirow{2}{*}{$0.984$} & \multirow{2}{*}{$0.829$} & \multirow{2}{*}{$0.00168$ or $0.00381$} \\ [-0.02in] \cline{4-6}
&  & & \multirow{2}{*}{$0.671$} & \multirow{2}{*}{$0.850$}  & $0.389$ &  &  &&&&& \\ [-0.02in] \cline{6-6}
&  & & &   & $0.026$ &  &  &&&&& \\   [-0.02in] \cline{3-8} \cline{10-13}

& &  &  \multirow{2}{*}{$0.331$} & \multirow{2}{*}{$0.149$} & $0.610$ & &&&&&&\\ [-0.02in] \cline{6-6}
&  & \multirow{2}{*}{IO} & &   & $0.973$ &  \multirow{2}{*}{$7.281$} & \multirow{2}{*}{$0.0274$} &  & \multirow{2}{*}{$0.393$} & \multirow{2}{*}{$0.984$} & \multirow{2}{*}{$0.773$} & \multirow{2}{*}{$0.0434$ or $0.0259$} \\ [-0.02in] \cline{4-6}
&  & & \multirow{2}{*}{$0.669$} & \multirow{2}{*}{$0.851$}  & $0.390$ &  &  &&&&& \\ [-0.02in] \cline{6-6}
&  & & &   & $0.027$ &  &  &&&&& \\  [-0.02in] \hline

\multirow{14}{*}{$U_{III,3}$}& \multirow{14}{*}{$\varrho_{1}=\frac{\pi}{3}$} & \multirow{7}{*}{NO} &     \multirow{2}{*}{$0.049$} &  \multirow{2}{*}{$0.040$} & $0.306$ &  \multirow{14}{*}{$0$} & \multirow{7}{*}{$0.0234$} &  \multirow{14}{*}{$0.308$} &  \multirow{7}{*}{$0.437$}   & \multirow{7}{*}{$0.873$} & \multirow{7}{*}{$0.852$}& \multirow{7}{*}{$0.00377$ or $0.00145$} \\ [-0.02in] \cline{6-6}
&  &   & &   & $0.681$ &  &   &  &     &  &&  \\ [-0.02in] \cline{4-6}
&  & & \multirow{2}{*}{$0.951$} & \multirow{2}{*}{$0.960$}  & $0.319$ &  &  &&&&& \\ [-0.02in] \cline{6-6}
&  & & &   & $0.694$ &  &  &&&&& \\   [-0.02in] \cline{4-6}

& &  &     \multirow{2}{*}{$0.281$} &  \multirow{2}{*}{$0.437$} & $0.035$ &  &  &   &     &  & &\\ [-0.02in] \cline{6-6}
&  &  & &   & $0.409$ &   &   &  &     & & & \\ [-0.02in] \cline{4-6}
&  & & \multirow{2}{*}{$0.719$} & \multirow{2}{*}{$0.563$}  & $0.591$ &  &  &&&&& \\ [-0.02in] \cline{6-6}
&  & & &   & $0.965$ &  &  &&&&& \\   [-0.02in] \cline{3-6}  \cline{8-8} \cline{10-13}

&  & \multirow{7}{*}{IO} &     \multirow{2}{*}{$0.050$} &  \multirow{2}{*}{$0.028$} & $0.308$ &  & \multirow{7}{*}{$0.024$} &  &  \multirow{7}{*}{$0.455$}   & \multirow{7}{*}{$0.870$} & \multirow{7}{*}{$0$} & \multirow{7}{*}{$0.0475$ or $0.0179$}\\ [-0.02in] \cline{6-6}
&  &   & &   & $0.683$ &  &   &  &     &  &&  \\ [-0.02in] \cline{4-6}
&  & & \multirow{2}{*}{$0.950$} & \multirow{2}{*}{$0.972$}  & $0.317$ &  &  &&&&& \\ [-0.02in] \cline{6-6}
&  & & &   & $0.692$ &  &  &&&&& \\   [-0.02in] \cline{4-6}

& &  &     \multirow{2}{*}{$0.334$} &  \multirow{2}{*}{$0.443$} & $0.356$ &  &  &   &     &&  & \\ [-0.02in] \cline{6-6}
&  &  & &   & $0.981$ &   &   &  &     & & & \\ [-0.02in] \cline{4-6}
&  & & \multirow{2}{*}{$0.666$} & \multirow{2}{*}{$0.557$}  & $0.019$ &  &  &&&&& \\ [-0.02in] \cline{6-6}
&  & & &   & $0.644$ &  &  &&&&& \\   [-0.02in]  \hline

\multirow{4}{*}{$U_{V,1}$}&  &  \multirow{2}{*}{NO} &     $0.454$ & $0.694$ & \multirow{2}{*}{$0.028$} &  \multirow{2}{*}{$3.327$} &  \multirow{2}{*}{$0.0233$} &  \multirow{2}{*}{$0.339$} &  \multirow{2}{*}{$0.433$}   & \multirow{2}{*}{$0.931$} &  \multirow{2}{*}{$0.072$} &  \multirow{2}{*}{$0.00210$ or $0.00385$} \\ [-0.02in] \cline{4-5}
&  $\varrho_{3}=0$, & & $0.468$ & $0.808$  &  &  &  &&&&&  \\  [-0.02in] \cline{3-13}

& $\varrho_{4}=\frac{\pi}{2}$ & \multirow{2}{*}{IO} &  $0.465$ & $0.692$ & \multirow{2}{*}{$0.021$} & \multirow{2}{*}{$3.599$} & \multirow{2}{*}{$0.0238$} & \multirow{2}{*}{$0.340$}   & \multirow{2}{*}{$0.449$} & \multirow{2}{*}{$0.961$} & \multirow{2}{*}{$0$} &  \multirow{2}{*}{$0.0149$ or $0.0475$} \\ [-0.02in] \cline{4-5}
&  & & $0.476$ &  $0.809$ &  &  &  &&&&& \\  [-0.02in] \hline

\multirow{7}{*}{$U_{V,1}$}&  &   &     \multirow{2}{*}{$0.448$} &  \multirow{2}{*}{$0.684$} & $0.017$ &  \multirow{7}{*}{$0$} &   &  \multirow{7}{*}{$0.308$} &  &$0.899$ & $0.929$ & $0.00153$ or $0.00374$ \\ [-0.02in] \cline{6-6} \cline{11-13}
&  & \multirow{2}{*}{NO} & &   & $0.746$ &  & \multirow{2}{*}{$0.0234$}  & & \multirow{2}{*}{$0.437$} & $0.703$& $0.848$ & $0.00366$ or $0.00173$ \\ [-0.02in] \cline{4-6} \cline{11-13}
&  & & \multirow{2}{*}{$0.465$} & \multirow{2}{*}{$0.798$}  & $0.003$ &  &  &&& $0.984$ & $0.892$ & $0.00274$ or $0.00298$ \\ [-0.02in] \cline{6-6} \cline{11-13}
& $\varrho_{3}=0$, & & &   & $0.740$ &  &  &&& $0.097$ & $0.999$ & $0.00259$ or $0.00310$ \\ [-0.02in]  \cline{3-6} \cline{8-8} \cline{10-13}

& $\varrho_{4}=\frac{\pi}{3}$ &  &     \multirow{2}{*}{$0.463$} &  \multirow{2}{*}{$0.684$} & $0.004$ &  &   & &     &$0.936$ & \multirow{2}{*}{$0.826$}&  \multirow{2}{*}{$0.0272$ or $0.0428$ } \\ [-0.02in] \cline{6-6} \cline{11-11}
&  & \multirow{2}{*}{IO} & &   & $0.736$ &  & \multirow{2}{*}{$0.024$} && \multirow{2}{*}{$0.455$} & $0.623$& &  \\ [-0.02in] \cline{4-6} \cline{11-13}
&  & & \multirow{2}{*}{$0.475$} & \multirow{2}{*}{$0.803$}  & $0.731$ &  &  &&& $0.191$ & \multirow{2}{*}{$0.806$} &  \multirow{2}{*}{$0.0267$ or $0.0431$ } \\[-0.02in]  \cline{6-6} \cline{11-11} \cline{11-11}
&  & & &   & $0.995$ &  &  &&& $0.997$ && \\ [-0.02in]  \hline

\multirow{4}{*}{$U_{VI}$}&  &  \multirow{2}{*}{NO} &   \multirow{2}{*}{$0.530$} & $0.075$ & \multirow{2}{*}{$0.487$} &  \multirow{2}{*}{$0.854$} &  \multirow{2}{*}{$0.0235$} &  \multirow{2}{*}{$0.323$} &  \multirow{2}{*}{$0.448$}   & \multirow{2}{*}{$0.894$} & \multirow{2}{*}{$0.016$}& \multirow{2}{*}{$0.00355$ or $0.00226$ } \\ [-0.02in] \cline{5-5}
& \multirow{2}{*}{$\varrho_{5}=\frac{\pi}{2}$}  & &  & $0.925$  &  &  &  &&&&&  \\  [-0.02in]\cline{3-13}

&  & \multirow{2}{*}{IO} &  \multirow{2}{*}{$0.508$} & $0.086$ & \multirow{2}{*}{$0.496$} & \multirow{2}{*}{$0.349$} & \multirow{2}{*}{$0.0240$} & \multirow{2}{*}{$0.317$}   & \multirow{2}{*}{$0.487$} & \multirow{2}{*}{$0.994$} & \multirow{2}{*}{$0$}& \multirow{2}{*}{$0.0170$ or $0.0475$ }   \\ [-0.02in]\cline{5-5}
&  & &  &  $0.914$ &  &  &  &&&&& \\ [-0.02in] \hline

\multirow{7}{*}{$U_{VIII,1}$}& \multirow{7}{*}{$-$} &   &     \multirow{2}{*}{$1$} &  \multirow{2}{*}{$0.862$} & $0.106$ &  &   &  \multirow{7}{*}{$0.308$} &  & && \\ [-0.02in] \cline{6-6}
&  & \multirow{2}{*}{NO} & & & $0.894$ & \multirow{2}{*}{$18.549$} & \multirow{2}{*}{$0.0244$} & &\multirow{2}{*}{$0.578$} & \multirow{2}{*}{$0.667$} & \multirow{2}{*}{$0.580$} & \multirow{2}{*}{$0.00228$ or $0.00337$ }\\ [-0.02in] \cline{4-6}
&  & & \multirow{2}{*}{$0$} & \multirow{2}{*}{$0.138$}  & $0.106$ &  &  &&&&& \\ [-0.02in]\cline{6-6}
&  & & &   & $0.894$ &  &  &&&&& \\  [-0.02in] \cline{3-8} \cline{10-13}

& &  &  \multirow{2}{*}{$1$} & \multirow{2}{*}{$0.861$} & $0.106$ & &&&&&& \\ [-0.02in]\cline{6-6}
&  & \multirow{2}{*}{IO} & &   & $0.894$ & \multirow{2}{*}{$0.794$} & \multirow{2}{*}{$0.024$} &    & \multirow{2}{*}{$0.579$}  &  \multirow{2}{*}{$0.668$} & \multirow{2}{*}{$0.616$}& \multirow{2}{*}{$0.0229$ or $0.0453$ }  \\ [-0.02in]\cline{4-6}
&  & & \multirow{2}{*}{$0$} & \multirow{2}{*}{$0.139$}  & $0.106$ &  &  &&&&& \\ [-0.02in]\cline{6-6}
&  & & &   & $0.894$ &  &  &&&&& \\  [-0.02in]\hline

\multirow{7}{*}{$U_{VIII,2}$}& \multirow{7}{*}{$-$} &   &     \multirow{2}{*}{$1$} &  \multirow{2}{*}{$0.861$} & $0.106$ &   &  &  \multirow{7}{*}{$0.308$} &  & && \\ [-0.02in] \cline{6-6}
&  & \multirow{2}{*}{NO} & & & $0.894$ & \multirow{2}{*}{$0.537$}& \multirow{2}{*}{$0.0236$} & & \multirow{2}{*}{$0.420$} & \multirow{2}{*}{$0.389$} & \multirow{2}{*}{$0.616$}& \multirow{2}{*}{$0.00228$ or $0.00334$ } \\ [-0.02in]\cline{4-6}
&  & & \multirow{2}{*}{$0$} & \multirow{2}{*}{$0.139$}  & $0.106$ &  &  &&&&& \\ [-0.02in]\cline{6-6}
&  & & &   & $0.894$ &  &  &&&&& \\ [-0.02in] \cline{3-8} \cline{10-13}

& &  &  \multirow{2}{*}{$0$} & \multirow{2}{*}{$0.138$} & $0.894$ & &&&&&& \\ [-0.02in]\cline{6-6}
&  & \multirow{2}{*}{IO} & &   & $0.106$ & \multirow{2}{*}{$1.182$} & \multirow{2}{*}{$0.0242$} &    & \multirow{2}{*}{$0.422$}  &  \multirow{2}{*}{$0.667$} & \multirow{2}{*}{$0.615$} & \multirow{2}{*}{$0.0229$ or $0.0453$ } \\ [-0.02in]\cline{4-6}
&  & & \multirow{2}{*}{$1$} & \multirow{2}{*}{$0.862$}  & $0.894$ &  &  &&&&& \\ [-0.02in]\cline{6-6}
&  & & &   & $0.106$ &  &  &&&&& \\ [-0.02in] \hline \hline

\end{tabular}
\caption{\label{tab:best_fit}
Results of the $\chi^2$ analysis for some representative mixing patterns which arise from the breaking of the $\Delta(6n^2)$ flavor group and CP to an abelian subgroup in the charged lepton sector and a single remnant CP transformation in the neutrino sector. The $\chi^2$ function has a global minimum $\chi^2_{\text{min}}$ at the best fit values $\theta^{\text{bf}}_{1}$, $\theta^{\text{bf}}_{2}$ and $\theta^{\text{bf}}_{3}$ for $\theta_{1}$, $\theta_{2}$ and $\theta_{3}$. We display the values of the mixing angles as well as $|\sin\delta|$ and $|\sin\phi|$ at the given $\theta^{\text{bf}}_{1,2,3}$. We also present the value of the effective Majorana neutrino mass $m_{ee}$ at the best fit points $\theta^{\rm{bf}}_{1,2,3}$. Notice that $m_{ee}$ can take two distinct values due to the CP parity matrix $\widehat{X}_{\nu}$.
}
\end{table}

\section{\label{sec:Conclusions} Conclusions}

The smallness of neutrino masses can be naturally explained by the seesaw mechanism in which two or three RH neutrinos are added in the SM. The 2RHN model can be regarded as the limiting case of the three RH neutrino model in which one of the RH neutrinos is very heavy.
The 2RHN model is more predictive than the three RH neutrino model because the number of parameters is greatly reduced. One remarkable feature is that the lightest neutrino is massless in the 2RHN model. Leptogenesis is a natural cosmological consequence of the seesaw mechanism, and it provides a simple explanation for the matter-antimatter asymmetry of the Universe.

Finite discrete flavor symmetry and CP symmetry which are broken to distinct subgroups in the charged lepton and neutrino sectors, is a quite powerful approach to explain the lepton mixing angles and CP violation phases. Other phenomena involving CP phases, such as neutrinoless double beta decay and leptogenesis, are also subject to strong constraint in this approach. In the present work, we study the interplay between residual  symmetry and leptogenesis in the 2RHN model, and we assume that the scale of flavor symmetry breaking is above the leptogenesis scale. In our method, only the residual symmetry is assumed, and we do not need to consider the possible dynamics which realizes the residual symmetry.

Without loss of generality we work in the basis in which both the charged lepton and RH neutrino mass matrices are diagonal. If two residual CP transformations or a cyclic residual flavor symmetry arising from the original flavor and CP symmetries are preserved by the seesaw Lagrangian, we find that each row of the $R$-matrix would have only one nonzero entry which is equal to $\pm1$. Hence the baryon asymmetry would be zero at leading order. Successful leptogenesis is possible only if the remnant symmetry is appropriately broken by subleading order contributions in concrete models~\cite{Jenkins:2008rb_leptogenesis}.

If a single residual CP transformation is preserved in the neutrino sector, the lepton mixing matrix contains three real free parameter $\theta_{1,2,3}$ in the range of $[0, \pi)$, the $R$-matrix is found to depend on only one real parameter $\vartheta$ and it can take three viable forms summarized in Eq.~\eqref{eq:R-matrix_three_cases}. Each entry of the $R$-matrix is real or purely imaginary in this case, consequently the total CP asymmetry $\epsilon_{1}$ vanishes unless the non-leading contributions are taken into account in a concrete model. Hence in this paper we discuss the flavored thermal leptogenesis in which the interactions mediated by the $\tau$ lepton Yukawa couplings are in equilibrium, and the lightest RH neutrino mass is typically in the interval of $10^9$ GeV $\leq M_{1}\leq10^{12}$ GeV. Then the baryon asymmetry is generated uniquely by the CP phases in the PMNS mixing matrix in this scenario. Therefore the observation of low energy leptonic CP violating phases would imply the existence of a baryon asymmetry. Moreover, we have performed a general analysis of leptogenesis in the 2RHN model with a residual CP transformation. For illustration, the numerical results of $Y_B$ for $\delta=0,-\pi/2$ are presented, as shown in figures~\ref{fig:YB_C11}, \ref{fig:YB_C22} and \ref{fig:YB_C23}.

We have performed a comprehensive study in which the single remnant CP transformation originates from the CP symmetry compatible with the $\Delta(6n^2)$ flavor group which is broken to an abelian subgroup in the charged lepton sector. All possible residual symmetries and
the resulting predictions for lepton flavor mixing and leptogenesis are studied. We find there are in total eight possible cases (from case I to case VIII). The case I and case IV give rise to the same lepton mixing pattern and the same results for leptogenesis. The cases III and VII are also the same after the shift of the free parameters $\theta_{1,2,3}$ is taken into account. The PMNS matrix in cases I and IV is real up to the CP parity of the neutrino states. As a consequence, although the experimental data on mixing angles can be accommodated in these cases, all the leptogenesis CP asymmetries are vanishing and a net baryon asymmetry can not be generated without corrections. For the remaining cases, the observed matter/antimatter asymmetry could be reproduced except for R-3rd with NO spectrum and R-1st of IO. Moreover, we find that small $\Delta(6n^2)$ group (e.g., $n=2,3,4$ etc) can describe the experimentally measured values of the mixing angles for certain choices of the parameter values. Our approach is very general and model independent, and the results of this paper should be helpful to discuss the phenomenology of leptogenesis in a specific 2RHN model based on flavor and CP symmetries.

\section*{Acknowledgements}

G.-J.\, D. acknowledges the support of the National Natural Science Foundation of China under Grant No 11522546. C.-C.\, L. is supported by  CPSF-CAS Joint Foundation for Excellent Postdoctoral Fellows No. 2017LH0003.

\section*{\label{sec:appendix}Appendix}

\begin{appendix}

\section{\label{sec:Basis_independence}Basis independence}

In this paper, we have worked in the leptogenesis basis in which both the charged lepton mass matrix and the RH neutrino mass matrix are diagonal. However, the conclusions of this paper don't depend on the basis. In a large class of models, the charged lepton mass matrix is diagonal while the RH neutrino mass matrix is not diagonal. Then the Lagrangian for the lepton masses is written as
\begin{equation}
\label{eq:lagragian}
\mathcal{L}^{mod}= -y_{\alpha}\bar{L}_\alpha H l_{\alpha R} -\lambda^{mod}_{i\alpha}\bar{N}_{iR}\widetilde{H}^\dag L_\alpha-\frac{1}{2}M^{mod}_{ij}\bar{N}_{iR}N_{jR}^c+h.c.~\,,
\end{equation}
where $M^{mod}_{ij}$ is a complex symmetric $2\times2$ matrix, and it can be
diagonalized by a unitary transformation $U_{N}$,
\begin{equation}\label{eq:dia_Mf}
U^{\dagger}_{N}M^{mod}U^{*}_{N}=\text{diag}(M_1, M_2)\equiv M\,.
\end{equation}
Similar to section~\ref{sec:LepG_one_CP}, we consider the scenario that the neutrino sector preserves one CP transformation, i.e.
\begin{equation}\label{eq:res_CP_flavor}
\nu_{L}\stackrel{\text{CP}}{\longmapsto} iX_{\nu}\gamma_0C\bar{\nu}^{T}_{L}\,,~\qquad N_{R}\stackrel{\text{CP}}{\longmapsto} iX_{N}\gamma_{0}C\bar{N}^{T}_{R}\,,
\end{equation}
where the CP transformation matrix $X_{N}$ is not diagonal for non-diagonal $M^{mod}$. The invariance of $\lambda^{mod}$ and $M^{mod}$ under the above residual CP transformation implies
\begin{equation}\label{eq:cons_rcp_flavor}
X^{\dagger}_{N}\lambda^{mod}X_{\nu}=\left(\lambda^{mod}\right)^{*}\,,\qquad X^{\dagger}_{N}M^{mod}X^{\ast}_{N}=\left(M^{mod}\right)^{*}\,.
\end{equation}
Inserting Eq.~\eqref{eq:dia_Mf} into Eq.~\eqref{eq:cons_rcp_flavor} we obtain
\begin{equation}
\label{eq:cons_Uf}U^{T}_{N}X^{\dagger}_{N}U_{N}=\text{diag}(\pm1, \pm1)\equiv\widehat{X}_{N}\,.
\end{equation}
In the leptogenesis basis, the neutrino Yukawa coupling $\lambda$ takes the form
\begin{equation}\label{eq:tb_trans}
\lambda=U^{\dagger}_{N}\lambda^{mod}\,.
\end{equation}
From Eqs.~(\ref{eq:cons_rcp_flavor}, \ref{eq:cons_Uf}, \ref{eq:tb_trans}) we can check that $\lambda$ and $M$ are subject to the following constraint
\begin{equation}
\widehat{X}^{\dagger}_{N}\lambda X_{\nu}=\lambda^{*},\qquad \widehat{X}^{\dagger}_{N}M\widehat{X}^{\ast}_{N}=M^{\ast}\,,
\end{equation}
which exactly coincides with Eq.~\eqref{eq:cons_rcp_v2}. Therefore the same predictions for leptogenesis are obtained as section~\ref{sec:LepG_one_CP},
the results don't change with the working basis.

\section{\label{sec:con_one_CP}General results of $\epsilon_{\alpha}$ and $\widetilde{m}_{\alpha}$}

In this appendix we shall present the explicit expressions of the CP asymmetry parameter $\epsilon_{\alpha}$ and the washout mass $\widetilde{m}_{\alpha}$ for the three viable forms of the $R-$matrix
shown in Eq.~\eqref{eq:R-matrix_three_cases}. Here we shall perform a general analysis, and the lepton mixing matrix is parameterized in the standard convention of Eq.~\eqref{eq:PMNS_parameterized}.

\begin{itemize}[labelindent=-0.7em, leftmargin=1.2em]

\item{R-1st}

In this case, the CP asymmetry parameter $\epsilon_{\alpha}$ for the NO case is given by
\begin{eqnarray}
\nonumber \epsilon_{e}&=&\frac{3M_1}{16\pi v^2}W_{\text{NO}}s_{12} c_{13}s_{13}\sin (\delta+\frac{\phi}{2})\,,\\
\nonumber \epsilon_{\mu}&=&-\frac{3M_1}{16\pi v^2}W_{\text{NO}}c_{13} s_{23}  \left[s_{12} s_{13} s_{23} \sin (\delta +\frac{\phi }{2})-c_{12}c_{23} \sin \frac{\phi }{2}\right]\,,\\
\epsilon_{\tau}&=&-\frac{3M_1}{16\pi v^2}W_{\text{NO}}c_{13} c_{23}\left[s_{12} s_{13} c_{23} \sin (\delta +\frac{\phi }{2})+c_{12}s_{23} \sin \frac{\phi }{2}\right]\,,
\end{eqnarray}
where the expression of $W_{\text{NO}}$ has been listed in table~\ref{tab:R_W_para}. It is easy to check the identity $\epsilon_{e}+\epsilon_{\mu}+\epsilon_{\tau}=0$ is fulfilled. Notice that the CP asymmetry $\epsilon_{\alpha}$ is closely related to the lower energy CP phases. If both the Dirac phase $\delta$ and the Majorana phase $\phi$ are trivially zero, all the asymmetry parameters $\epsilon_{e}$, $\epsilon_{\mu}$ and $\epsilon_{\tau}$ would be vanishing such that a nonzero baryon asymmetry can not be generated. The washout mass $\widetilde{m}_{\alpha}$ for NO takes the form
\begin{eqnarray}
\nonumber&&\hskip-0.2in\widetilde{m}_{e}=\left|\sqrt{m_2}s_{12}c_{13} e^{\frac{i \phi }{2}} \cos\vartheta+ \xi\sqrt{m_3}  s_{13}e^{-i \delta } \sin\vartheta \right|^2\,,\\
&&\hskip-0.2in\nonumber \widetilde{m}_{\mu}=\left|\sqrt{m_2}  \left(c_{12} c_{23}- s_{12} s_{13} s_{23}e^{i \delta }\right)e^{\frac{i \phi }{2}} \cos\vartheta+ \xi\sqrt{m_3} c_{13}s_{23}\sin\vartheta\right|^2\,,\\
&&\hskip-0.2in\widetilde{m}_{\tau}=\left|\sqrt{m_2} \left(c_{12} s_{23}+s_{12} s_{13}c_{23}e^{i \delta }\right)e^{\frac{i \phi }{2}} \cos\vartheta- \xi\sqrt{m_3}c_{13} c_{23}  \sin\vartheta \right|^2\,.
\end{eqnarray}
In the same manner, we find $\epsilon_{\alpha}$ for IO spectrum  is
\begin{eqnarray}
\nonumber \hskip-0.3in \epsilon_{e}&=&-\frac{3M_1}{16\pi v^2}W_{\text{IO}}\,c_{12}s_{12}c_{13}^2\sin\frac{\phi }{2}\,,\\
\hskip-0.3in \nonumber \epsilon_{\mu}&=&\frac{-3M_1}{16\pi v^2}W_{\text{IO}}\left[s_{13}c_{23} s_{23} (c_{12}^2 \sin(\delta-\frac{\phi }{2})+s_{12}^2 \sin(\delta +\frac{\phi }{2}))-c_{12} s_{12}  (c_{23}^2-s_{13}^2 s_{23}^2)\sin\frac{\phi }{2}\right],\\
\hskip-0.3in \epsilon_{\tau}&=&\frac{3M_1}{16\pi v^2}W_{\text{IO}}\left[s_{13}c_{23}s_{23} (c_{12}^2 \sin(\delta -\frac{\phi }{2})+s_{12}^2 \sin(\delta +\frac{\phi }{2}))+c_{12} s_{12}(s_{23}^2-s_{13}^2c_{23}^2 )\sin \frac{\phi }{2}\right]
\end{eqnarray}
and for the washout mass $\widetilde{m}_{\alpha}$ we get
\begin{eqnarray}
\nonumber&&\hskip-0.2in\widetilde{m}_{e}=c^2_{13}\left|\sqrt{m_{1}}c_{12}\cos \vartheta+ \xi\sqrt{m_{2}}s_{12} e^{\frac{i \phi }{2}} \sin \vartheta  \right|^2\,,\\
&&\hskip-0.2in \nonumber \widetilde{m}_{\mu}=\left|\sqrt{m_{1}} (s_{12}c_{23} +c_{12}s_{13} s_{23} e^{i \delta } )\cos\vartheta-\xi\sqrt{m_{2}}(c_{12} c_{23}-s_{12} s_{13} s_{23}e^{i \delta } )e^{\frac{i\phi}{2}} \sin\vartheta\right|^2\,,\\
&&\hskip-0.2in\widetilde{m}_{\tau}=\left|\sqrt{m_{1}}(s_{12} s_{23}-c_{12}s_{13}c_{23} e^{i \delta } )\cos\vartheta-\xi\sqrt{m_{2}} (c_{12} s_{23}+s_{12} s_{13}c_{23} e^{i \delta } )e^{\frac{i \phi }{2}}\sin \vartheta \right|^2\,.
\end{eqnarray}
We see that both $\epsilon_{\alpha}$ and $\widetilde{m}_{\alpha}$ depend on the CP violating phases $\delta$, $\phi$ and the free parameter $\vartheta$.

\item{R-2nd}

In this case, $\epsilon_{\alpha}$ for NO is found to be,
\begin{eqnarray}
\nonumber \epsilon_{e}&=&-\frac{3M_1}{16\pi v^2}W_{\text{NO}}s_{12}c_{13}s_{13} \cos(\delta +\frac{\phi }{2})\,,\\
\nonumber \epsilon_{\mu}&=&-\frac{3M_1}{16\pi v^2}W_{\text{NO}}c_{13} s_{23}  \left[c_{12}c_{23} \cos \frac{\phi }{2}-s_{12} s_{13} s_{23} \cos (\delta+\frac{\phi }{2})\right]\,,\\
\epsilon_{\tau}&=&\frac{3M_1}{16\pi v^2}W_{\text{NO}}c_{13}c_{23}\left[c_{12}s_{23} \cos \frac{\phi}{2}+s_{12} s_{13} c_{23}\cos(\delta+\frac{\phi}{2})\right]\,.
\end{eqnarray}
The washout mass $\widetilde{m}_{\alpha}$ is of the following form
\begin{eqnarray}
\nonumber&&\hskip-0.2in\widetilde{m}_{e}=\left|\sqrt{m_2}s_{12}c_{13}e^{\frac{i \phi }{2}} \cosh\vartheta- i\xi\sqrt{m_3}  s_{13}e^{-i \delta } \sinh\vartheta \right|^2\,,\\
&&\hskip-0.2in\nonumber\widetilde{m}_{\mu}=\left|\sqrt{m_2}  \left(c_{12} c_{23}- s_{12} s_{13} s_{23}e^{i \delta }\right)e^{\frac{i \phi }{2}} \cosh\vartheta- i\xi\sqrt{m_3}c_{13}s_{23} \sinh\vartheta\right|^2\,,\\
&&\hskip-0.2in\widetilde{m}_{\tau}=\left|\sqrt{m_2} \left(c_{12} s_{23}+s_{12} s_{13}c_{23}e^{i \delta }\right)e^{\frac{i \phi }{2}} \cosh\vartheta+ i\xi\sqrt{m_3}c_{13} c_{23}  \sinh\vartheta \right|^2\,.
\end{eqnarray}
Similarly for IO mass spectrum, we have
\begin{eqnarray}
\nonumber \hskip-0.3in \epsilon_{e}&=&-\frac{3M_1}{16\pi v^2}W_{\text{IO}}\,c_{12} s_{12} c_{13}^2\cos\frac{\phi }{2}\,,\\
\hskip-0.3in \nonumber \epsilon_{\mu}&=&\frac{3M_1}{16\pi v^2}W_{\text{IO}}\left[s_{13}c_{23}s_{23} (c_{12}^2 \cos (\delta -\frac{\phi }{2})-s_{12}^2 \cos (\delta +\frac{\phi }{2}))+c_{12} s_{12}(c_{23}^2-s_{13}^2 s_{23}^2)\cos\frac{\phi }{2}\right],\\
\hskip-0.3in \epsilon_{\tau}&=&-\frac{3M_1}{16\pi v^2}W_{\text{IO}}\left[s_{13}c_{23}s_{23} (c_{12}^2 \cos (\delta -\frac{\phi }{2})-s_{12}^2 \cos (\delta +\frac{\phi }{2}))+c_{12} s_{12}(s_{13}^2c_{23}^2-s_{23}^2)\cos\frac{\phi}{2}\right]\,.
\end{eqnarray}
and
\begin{eqnarray}
\nonumber&&\hskip-0.2in\widetilde{m}_{e}=c^2_{13}\left|\sqrt{m_{1}}c_{12} \cosh\vartheta-i\xi\sqrt{m_{2}} s_{12} e^{\frac{i \phi }{2}}\sinh\vartheta  \right|^2\,,\\
&&\hskip-0.2in\nonumber\widetilde{m}_{\mu}=\left|\sqrt{m_{1}} (s_{12}c_{23}+c_{12}s_{13}s_{23} e^{i \delta})\cosh\vartheta +i\xi\sqrt{m_{2}}(c_{12} c_{23}-s_{12} s_{13} s_{23}e^{i\delta} )e^{\frac{i \phi }{2}} \sinh\vartheta\right|^2\,,\\
&&\hskip-0.2in\widetilde{m}_{\tau}=\left|\sqrt{m_{1}}(s_{12} s_{23}-c_{12}s_{13}c_{23}e^{i \delta})\cosh\vartheta+i\xi\sqrt{m_{2}}(c_{12} s_{23}+s_{12} s_{13}c_{23} e^{i \delta})e^{\frac{i \phi }{2}} \sinh\vartheta \right|^2
\end{eqnarray}

\item{R-3rd}

In the case of NO, we find the flavored CP asymmetry $\epsilon_{\alpha}$ is
\begin{eqnarray}
\nonumber \epsilon_{e}&=&\frac{3M_1}{16\pi v^2}W_{\text{NO}}s_{12}c_{13} s_{13} \cos (\delta +\frac{\phi }{2})\,,\\
\nonumber \epsilon_{\mu}&=&\frac{3M_1}{16\pi v^2}W_{\text{NO}}c_{13} s_{23}  \left[c_{12}c_{23} \cos \frac{\phi }{2}-s_{12} s_{13} s_{23} \cos(\delta +\frac{\phi }{2})\right]\,,\\
\epsilon_{\tau}&=&-\frac{3M_1}{16\pi v^2}W_{\text{NO}}c_{13} c_{23}\left[c_{12}s_{23} \cos \frac{\phi }{2}+s_{12} s_{13} c_{23} \cos(\delta +\frac{\phi }{2})\right]\,.
\end{eqnarray}
It is easy to check the equality $\epsilon_{2}\equiv\epsilon_{e}+\epsilon_{\mu}=-\epsilon_{\tau}$ is satisfied. The washout mass $\widetilde{m}_{\alpha}$ takes the form
\begin{eqnarray}
\nonumber&&\hskip-0.2in\widetilde{m}_{e}=\left|i\sqrt{m_2}s_{12}c_{13}e^{\frac{i \phi }{2}} \sinh\vartheta+\xi\sqrt{m_3}  s_{13}e^{-i \delta } \cosh\vartheta \right|^2\,,\\
&&\hskip-0.2in\nonumber\widetilde{m}_{\mu}=\left|i\sqrt{m_2}  \left(c_{12} c_{23}- s_{12} s_{13} s_{23}e^{i \delta }\right)e^{\frac{i \phi }{2}} \sinh\vartheta+\xi\sqrt{m_3} c_{13}s_{23} \cosh\vartheta\right|^2\,,\\
&&\hskip-0.2in\widetilde{m}_{\tau}=\left|i\sqrt{m_2} \left(c_{12} s_{23}+s_{12} s_{13}c_{23}e^{i \delta }\right)e^{\frac{i \phi }{2}} \sinh\vartheta-\xi\sqrt{m_3}c_{13} c_{23}  \cosh\vartheta \right|^2\,.
\end{eqnarray}
For the IO case, we can read out $\epsilon_{\alpha}$ as
\begin{eqnarray}
\nonumber \hskip-0.3in \epsilon_{e}&=&\frac{3M_1}{16\pi v^2}W_{\text{IO}}c_{12} s_{12}c_{13}^2\cos\frac{\phi}{2}\,,\\
\hskip-0.3in\nonumber \epsilon_{\mu}&=&\frac{-3M_1}{16\pi v^2}W_{\text{IO}}\left(s_{13}c_{23}  s_{23} (c_{12}^2 \cos (\delta -\frac{\phi }{2})-s_{12}^2 \cos(\delta +\frac{\phi }{2}))+c_{12} s_{12} (c_{23}^2-s_{13}^2 s_{23}^2)\cos\frac{\phi }{2}\right),\\
\hskip-0.3in  \epsilon_{\tau}&=&\frac{3M_1}{16\pi v^2}W_{\text{IO}}\left(s_{13}c_{23}s_{23}(c_{12}^2 \cos (\delta -\frac{\phi }{2})-s_{12}^2 \cos (\delta +\frac{\phi }{2}))-c_{12} s_{12}(s_{23}^2-s_{13}^2c_{23}^2 )\cos\frac{\phi }{2} \right)\,.
\end{eqnarray}
Furthermore the washout mass $\widetilde{m}_{\alpha}$ for IO turns out to be
\begin{eqnarray}
\nonumber&&\hskip-0.2in\widetilde{m}_{e}=c_{13}^2\left|i\sqrt{m_{1}}c_{12} \sinh\vartheta+ \xi\sqrt{m_{2}}s_{12} e^{\frac{i \phi }{2}} \cosh\vartheta\right|^2\,,\\
&&\hskip-0.2in\nonumber \widetilde{m}_{\mu}=\left|i \sqrt{m_{1}}(s_{12}c_{23} +c_{12}s_{13} s_{23} e^{i \delta })\sinh\vartheta-\xi\sqrt{m_{2}}(c_{12} c_{23}-s_{12} s_{13} s_{23}e^{i \delta } )e^{\frac{i \phi }{2}} \cosh\vartheta\right|^2\,,\\
&&\hskip-0.2in\widetilde{m}_{\tau}=\left|i \sqrt{m_{1}} (s_{12} s_{23}-c_{12}s_{13} c_{23} e^{i \delta })\sinh\vartheta-\xi\sqrt{m_{2}} (c_{12} s_{23}+s_{12} s_{13}c_{23} e^{i \delta})e^{\frac{i \phi }{2}}\cosh\vartheta \right|^2\,.
\end{eqnarray}

\end{itemize}

\section{\label{sec:D6n2_example}Lepton mixing patterns and leptogenesis from $\Delta(6n^2)$ and CP}

As shown in section~\ref{sec:example}, it is sufficient to only consider eight possible residual symmetries in the scenario that the discrete flavor group $\Delta(6n^2)$ and CP symmetry are broken down to an abelian subgroup $G_{l}$ in the charged lepton sector and to a single remnant CP transformation $X_{\nu}$ in the neutrino sector. In the following, we shall investigate the predictions for lepton flavor mixing and matter-antimatter asymmetry via leptogenesis in each possible case.

\begin{description}[labelindent=-0.8em, leftmargin=0.3em]

\item[~~(\uppercase\expandafter{\romannumeral1})]{$G_{l}=\left\langle c^{s}d^{t}\right\rangle$, $X_{\nu}=\rho_{\mathbf{3}}(c^{x}d^{y})$}

From table~\ref{tab:cle_diagonal_matrix} and Eq.~\eqref{eq:Unu_1} we find that the lepton mixing matrix is given by
\begin{equation}\label{eq:PMNS_I}
U_{I}=P_{l}O_{3\times3}(\theta_{1},\theta_{2},\theta_{3})\widehat{X}^{-\frac{1}{2}}_{\nu}\,.
\end{equation}
The permutation matrix $P_{l}$ can be absorbed into the orthogonal matrix $O_{3\times3}$, hence we can choose $P_{l}=P_{123}=1_{3\times3}$ without loss of generality. Thus the three lepton mixing angles read
\begin{eqnarray}\label{eq:mixing_para_caseI}
 \sin^2\theta_{12}=\sin^2\theta_{3}, \quad\sin^{2}\theta_{13}=\sin^2\theta_{2}, \quad \sin^{2}\theta_{23}=\sin^2\theta_{1}
\end{eqnarray}
and the Jarlskog invariant $J_{CP}$ is vanishing
\begin{equation}
J_{CP}=0\,,
\end{equation}
where $J_{CP}$ is defined as~\cite{Jarlskog:1985ht}
\begin{equation}
J_{CP}=\Im\left(U_{11}U_{33}U^{*}_{13}U^{*}_{31}\right)=\frac{1}{8}\sin2\theta_{12}\sin2\theta_{13}\sin2\theta_{23}\cos\theta_{13}\sin\delta\,.
\end{equation}
Consequently the Dirac CP phase $\delta$ is either 0 or $\pi$. Moreover, we can easily check that both the rephase invariants $I_{\text{NO}}^{\alpha}$, $I_{\text{IO}}^{\alpha}$ and the CP asymmetry $\epsilon_{\alpha}$ in leptogenesis are vanishing as well,
\begin{equation}
I_{\text{NO}}^{\alpha}=I_{\text{IO}}^{\alpha}=\epsilon_{\alpha}=0\,.
\end{equation}
Therefore a net baryon asymmetry can not be generated in this case, and moderate subleading corrections are necessary in order to make the leptogenesis viable.

\item[~~(\uppercase\expandafter{\romannumeral2})]{$G_{l}=\left\langle c^{s}d^{t}\right\rangle$, $X_{\nu}=\rho_{\mathbf{3}}(bc^{x}d^{-x})$}

In this case, the PMNS mixing matrix is determined to be of the form
\begin{eqnarray}
\nonumber U_{II}&=&\frac{1}{\sqrt{2}}\begin{pmatrix}
0 &~ -i ~& 1 \\
\sqrt{2} &~ 0 ~& 0 \\
0&~ i ~& 1
\end{pmatrix}
O_{3\times3}(\theta_{1},\theta_{2},\theta_{3})\widehat{X}^{-\frac{1}{2}}_{\nu}  \\
\label{eq:PMNS_II} &=&\text{diag}(e^{-i\theta_{1}},1,e^{i\theta_{1}})\frac{1}{\sqrt{2}}\begin{pmatrix}
0 &~ -i ~& 1 \\
\sqrt{2} &~ 0 ~& 0 \\
0&~ i ~& 1
\end{pmatrix}
O_{3\times3}(0,\theta_{2},\theta_{3})\widehat{X}^{-\frac{1}{2}}_{\nu}
\end{eqnarray}
up to possible permutations of rows. The diagonal phase matrix $\text{diag}(e^{-i\theta_{1}},1,e^{i\theta_{1}})$ can be absorbed into the charged lepton fields. Moreover, it is easy to check that the following identity is fulfilled
\begin{equation}
P_{321}U_{II}(\theta_{1},\theta_{2},\theta_{3})=U_{II}(-\theta_{1},\theta_{2},-\theta_{3})\text{diag}(1,-1,1)\,.
\end{equation}
Consequently the six possible row permutations lead to three independent mixing patterns,
\begin{equation}\label{eq:mix_II_ind}
U_{II,1}=U_{II}, \quad U_{II,2}=P_{132}U_{II}, \quad U_{II,3}=P_{213}U_{II}\,.
\end{equation}
We find that $U_{II,1}$ and $U_{II,2}$ predict $\tan\theta_{13}=\cos\theta_{23}$ and $\tan\theta_{13}=\sin\theta_{23}$ respectively such that the experimental data~\cite{Capozzi:2013csa} of the mixing angles $\theta_{13}$ and $\theta_{23}$ can not be accommodated simultaneously. For the mixing matrix $U_{II,3}$, the lepton mixing parameters are given by
\begin{eqnarray}
\nonumber&&\sin^2\theta_{13}=\sin ^2\theta_{2}, \quad \sin^2\theta_{12}=\sin ^2\theta_{3} , \quad \sin^2\theta_{23}=\frac{1}{2}\,,\\
\label{eq:mixing_para_caseII}&&J_{CP}=\frac{1}{8}\cos \theta_{2}\sin 2\theta_{2} \sin 2\theta_{3}, \quad |\sin\delta|=1\,.
\end{eqnarray}
 Furthermore, we find that the rephasing bilinear invariants take the form
\begin{eqnarray}
\nonumber&&I^{e}_{\text{NO}}=0,\quad I^{\mu}_{\text{NO}}=-I^{\tau}_{\text{NO}}=\frac{1}{2}\cos\theta_{2}\cos\theta_{3},\\
&&I^{e}_{\text{IO}}=0,\quad I^{\mu}_{\text{IO}}=-I^{\tau}_{\text{IO}}=\frac{1}{2}\sin\theta_{2}\,.
\end{eqnarray}
Hence only the muon and tau flavored asymmetries in heavy neutrino decay contribute to the leptogenesis.

\item[~~(\uppercase\expandafter{\romannumeral3})]{$G_{l}=\left\langle bc^{s}d^{t}\right\rangle$, $X_{\nu}=\rho_{\mathbf{3}}(c^{x}d^{y})$}

Using table~\ref{tab:cle_diagonal_matrix} and Eq.~\eqref{eq:Unu_1}, we find that the lepton mixing matrix up to possible permutations of rows is fixed to be
\begin{equation}\label{eq:PMNS_III}
U_{III}=\frac{1}{\sqrt{2}}
\begin{pmatrix}
 1 &~ 0 ~& -e^{i\varrho_{1}} \\
 0 &~ \sqrt{2} ~& 0 \\
 1 &~ 0 ~& e^{i \varrho_{1}} \\
\end{pmatrix}O_{3\times3}(\theta_{1},\theta_{2},\theta_{3})\widehat{X}^{-\frac{1}{2}}_{\nu}\,,
\end{equation}
with
\begin{equation}
\varrho_{1}=-\frac{  (s+t+x+y)}{n}\pi\,,
\end{equation}
which can take the following values
\begin{equation}
\varrho_{1}~(\mathrm{mod}~2\pi)=0, \frac{1}{n}\pi, \frac{2}{n}\pi, \cdots, \frac{2n-1}{n}\pi\,.
\end{equation}
We can easily check that the mixing matrix $U_{III}$ has the properties
\begin{eqnarray}
\nonumber && U_{III}(\varrho_{1}+\pi,\theta_{1},\theta_{2},\theta_{3})=U_{III}(\varrho_{1},-\theta_{1},-\theta_{2},\theta_{3})\text{diag}(1,1,-1), \\
\label{eq:symmetry_caseIII_1st}&& U_{III}(\pi-\varrho_{1},\theta_{1},\theta_{2},\theta_{3})=\text{diag}(-e^{-i\varrho_{1}},1,e^{-i\varrho_{1}})
U_{III}(\varrho_{1},\theta^{\prime}_{1},\theta^{\prime}_{2},\theta^{\prime}_{3})\text{diag}(1,1,-1),
\end{eqnarray}
where the parameters $\theta'_{1,2,3}$ fulfill $O_{3\times3}(\theta^{\prime}_{1},\theta^{\prime}_{2},\theta^{\prime}_{3})=P_{321}O_{3\times3}(-\theta_{1},-\theta_{2},\theta_{3})$. As a consequence, the fundamental interval of the parameter $\varrho_1$ can be chosen to be $0\leq\varrho_1\leq\frac{\pi}{2}$. The mixing pattern arising from the multiplication of the permutation matrix $P_{321}$ from the left-hand side, is related to $U_{III}$ through shifts of the continuous parameters $\theta_{1,2,3}$ and redefining $\hat{X}_{\nu}$ as follow,
\begin{equation}
P_{321}U_{III}(\varrho_{1},\theta_{1},\theta_{2},\theta_{3})=U_{III}(\varrho_{1},-\theta_{1},-\theta_{2},\theta_{3})\text{diag}(1,1,-1)\,.
\end{equation}
Hence three mixing patterns are obtained after all the six row permutations are considered,
\begin{equation}\label{eq:mix_III_ind}
U_{III,1}=U_{III}\,, \quad U_{III,2}=P_{132}U_{III}, \quad U_{III,3}=P_{213}U_{III}\,.
\end{equation}
For the mixing matrix $U_{III,1}$, we can extract the mixing parameters in the usual way and find
\begin{equation}
\label{eq:mixing_para_caseIII1}
\begin{aligned}
&\sin^2\theta_{13}=\frac{1}{2} \left(\sin ^2\theta_{2}+\cos ^2\theta_{1} \cos ^2\theta_{2}- \cos \theta_{1} \sin 2\theta_{2}  \cos \varrho_{1}\right),\\
&\sin^2\theta_{12}=\sin^2\theta_{3}+\frac{\sin2\theta_{3} (2 \sin\theta_{1} \cos \theta_{2} \cos \varrho_1+\sin2 \theta_{1} \sin\theta_{2})+2 \sin ^2\theta_{1} \cos 2 \theta_{3}}{2-\sin ^2\theta_{2}-\cos ^2\theta_{1} \cos ^2\theta_{2}+\cos\theta_{1} \sin 2\theta_{2}  \cos \varrho_{1}},\\
&\sin^2\theta_{23}=\frac{2 \sin ^2\theta_{1} \cos ^2\theta_{2}}{2-\sin ^2\theta_{2}-\cos ^2\theta_{1} \cos ^2\theta_{2}+\cos\theta_{1} \sin 2\theta_{2}  \cos \varrho_{1}}\,, \\
& J_{CP}=\frac{1}{16}\sin \theta_{1} \cos \theta_{2} \sin \varrho_{1}\left[4\sin 2\theta_{1} \sin \theta_{2}\cos2\theta_{3}+(1+3\cos2\theta_1+2\sin^2\theta_1\cos2\theta_{2})\sin2\theta_{3}\right]\,,
\end{aligned}
\end{equation}
which have the symmetry transformation $(\theta_{1}, \theta_{2}, \theta_{3})\rightarrow (\pi-\theta_{1}, \pi-\theta_{2}, \pi-\theta_{3})$. As regards the leptogenesis, the relevant CP invariants are of the form
\begin{eqnarray}
\nonumber && I^{e}_{\text{NO}}=-I^{\tau}_{\text{NO}}=-\frac{1}{2}  (\cos \theta_{1} \sin \theta_{3}+\sin \theta_{1} \sin \theta_{2} \cos \theta_{3})\sin \varrho_{1}, \quad I^{\mu}_{\text{NO}}=0\,, \\
\label{eq:Ia_III1}&& I^{e}_{\text{IO}}=-I^{\tau}_{\text{IO}}=\frac{1}{2} \sin \theta_{1} \cos \theta_{2} \sin \varrho_{1}, \quad I^{\mu}_{\text{IO}}=0\,,
\end{eqnarray}
One sees that the all the lepton asymmetry $\epsilon_{\alpha}$ would be vanishing for $\varrho_{1}=0$ such that the cosmological baryon asymmetry can not be generated. For the second mixing pattern $U_{III,2}$, the three lepton mixing angles and Jarlskog invariant are determined to be
\begin{equation}
\label{eq:mixing_para_caseIII2}
\begin{aligned}
&\sin^2\theta_{13}=\frac{1}{2} \left(\sin ^2\theta_{2}+\cos ^2\theta_{1} \cos ^2\theta_{2}- \cos \theta_{1} \sin 2\theta_{2}  \cos \varrho_{1}\right),\\
&\sin^2\theta_{12}=\sin^2\theta_{3}+\frac{\sin2\theta_{3} (2 \sin\theta_{1} \cos \theta_{2} \cos \varrho_1+\sin2 \theta_{1} \sin\theta_{2})+2 \sin ^2\theta_{1} \cos 2 \theta_{3}}{2-\sin ^2\theta_{2}-\cos^2\theta_{1} \cos ^2\theta_{2}+ \cos \theta_{1} \sin 2\theta_{2}  \cos \varrho_{1}},\\
&\sin^2\theta_{23}=\frac{1-\sin ^2\theta_{1}\cos^2\theta_2+ \cos \theta_{1} \sin 2\theta_{2}  \cos \varrho_{1}}{2-\sin ^2\theta_{2}-\cos ^2\theta_{1} \cos ^2\theta_{2}+ \cos \theta_{1} \sin 2\theta_{2}  \cos \varrho_{1}}\,, \\
& J_{CP}=-\frac{1}{16}\sin \theta_{1} \cos \theta_{2} \sin \varrho_{1} \left[4\sin 2\theta_{1} \sin \theta_{2} \cos 2\theta_{3}+\left(1+3\cos2\theta_1+2\sin^2\theta_1\cos2\theta_2\right)\sin 2\theta_{3}\right]\,.
\end{aligned}
\end{equation}
Regarding the CP invariants in leptopgenesis, we get
\begin{eqnarray}
\nonumber && I^{e}_{\text{NO}}=-I^{\mu}_{\text{NO}}=-\frac{1}{2} (\cos \theta_{1} \sin \theta_{3}+\sin \theta_{1} \sin \theta_{2} \cos \theta_{3}) \sin \varrho_{1}, \quad I^{\tau}_{\text{NO}}=0\,, \\
\label{eq:Ia_III2} && I^{e}_{\text{IO}}=-I^{\mu}_{\text{IO}}=\frac{1}{2} \sin \theta_{1} \cos \theta_{2} \sin \varrho_{1}, \quad I^{\tau}_{\text{IO}}=0\,,
\end{eqnarray}
which implies  $I^{e}_{\text{NO}}+I^{\mu}_{\text{NO}}=0$ and $I^{e}_{\text{IO}}+I^{\mu}_{\text{IO}}=0$.
Hence the summation of the CP asymmetry in the electron and muon flavors would vanish, i.e., $\epsilon_2\equiv\epsilon_{e}+\epsilon_{\mu}=0$. As a consequence, $Y_{B}$ would be predicted to be zero in the mass window $10^{9}\,\mathrm{GeV}\leq M_1\leq 10^{12}$ GeV unless the postulated residual symmetry is broken by non-leading order corrections arising from higher dimensional operators. For the third possible PMNS mixing matrix $U_{III,3}$, the lepton mixing parameters read as
\begin{equation}
\label{eq:mixing_para_caseIII3}
\begin{aligned}
&\sin^2\theta_{13}=\sin ^2\theta_{1} \cos ^2\theta_{2},\\
&\sin^2\theta_{12}=\frac{(\cos \theta_{1} \cos \theta_{3}-\sin \theta_{1} \sin \theta_{2} \sin \theta_{3})^2}{1-\sin ^2\theta_{1} \cos ^2\theta_{2}},\\
&\sin^2\theta_{23}=\frac{1}{2}-\frac{\cos \theta_{1} \sin 2 \theta_{2} \cos \varrho_{1}}{2-2 \sin ^2\theta_{1} \cos ^2\theta_{2}}\,, \\
&J_{CP}=-\frac{1}{16}\sin\theta_{1} \cos \theta_{2} \sin \varrho_{1}\left[4\sin2\theta_{1}\sin\theta_{2}\cos2\theta_3+\left(1+3\cos2\theta_1+2\sin^2\theta_1\cos2\theta_2\right)\sin2\theta_3\right]\,.
\end{aligned}
\end{equation}
The rephase invariants $I^{\alpha}_{\text{NO}}$ and $I^{\alpha}_{\text{IO}}$ are of the following form
\begin{eqnarray}
\nonumber && I^{\mu}_{\text{NO}}=-I^{\tau}_{\text{NO}}=-\frac{1}{2} (\cos \theta_{1} \sin \theta_{3}+\sin \theta_{1} \sin \theta_{2} \cos \theta_{3})\sin \varrho_{1}, \quad I^{e}_{\text{NO}}=0\,, \\
\label{eq:Ia_III3} && I^{\mu}_{\text{IO}}=-I^{\tau}_{\text{IO}}=\frac{1}{2} \sin \theta_{1} \cos \theta_{2} \sin \varrho_{1}, \quad I^{e}_{\text{IO}}=0\,.
\end{eqnarray}

\item[~~(\uppercase\expandafter{\romannumeral4})]{$G_{l}=\left\langle bc^{s}d^{t}\right\rangle$, $X_{\nu}=\rho_{\mathbf{3}}(bc^{x}d^{-x})$}

In the same manner as previous cases, we find the lepton mixing matrix is given by
\begin{eqnarray}\label{eq:PMNS_IV}
\nonumber U_{IV}&=&P_{l}\left(
\begin{array}{ccc}
 0 &~ \cos \varrho_{2} ~& \sin \varrho_{2} \\
 1 &~ 0 ~& 0 \\
 0 &~ -\sin \varrho_{2} ~& \cos \varrho_{2} \\
\end{array}
\right)O_{3\times3}(\theta_{1},\theta_{2},\theta_{3})\widehat{X}^{-\frac{1}{2}}_{\nu}\\
&=&P_{l}P_{213}O_{3\times3}(\theta_{1}+\varrho_{2},\theta_{2},\theta_{3})
\widehat{X}^{-\frac{1}{2}}_{\nu}\,,
\end{eqnarray}
where $P_{l}$ is a generic $3\times3$ permutation matrix, and the contributions of $P_{l}$ and $P_{213}$ can be absorbed into the real orthogonal matrix $O_{3\times3}$. The parameter $\varrho_2$ is fixed by the chosen residual symmetry as
\begin{equation}
\varrho_{2}=-\frac{s+t}{2n}\pi
\end{equation}
whose possible values are
\begin{equation}
\varrho_{2}~(\mathrm{mod}~2\pi)=0, \frac{1}{2n}\pi, \frac{2}{2n}\pi, \cdots, \frac{4n-1}{2n}\pi\,.
\end{equation}
After the relabeling of $P_{l}P_{213}\rightarrow P_{l}$ and $\theta_{1}+\varrho_{1}\rightarrow\theta_{1}$ is taken into account, the mixing matrix $U_{IV}$ would coincide with $U_I$ shown in Eq.~\eqref{eq:PMNS_I}. As a result, the predictions for mixing parameters and leptogenesis are exactly the same as case I. The experimentally preferred values of the lepton mixing angles can be accommodated, the Dirac CP phase $\delta$ is trivial, and the cosmic baryon asymmetry $Y_{B}$ is predicted to be vanishing without higher order corrections.

\item[~~(\uppercase\expandafter{\romannumeral5})]{$G_{l}=\left\langle ac^{s}d^{t}\right\rangle$, $X_{\nu}=\rho_{\mathbf{3}}(c^{x}d^{y})$}

Combining the unitary transformations $U_{l}$ for $G_{l}=\left\langle ac^{s}d^{t}\right\rangle$ shown in table~\ref{tab:cle_diagonal_matrix} and $U_{\nu}$ in Eq.~\eqref{eq:Unu_1}, we find that the PMNS mixing matrix is of the form
\begin{equation}\label{eq:PMNS_V}
U_{V}=\frac{1}{\sqrt{3}}
\begin{pmatrix}
 e^{i \varrho_{3}} &~ 1 ~& e^{i \varrho_{4}} \\
 \omega ^2e^{i \varrho_{3}}  &~ 1 ~& \omega  e^{i \varrho_{4}}  \\
 \omega e^{i \varrho_{3}}   &~ 1 ~& \omega ^2e^{i \varrho_{4}}  \\
\end{pmatrix}O_{3\times3}(\theta_{1},\theta_{2},\theta_{3})\widehat{X}^{-\frac{1}{2}}_{\nu}\,,
\end{equation}
up to permutations of rows, where $\varrho_{3}$ and $\varrho_{4}$ are determined by residual symmetry,
\begin{equation}
\varrho_{3}=\frac{ 2 s-2 t+2 x-y}{n}\pi\,, \quad \varrho_{4}=\frac{-2 t+x-2 y}{n}\pi\,,
\end{equation}
which can independently take the values
\begin{equation}
\varrho_{3},\varrho_{4}~(\mathrm{mod}~2\pi)=0, \frac{1}{n}\pi, \frac{2}{n}\pi, \cdots, \frac{2n-1}{n}\pi\,.
\end{equation}
We observe that the mixing matrix $U_{V}$ has the following properties
\begin{eqnarray}
\nonumber U_{V}(\varrho_3+\pi,\varrho_4,\theta_1, \theta_2, \theta_3)=U_{V}(\varrho_3,\varrho_4,\theta_1, -\theta_2, -\theta_3)\text{diag}(-1, 1, 1),\\
U_{V}(\varrho_3,\varrho_4+\pi, \theta_1, \theta_2, \theta_3)=U_{V}(\varrho_3,\varrho_4, -\theta_1, -\theta_2, \theta_3)\text{diag}(1, 1, -1)\,.
\end{eqnarray}
Consequently the fundamental regions of the parameters $\varrho_3$ and $\varrho_4$ can be taken to be $[0, \pi)$. Exchanging the second and the third rows of $U_{V}$ leads to the same mixing pattern as swapping $\varrho_{3}$ and $\varrho_{4}$, i.e.
\begin{equation}
P_{132}U_{V}(\varrho_{3},\varrho_{4},\theta_{1},\theta_{2},\theta_{3})
=U_{V}(\varrho_{4},\varrho_{3},\theta^{\prime}_{1},\theta^{\prime}_{2},\theta^{\prime}_{3})\,,
\end{equation}
where $\theta^{\prime}_{1, 2, 3}$ fulfill $O_{3\times3}(\theta^{\prime}_{1},\theta^{\prime}_{2},\theta^{\prime}_{3})=P_{321}O_{3\times3}(\theta_{1},\theta_{2},\theta_{3})$. Hence it is enough to only consider three out of the six possible row permutations if all possible values of $\varrho_3$ and $\varrho_4$ are considered,
\begin{equation}\label{eq:all_PMNS_V}
U_{V,1}=U_{V}, \quad U_{V,2}=P_{213}U_{V}, \quad U_{V,3}=P_{231}U_{V}\,.
\end{equation}
For the case of $U_{V,1}$, we can obtain the following expressions for the mixing angles and the Jarlskog invariant,
\begin{eqnarray}
\nonumber&&\sin^{2}\theta_{13}=\frac{1}{3} \left[\sin 2\theta_{2} (\cos \theta_{1} \cos (\varrho_{3}-\varrho_{4})+\sin \theta_{1} \cos\varrho_{3})+\sin 2\theta_{1} \cos ^2\theta_{2} \cos\varrho_{4}+1\right],\\
\nonumber&&\sin^2\theta_{12}=\sin^2\theta_{3}+\frac{\sin 2 \theta_{3} (\cos \theta_{1} \cos \theta_{2} \cos \varrho_{3}-\sin \theta_{1} \cos \theta_{2} \cos (\varrho_{3}-\varrho_{4})-\cos 2 \theta_{1} \sin \theta_{2} \cos \varrho_{4})}{2-\sin 2\theta_{2} (\cos \theta_{1} \cos (\varrho_{3}-\varrho_{4})+\sin \theta_{1} \cos\varrho_{3})-\sin 2\theta_{1} \cos ^2\theta_{2} \cos \varrho_{4}}\\
\nonumber &&\qquad\qquad+\frac{\cos 2 \theta_{3} (1- \sin 2\theta_{1}  \cos \varrho_{4})}{2-\sin 2\theta_{2} (\cos \theta_{1} \cos (\varrho_{3}-\varrho_{4})+\sin \theta_{1} \cos\varrho_{3})-\sin 2\theta_{1} \cos ^2\theta_{2} \cos \varrho_{4}},\\
\nonumber &&\sin^2\theta_{23}=\frac{\sin 2\theta_{2} \left(\cos \theta_{1}  \sin (\varrho_{3}-\varrho_{4}+\frac{\pi}{6})+\sin \theta_{1} \cos(\varrho_{3}+\frac{\pi}{3})\right)+\cos ^2\theta_{2} \sin 2\theta_{1}  \cos(\varrho_{4}-\frac{\pi}{3})-1}{\sin 2\theta_{2} (\cos \theta_{1} \cos (\varrho_{3}-\varrho_{4})+\sin \theta_{1} \cos\varrho_{3})+\sin 2\theta_{1} \cos ^2\theta_{2} \cos\varrho_{4}-2}\,, \\
\nonumber && J_{CP}=\frac{1}{6\sqrt{3}}\Big[\cos 2 \theta_{1} \cos 2 \theta_{2} \cos 2 \theta_{3}+\sin2 \theta_{1}\sin\theta_2(1-3/2\cos^2\theta_2)\sin 2 \theta_{3} \\
\nonumber &&\qquad~+\cos \theta_{2} \left(\sin 2 \theta_{3} \left(\cos ^2\theta_{1}-\sin ^2\theta_{1} \sin ^2\theta_{2}\right)+\sin 2 \theta_{1} \sin \theta_{2} \cos 2 \theta_{3}\right) [\sin \theta_{1} \cos (\varrho_{3}+\varrho_{4}) \\
\label{eq:mixing_para_caseV1} &&\qquad-\cos \theta_{1} \cos (\varrho_{3}-2 \varrho_{4})]+\cos ^3\theta_{2} \sin 2 \theta_{3} (\cos \theta_{1} \cos (\varrho_{3}-2 \varrho_{4})-\tan \theta_{2} \cos (2 \varrho_{3}-\varrho_{4}))\Big]\,.
\end{eqnarray}
Moreover, the CP invariants $I^{\alpha}_{\text{NO}}$ and $I^{\alpha}_{\text{IO}}$ are given by
\begin{eqnarray}
\nonumber && I^{e}_{\text{NO}}=\frac{1}{3} \left[\sin (\varrho_{4}-\varrho_{3})(\sin \theta_{1} \sin \theta_{2} \cos \theta_{3}+\cos\theta_{1}\sin\theta_{3})+\cos \theta_{2} \cos \theta_{3} \sin \varrho_{4}\right. \\
\nonumber &&~~~~~~~ \left.+\sin\varrho_{3}(\cos \theta_{1} \sin \theta_{2} \cos \theta_{3}-\sin\theta_{1}\sin\theta_{3})\right]\, \,,\\
\nonumber && I^{\mu}_{\text{NO}}=\frac{1}{3} \left[\sin (\varrho_{3}-\varrho_{4}-\frac{\pi}{3})(\sin \theta_{1} \sin \theta_{2} \cos \theta_{3}+\cos\theta_{1}\sin\theta_{3})+\cos \theta_{2} \cos \theta_{3} \sin (\frac{\pi}{3}-\varrho_{4})\right.  \\
\nonumber &&~~~~~~~ \left.+\sin (\varrho_{3}+\frac{\pi}{3})(\sin\theta_{1}\sin\theta_{3}-\cos \theta_{1} \sin \theta_{2} \cos \theta_{3})\right]\,\,, \\
\nonumber && I^{e}_{\text{IO}}=\frac{1}{3} (\sin \theta_{1} \cos \theta_{2} \sin (\varrho_{3}-\varrho_{4})-\cos \theta_{1} \cos \theta_{2} \sin\varrho_{3}+\sin \theta_{2} \sin \varrho_{4})\,, \\
\nonumber && I^{\mu}_{\text{IO}}=\frac{1}{3} \left(\sin \theta_{1} \cos \theta_{2} \sin (\frac{\pi}{3}-\varrho_{3}+\varrho_{4})+\cos \theta_{1} \cos \theta_{2} \sin (\varrho_{3}+\frac{\pi}{3})+\sin \theta_{2} \sin (\frac{\pi}{3}-\varrho_{4})\right)\,, \\
\label{eq:Ia_V}&& I^{\tau}_{\text{NO}}=-(I^{e}_{\text{NO}}+I^{\mu}_{\text{NO}}), \qquad I^{\tau}_{\text{IO}}=-(I^{e}_{\text{IO}}+I^{\mu}_{\text{IO}})\,.
\end{eqnarray}
Then we proceed to discuss the second permutation $U_{V,2}$, we can straightforwardly extract the mixing parameters and find
\begin{eqnarray}
\nonumber  \sin^{2}\theta_{13}&=&\frac{1}{3} \left[1-\sin 2 \theta_{2} (\cos \theta_{1} \cos (\varrho_{3}-\varrho_{4}-\frac{\pi }{3})+\sin \theta_{1} \cos (\varrho_{3}+\frac{\pi }{3}))-\sin2 \theta_{1} \cos ^2\theta_{2} \cos (\varrho_{4}-\frac{\pi }{3})\right],\\
\nonumber \sin^2\theta_{12}&=&\sin^2\theta_{3}+\frac{\cos 2 \theta_{3} \left(\sin 2 \theta_{1} \cos \left(\varrho_{4}-\frac{\pi }{3}\right)+1\right)}{2+\sin 2 \theta_{2} (\cos \theta_{1} \cos (\varrho_{3}-\varrho_{4}-\frac{\pi }{3})+\sin \theta_{1} \cos (\varrho_{3}+\frac{\pi }{3}))+\sin2 \theta_{1} \cos ^2\theta_{2} \cos (\varrho_{4}-\frac{\pi }{3})},\\
\nonumber && +\frac{\sin 2 \theta_{3} \left(\sin \theta_{1} \cos \theta_{2} \cos (\varrho_{3}-\varrho_{4}-\frac{\pi }{3})-\cos \theta_{1} \cos \theta_{2} \cos (\varrho_{3}+\frac{\pi }{3})+\cos 2 \theta_{1} \sin \theta_{2} \cos (\varrho_{4}-\frac{\pi }{3})\right)}{2+\sin 2 \theta_{2} (\cos \theta_{1} \cos (\varrho_{3}-\varrho_{4}-\frac{\pi }{3})+\sin \theta_{1} \cos (\varrho_{3}+\frac{\pi }{3}))+\sin2 \theta_{1} \cos ^2\theta_{2} \cos (\varrho_{4}-\frac{\pi }{3})},\\
 \nonumber \sin^{2}\theta_{23}&=&\frac{\sin2 \theta_{2} (\cos \theta_{1} \cos (\varrho_{3}-\varrho_{4})+\sin \theta_{1} \cos \varrho_{3})+\sin 2 \theta_{1} \cos ^2\theta_{2} \cos \varrho_{4}+1}{2+\sin 2 \theta_{2} (\cos \theta_{1} \cos (\varrho_{3}-\varrho_{4}-\frac{\pi }{3})+\sin \theta_{1} \cos (\varrho_{3}+\frac{\pi }{3}))+\sin2 \theta_{1} \cos ^2\theta_{2} \cos (\varrho_{4}-\frac{\pi }{3})}\,, \\
\nonumber J_{CP}&=&-\frac{1}{6\sqrt{3}}\Big[\cos 2 \theta_{1} \cos 2 \theta_{2} \cos 2 \theta_{3}+\sin2 \theta_{1}\sin\theta_2(1-3/2\cos^2\theta_2)\sin 2 \theta_{3} \\
\nonumber &&\qquad~+\cos \theta_{2} \left(\sin 2 \theta_{3} \left(\cos ^2\theta_{1}-\sin ^2\theta_{1} \sin ^2\theta_{2}\right)+\sin 2 \theta_{1} \sin \theta_{2} \cos 2 \theta_{3}\right) [\sin \theta_{1} \cos (\varrho_{3}+\varrho_{4}) \\
\label{eq:mixing_para_caseV2} &&\qquad-\cos \theta_{1} \cos (\varrho_{3}-2 \varrho_{4})]+\cos ^3\theta_{2} \sin 2 \theta_{3} (\cos \theta_{1} \cos (\varrho_{3}-2 \varrho_{4})-\tan \theta_{2} \cos (2 \varrho_{3}-\varrho_{4}))\Big].
\end{eqnarray}
Since $U_{V,2}$ and $U_{V,1}$ are related through the permutation of the first and second rows, the rephasing invariants $I^{\alpha}_{\text{NO}, \text{IO}}$ for $U_{V,2}$ can be obtained from Eq.~\eqref{eq:Ia_V} by interchanging the expressions of $I^{e}_{\text{NO}, \text{IO}}$ and $I^{\mu}_{\text{NO}, \text{IO}}$. The third mixing matrix $U_{V,3}$ can be easily obtained by exchanging the second and third rows of $U_{V, 2}$. As a consequence, $U_{V, 2}$ and $U_{V, 3}$ lead to the same reactor and solar mixing angles, while the atmospheric one changes from $\theta_{23}$ to $\pi/2-\theta_{23}$, i.e., $\sin^2\theta_{23}$ is replaced by $\cos^2\theta_{23}$ in Eq.~\eqref{eq:mixing_para_caseV2}, and the Dirac phase changes from $\delta$ to $\pi+\delta$ such that the overall sign of the Jarlskog invariant $J_{CP}$ becomes opposite. Furthermore, the CP invariants can be  obtained from Eq.~\eqref{eq:Ia_V} by replacing $I^{e}_{\text{NO}, \text{IO}}\rightarrow I^{\tau}_{\text{NO}, \text{IO}}$, $I^{\mu}_{\text{NO}, \text{IO}}\rightarrow I^{e}_{\text{NO}, \text{IO}}$ and $I^{\tau}_{\text{NO}, \text{IO}}\rightarrow I^{\mu}_{\text{NO}, \text{IO}}$.

\item[~~(\uppercase\expandafter{\romannumeral6})]{$G_{l}=\left\langle ac^{s}d^{t}\right\rangle$, $X_{\nu}=\rho_{\mathbf{3}}(bc^{x}d^{-x})$}

In this case, the PMNS mixing matrix takes the following form
\begin{eqnarray}
\nonumber U_{VI}&=&\sqrt{\frac{2}{3}}
\begin{pmatrix}
 \frac{e^{i \varrho_{5}}}{\sqrt{2}} &~ \sin\varrho_{6} ~& \cos \varrho_{6} \\
 -\frac{e^{i \varrho_{5}}}{\sqrt{2}} &~ \cos \left(\frac{\pi }{6}-\varrho_{6}\right) ~& \sin \left(\frac{\pi }{6}-\varrho_{6}\right) \\
 \frac{e^{i \varrho_{5}}}{\sqrt{2}} &~ \cos \left(\varrho_{6}+\frac{\pi }{6}\right) ~& -\sin \left(\varrho_{6}+\frac{\pi }{6}\right) \\
\end{pmatrix}O_{3\times3}(\theta_{1},\theta_{2},\theta_{3})\widehat{X}^{-\frac{1}{2}}_{\nu}  \\
\label{eq:PMNS_VI} &=& \sqrt{\frac{2}{3}}
\begin{pmatrix}
 \frac{e^{i \varrho_{5}}}{\sqrt{2}} &~ 0 ~& 1\\
 -\frac{e^{i \varrho_{5}}}{\sqrt{2}} &~ \cos \frac{\pi }{6} ~& \sin \frac{\pi }{6} \\
 \frac{e^{i \varrho_{5}}}{\sqrt{2}} &~ \cos \frac{\pi }{6} ~& -\sin \frac{\pi }{6} \\
\end{pmatrix}O_{3\times3}(\theta_{1}-\varrho_{6},\theta_{2},\theta_{3})\widehat{X}^{-\frac{1}{2}}_{\nu}\,,
\end{eqnarray}
where the discrete parameters $\varrho_{5}$ and $\varrho_{6}$ depend on the choice of the residual symmetry as
\begin{equation}
\varrho_{5}=\frac{-s+2t-3 x}{n}\pi\,, \qquad \varrho_{6}=\frac{s}{n}\pi\,,
\end{equation}
whose values can be
\begin{equation}
\varrho_{5},\varrho_{6}~(\mathrm{mod}~2\pi)=0, \frac{1}{n}\pi, \frac{2}{n}\pi, \cdots, \frac{2n-1}{n}\pi\,.
\end{equation}
From Eq.~\eqref{eq:PMNS_VI} we can see that the parameter $\varrho_6$ is irrelevant since it can be absorbed into the free parameter $\theta_1$. Furthermore we find that $U_{VI}$ has several symmetry properties,
\begin{eqnarray}
\nonumber && P_{132}U_{VI}(\varrho_{5},\varrho_{6},\theta_{1},\theta_{2},\theta_{3})=\text{diag}(1,-1,-1)U_{VI}(\varrho_{5},-\varrho_{6}
,-\theta_{1},\theta_{2},-\theta_{3})\text{diag}(1,-1,1)\,, \\
\nonumber && P_{312}U_{VI}(\varrho_{5},\varrho_{6},\theta_{1},\theta_{2},\theta_{3})=\text{diag}(1,-1,-1)U_{VI}(\varrho_{5},\varrho_{6}+\frac{2\pi}{3}
,\theta_{1},\theta_{2},\theta_{3})\,, \\
\label{eq:permutations_VI}&&P_{231}U_{VI}(\varrho_{5},\varrho_{6},\theta_{1},\theta_{2},\theta_{3})=\text{diag}(-1,-1,1)U_{VI}(\varrho_{5},\varrho_{6}-\frac{2\pi}{3}
,\theta_{1},\theta_{2},\theta_{3})\,,
\end{eqnarray}
and
\begin{equation}
\label{eq:fund_int_rho5}U_{VI}(\varrho_{5}+\pi,\varrho_{6},\theta_{1},\theta_{2},\theta_{3})=U_{VI}(\varrho_{5},\varrho_{6},\theta_{1},-\theta_{2},-\theta_{3})
\text{diag}(-1,1,1)\,.
\end{equation}
Eq.~\eqref{eq:permutations_VI} implies that the six possible row permutations lead to the same mixing pattern, and Eq.~\eqref{eq:fund_int_rho5} indicates that the fundamental region of $\varrho_5$ is $[0, \pi)$. We can read off the mixing parameters from the mixing matrix $U_{VI}$ in Eq.~\eqref{eq:PMNS_VI} as follows
\begin{small}
\begin{eqnarray}
\nonumber && \sin^{2}\theta_{13}=\frac{1}{3} \left(1+\cos2\theta_{1}\cos^2\theta_{2}+\sqrt{2}\cos\theta_{1}\sin2\theta_{2}\cos\varrho_{5}\right),\\
\nonumber&&\sin^2\theta_{12}=\sin^2\theta_{3}+\frac{\sin\theta_{1}\left(2 \sin \theta_{1}\cos2\theta_{3}-\sin2\theta_{3}\left(\sqrt{2} \cos \theta_{2} \cos \varrho_5-2 \cos \theta_{1} \sin \theta_{2}\right)\right)}{2-\cos2\theta_{1}\cos^2\theta_{2}-\sqrt{2}\cos\theta_{1}\sin2\theta_{2}\cos\varrho_{5}}\,,\\
\nonumber &&\sin^2\theta_{23}=\frac{1-\cos\left(2\theta_{1}+\pi/3\right) \cos ^2\theta_{2}- \sqrt{2} \sin(\theta_{1}+\pi/6) \sin 2\theta_{2}  \cos \varrho_5}{2-\cos2\theta_{1}\cos^2\theta_{2}-\sqrt{2}\cos\theta_{1}\sin2\theta_{2}\cos\varrho_{5}}\,, \\
\label{eq:mixing_para_caseVI}&& J_{CP}=\frac{\cos\theta_{2}\sin 2 \theta_{3} \sin \varrho_{5} \left[4 \sin 3\theta_{1} \sin\theta_{2} \cot 2 \theta_{3}-\cos 3\theta_{1}  (\cos 2 \theta_{2}-3) -2 \sqrt{2} \sin 2 \theta_{2}  \cos \varrho_{5}\right]}{12 \sqrt{6}}\,,
\end{eqnarray}
\end{small}
where the redefinition of $\theta_{1}\rightarrow\theta_1+\varrho_6$ is used. Moreover the rephasing invariants involved in leptogenesis are found to be of the form
\begin{eqnarray}
\nonumber && I^{e}_{\text{NO}}=-\frac{\sqrt{2}}{3}  \sin\varrho_{5} (\cos\theta_{1}\sin\theta_{3}+\sin\theta_{1}\sin \theta_{2}\cos \theta_{3}),\\
\nonumber && I^{\mu}_{\text{NO}}=\frac{\sqrt{2}}{3}\sin\varrho_{5} \left[ \sin (\theta_{1}+\pi/6)\sin\theta_{3}+\sin (\theta_{1}-\pi/3)\sin \theta_{2} \cos \theta_{3}\right]\, \\
\nonumber && I^{e}_{\text{IO}}=\frac{\sqrt{2}}{3}\sin\theta_{1}\cos\theta_{2}\sin\varrho_{5}\,,\\
\nonumber && I^{\mu}_{\text{IO}}=-\frac{\sqrt{2}}{3}\sin (\theta_{1}-\pi/3)\cos\theta_{2}\sin\varrho_{5}\,, \\
\label{eq:Ia_VI}&& I^{\tau}_{\text{NO}}=-(I^{e}_{\text{NO}}+I^{\mu}_{\text{NO}}), \qquad I^{\tau}_{\text{IO}}=-(I^{e}_{\text{IO}}+I^{\mu}_{\text{IO}})\,.
\end{eqnarray}

\item[~~(\uppercase\expandafter{\romannumeral7})]{$G_{l}=\left\langle abc^{s}d^{t}\right\rangle$, $X_{\nu}=\rho_{\mathbf{3}}(c^{x}d^{y})$}

Similar to previous cases, the lepton mixing matrix is given by, up to permutations of rows and unphysical phases,
\begin{eqnarray}
\nonumber U_{VII}&=&\frac{1}{\sqrt{2}}
\begin{pmatrix}
e^{i \varrho_{7}} &~ -1 ~& 0 \\
e^{i \varrho_{7}} &~ 1 ~& 0 \\
 0 &~ 0 ~&\sqrt{2} \\
\end{pmatrix}O_{3\times3}(\theta_{1},\theta_{2},\theta_{3})\widehat{X}^{-\frac{1}{2}}_{\nu}\,,\\
\label{eq:PMNS_VII} &=&\text{diag}(-1,1,1)P_{132}\frac{1}{\sqrt{2}}
\begin{pmatrix}
 1 &~ 0 ~& -e^{i\varrho_{7}} \\
 0 &~ \sqrt{2} ~& 0 \\
 1 &~ 0 ~& e^{i \varrho_{7}} \\
\end{pmatrix}\left[P_{231}O_{3\times3}(\theta_{1},\theta_{2},\theta_{3})\right]\widehat{X}^{-\frac{1}{2}}_{\nu}\,,
\end{eqnarray}
with
\begin{equation}
\varrho_{7}=\frac{2 s-t+2 x-y}{n}\pi\,.
\end{equation}
Comparing Eq.~\eqref{eq:PMNS_VII} with Eq.~\eqref{eq:PMNS_III}, we can see this case gives rise to the same mixing pattern and the baryon asymmetry $Y_{B}$ as case III if all possible row permutations are taken into account.

\item[~~(\uppercase\expandafter{\romannumeral8})]{$G_{l}=\left\langle abc^{s}d^{t}\right\rangle$, $X_{\nu}=\rho_{\mathbf{3}}(bc^{x}d^{-x})$}

In this case, we find that the PMNS mixing matrix takes the form
\begin{eqnarray}
\nonumber U_{VIII}&=& \frac{1}{2}
\begin{pmatrix}
 -\sqrt{2}  &~ -ie^{i \varrho_{8}} ~& e^{i \varrho_{8}} \\
 \sqrt{2}  &~ -ie^{i \varrho_{8}} ~& e^{i \varrho_{8}} \\
 0 &~ i \sqrt{2}e^{-i \varrho_{8}} ~& \sqrt{2}e^{-i \varrho_{8}} \\
\end{pmatrix}O_{3\times3}(\theta_{1},\theta_{2},\theta_{3})\widehat{X}^{-\frac{1}{2}}_{\nu} \\
\label{eq:PMNS_VIII} &=& \frac{1}{2}
\begin{pmatrix}
 -\sqrt{2} &~ -i ~& 1 \\
 \sqrt{2} &~ -i ~& 1 \\
 0 &~ i \sqrt{2} ~& \sqrt{2} \\
\end{pmatrix}O_{3\times3}(\theta_{1}-\varrho_{8},\theta_{2},\theta_{3})\widehat{X}^{-\frac{1}{2}}_{\nu}\,,
\end{eqnarray}
with
\begin{equation}
\varrho_{8}=\frac{2 s-t+3 x}{n}\pi\,.
\end{equation}
Obviously the value of $\varrho_8$ is irrelevant since it can be absorbed into the free parameter $\theta_1$. Furthermore, the six possible row permutations lead to three independent mixing patterns which can be chosen as
\begin{equation}\label{eq:all_PMNS_VIII}
U_{VIII,1}=U_{VIII}, \qquad U_{VIII,2}=P_{132}U_{VIII}, \qquad U_{VIII,3}=P_{312}U_{VIII}\,.
\end{equation}
The reason is because $U_{VIII}$ fulfills the equality
\begin{equation}
P_{213}U_{VIII}(\varrho_{8},\theta_{1},\theta_{2},\theta_{3})=U_{VIII}(\varrho_{8},\theta_{1},-\theta_{2},-\theta_{3})\text{diag}(-1,1,1)\,.
\end{equation}
For the mixing matrix $U_{VIII,1}$, after the parameter $\theta_1$ is shifted into $\theta_1+\varrho_{8}$, we can read off the mixing parameters as
\begin{eqnarray}
\nonumber && \sin^{2}\theta_{13}=\frac{1}{8} \left(3-\cos 2 \theta_{2}-2\sqrt{2}\cos\theta_{1}\sin2\theta_{2}\right),\\
\nonumber&&\sin^2\theta_{12}=\sin^2\theta_{3}+\frac{2\left(\cos 2 \theta_{3}+\sqrt{2}\sin\theta_{1} \cos \theta_{2}\sin2\theta_{3}\right)}{5+\cos 2 \theta_{2}+2 \sqrt{2} \cos \theta_{1} \sin 2 \theta_{2}},\\
\nonumber  &&\sin^2\theta_{23}=\frac{3-\cos 2 \theta_{2}+2\sqrt{2}\cos \theta_{1}\sin2\theta_{2}}{5+\cos 2 \theta_{2}+2\sqrt{2}\cos \theta_{1}\sin2\theta_{2}}\,, \\
\label{eq:mixing_para_caseVIII1} && J_{CP}=\frac{1}{32\sqrt{2}}\left[4\sin\theta_1\sin2\theta_2\cos2\theta_3-\cos\theta_1\left(\cos\theta_2+3\cos3\theta_2\right)\sin2\theta_3\right]\,.
\end{eqnarray}
The CP invariants $I^{\alpha}_{\text{NO}, \text{IO}}$ $(\alpha=e, \mu, \tau)$ turn out to take the form
\begin{eqnarray}
\nonumber && I^{e}_{\text{NO}}=\frac{1}{4} \left[\cos \theta_{2} \cos \theta_{3}+\sqrt{2} (\sin \theta_{1}\sin \theta_{3}-\cos \theta_{1}\sin\theta_{2}\cos \theta_{3})\right],\\
\nonumber && I^{\mu}_{\text{NO}}=\frac{1}{4} \left[\cos \theta_{2} \cos \theta_{3}-\sqrt{2} (\sin \theta_{1}\sin \theta_{3}-\cos \theta_{1}\sin \theta_{2} \cos \theta_{3} )\right]\, \\
\nonumber && I^{e}_{\text{IO}}=\frac{1}{4} \left(\sin\theta_{2}+\sqrt{2}\cos\theta_{1}\cos\theta_{2}\right)\,, \quad I^{\mu}_{\text{IO}}=\frac{1}{4} \left(\sin \theta_{2}-\sqrt{2}\cos \theta_{1} \cos \theta_{2} \right)\,, \\
\label{eq:Ia_VIII}&& I^{\tau}_{\text{NO}}=-(I^{e}_{\text{NO}}+I^{\mu}_{\text{NO}})=-\frac{1}{2} \cos \theta_{2} \cos \theta_{3}, \quad
I^{\tau}_{\text{IO}}=-(I^{e}_{\text{IO}}+I^{\mu}_{\text{IO}})=-\frac{1}{2}\sin\theta_{2}\,.
\end{eqnarray}
The second mixing matrix $U_{VIII, 2}$ is related to $U_{VIII, 1}$ through the permutation of the second and third rows. As a consequence, the expressions for $\theta_{12}$ and $\theta_{13}$ coincide with Eq.~\eqref{eq:mixing_para_caseVIII1}, the overall sign of $J_{CP}$ is reversed, while the atmospheric mixing angle $\theta_{23}$ changes into
\begin{equation}
\sin^2\theta_{23}=\frac{4\cos^2\theta_{2}}{5+\cos2\theta_{2}+2\sqrt{2}\cos\theta_{1}\sin2\theta_{2}}\,,
\end{equation}
Moreover the rephase invariants can be obtained from Eq.~\eqref{eq:Ia_VIII} by interchanging $I^{\mu}_{\text{NO, IO}}$ and $I^{\tau}_{\text{NO, IO}}$. Finally we proceed to the third mixing matrix $U_{VIII, 3}$, we can extract the following results for the mixing angles
\begin{equation}
\sin^2\theta_{13}=\frac{1}{2}\cos^2\theta_2,\quad \sin^2\theta_{12}=\frac{1}{2}+\frac{\cos^2\theta_2\cos2\theta_3}{3-\cos2\theta_2},\quad \sin^2\theta_{23}=\frac{1}{2}-\frac{\sqrt{2}\cos\theta_1\sin2\theta_2}{3-\cos2\theta_2}\,,
\end{equation}
which implies
\begin{equation}
\label{eq:caseVIII_sum}\left|\sin^2\theta_{12}-\frac{1}{2}\right|\leq\frac{1}{2}\tan^2\theta_{13},\quad \left|\sin^2\theta_{23}-\frac{1}{2}\right|\leq\tan\theta_{13}\sqrt{1-\tan^2\theta_{13}}~\,.
\end{equation}
Hence the experimental data~\cite{Capozzi:2013csa} on $\theta_{13}$ and $\theta_{12}$ can not be accommodated simultaneously without higher order corrections for this mixing matrix.

\end{description}

\end{appendix}

\end{document}